\RequirePackage{fix-cm}

\documentclass{svjour3}     
\smartqed 
\usepackage{setspace}
\usepackage{tabularx}
\usepackage{graphicx}
\usepackage{blindtext}
\usepackage[utf8]{inputenc}
\usepackage[table]{xcolor}
\usepackage{algorithm,setspace} 
\usepackage{algpseudocode} 
\usepackage{tcolorbox}
\usepackage{framed}
\usepackage{mdframed,lipsum}
\usepackage{amssymb, amsmath} 
\usepackage{caption}
\usepackage{subcaption}
\usepackage{url}
\usepackage[all]{nowidow}
\usepackage{makecell}
\usepackage{listings}

\begin{document}

\title{Learning How to Search: Generating Effective Test Cases Through Adaptive Fitness Function Selection\thanks{This research was supported by Vetenskapsr{\aa}det grant 2019-05275.}}

\titlerunning{Adaptive Fitness Function Selection}     

\author{Hussein Almulla         \and
        Gregory Gay
}

\institute{Hussein Almulla \at
              University of South Carolina, Columbia, SC, USA \\
              \email{halmulla@email.sc.edu}          
           \and
           Gregory Gay \at
              Chalmers and the University of Gothenburg, Gothenburg, Sweden \\
              \email{greg@greggay.com}
}

\date{Received: date / Accepted: date}

\maketitle

\begin{abstract}
Search-based test generation is guided by feedback from one or more fitness functions---scoring functions that judge solution optimality. Choosing informative fitness functions is crucial to meeting the goals of a tester. Unfortunately, many goals---such as forcing the class-under-test to throw exceptions, increasing test suite diversity, and attaining Strong Mutation Coverage---\textit{do not} have effective fitness function formulations. We propose that meeting such goals requires treating fitness function identification as a secondary optimization step. An \textit{adaptive} algorithm that can vary the selection of fitness functions could adjust its selection throughout the generation process to maximize goal attainment, based on the current population of test suites. To test this hypothesis, we have implemented two reinforcement learning algorithms in the EvoSuite unit test generation framework, and used these algorithms to dynamically set the fitness functions used during generation for the three goals identified above. 

We have evaluated our framework, EvoSuiteFIT, on a set of Java case examples. EvoSuiteFIT techniques attain significant improvements for two of the three goals, and show limited improvements on the third when the number of generations of evolution is fixed. Additionally, for two of the three goals, EvoSuiteFIT detects faults missed by the other techniques. The ability to adjust fitness functions allows strategic choices that efficiently produce more effective test suites, and examining these choices offers insight into how to attain our testing goals. We find that adaptive fitness function selection is a powerful technique to apply when an effective fitness function does not already exist for achieving a testing goal. 
\keywords{Automated Test Generation \and Search-Based Test Generation \and Reinforcement Learning \and Hyperheuristic Search}
\end{abstract}



\section{Introduction}\label{sec:intro}

The testing of software is crucial, as testing is our primary means of ensuring that complex software is robust and operates correctly~\cite{Pezze06:testing}. However, testing is an expensive task that can consume much of the development budget~\cite{Pezze06:testing}. Test creation is an effort-intensive task that requires the selection of sequences of program input and the creation of oracles that judge the correctness of the resulting execution~\cite{Harman13:oraclesurvey}. If test creation could be even partially automated, the effort and cost of testing could be significantly reduced.

One promising method of automating test creation is search-based test generation~\cite{Anand13:Orchestrated,McMinn04:SBTesting}. Test input selection can naturally be seen as a search problem~\cite{Harman01:SBSE}. Testers approach input selection with a \textbf{goal} in mind---perhaps they would like to cause the program to crash, maximize code coverage, detect a set of known faults, or any number of other potential goals. Of the near-infinite number of possible inputs that could be provided to a program, the tester seeks those that meets their chosen goal. This search can then be automated. Given a measurable goal, a metaheuristic optimization \textbf{algorithm} can systematically sample the space of possible test input and manipulate those samples, guided by feedback from one or more \textbf{fitness functions}---scoring functions that judge the optimality of the chosen input~\cite{Gay19:fitness}. In other words: \textit{\textbf{algorithm + fitness functions $\implies$ goal}}.

Effective search-based generation relies on the selection of the correct sampling mechanism---the algorithm---and, perhaps more importantly, the proper feedback mechanisms---the fitness functions. Fitness functions shape the test suites generated by the search process to have properties promoted by those functions. The fitness functions chosen, in \textit{normal} use, are expected to embody the overall goals of the tester. By offering feedback on the quality of the generated solutions, they ensure that test suites converge on these goals. The best fitness functions rapidly increase attainment of the goal by both differentiating \textit{good} solutions from \textit{bad} solutions and by offering the feedback needed to locate even better solutions.

Consider, for example, Branch Coverage---a measurement of \textit{how many parts} of the code have been executed. For each program statement that can cause the execution path to diverge---such as \texttt{if} and \texttt{case} statements---test input should ensure that all potential outcomes are executed~\cite{Pezze06:testing}. If our \textit{goal} is to achieve 100\% Branch Coverage, there are multiple fitness functions that could be used to guide the algorithm towards meeting that goal. A simple fitness function could measure the attained coverage. A test suite that attains 75\% Branch Coverage is inherently better than one that attains 50\% Branch Coverage. This tells the algorithm which test suites to favor, leading to higher and higher attainment of Branch Coverage. 

However, there is a more informative fitness function that leads to faster attainment. Instead, we could take each branch outcome we wish to cover and judge \textit{how close} the chosen test input was to achieving that outcome. If we execute an expression ``\texttt{if (x == 5)}'' with the value of \texttt{x} set to 3, and we seek a \texttt{true} outcome, then \texttt{x} needs to be incremented by 2. This suggests the magnitude of change needed to reach the desired outcome~\cite{McMinn04:SBTesting}. Fitness functions based on this concept, the \textit{branch distance}~\cite{Arcuri13:Normalize}, offer additional feedback by offering both a measurement of \textit{how much} of the goal has been met as well as clues on how to \textit{attain further coverage}. If our goal is Branch Coverage, we have known and effective fitness functions that enable attainment of that goal. Unfortunately, many goals \textit{do not} have a known or effective fitness function. In fact, many goals do not inherently lend themselves to such a formulation. 

To illustrate, consider the following three goals:
\begin{itemize}
\setlength{\itemsep}{1pt}
  \setlength{\parskip}{0pt}
  \setlength{\parsep}{0pt}
    \item \textbf{Exception Discovery:} ``Causing the program to crash'' is a common goal in testing. The number of crashes discovered is often measured by counting the number of \textit{exceptions}---program-interrupting error messages---thrown during test execution~\cite{Robillard00:DRJ}. Exceptions indicate fault and abnormal operating conditions in programs. Thus, tests that trigger exceptions are valuable. 
    \item \textbf{Test Suite Diversity:}  When testing, it is generally impossible to try every input. It follows, then, that \textit{diverse} test input is more effective than similar input~\cite{Neto18:Visual,Shahbazi15:Diversity}. This intuition has led to effective automated test generation, prioritization, and reduction~\cite{Neto18:Visual}.
    \item \textbf{Strong Mutation Coverage:} Mutation testing is a practice where synthetic faults (mutants) are injected into the code. If test suites detect these faults, they are thought to be more robust to real faults. In \textit{Weak} Mutation Coverage, a mutant is detected if execution reaches the infected expression and the outcome of that expression differs from the original program---i.e., the state is infected. \textit{Strong} Mutation Coverage requires that the infected state propagates to the program output, offering clear evidence that the fault was detected~\cite{Lind7528953}. 
\end{itemize}

All three are valid, \textit{measurable}, goals for test suite generation. In principle, all three should be reasonable targets for search-based test generation. However, all three have properties that make them difficult to optimize directly:
\begin{itemize}
\setlength{\itemsep}{1pt}
  \setlength{\parskip}{0pt}
  \setlength{\parsep}{0pt}
    \item As we cannot know how many or what exceptions are possible to throw, ``throw more exceptions'' is not a goal that translates into an informative fitness representation. Prior work has proposed counting thrown exceptions as a fitness function~\cite{Rojas15:Combining}. Unfortunately, this count yields poor results, as it offers the algorithm no guidance for improving its guesses~\cite{Gay19:fitness,Gay17:ICST,Gay17:Combos}.
    \item While numerous diversity metrics exist---for example, the \textit{Levenshtein distance}~\cite{Shahbazi15:Diversity} measures the number of operations needed to convert one string to another---these metrics tend to serve as poor fitness functions, as little feedback is offered to suggest how to gain \textit{more} diversity. 
    \item Weak Mutation Coverage can be optimized using a variant of the branch distance, which measures how close execution came to reaching the mutated line and corrupting the program state~\cite{Fraser14:Mutation}. It is more difficult to offer feedback on how to propagate corruption to the output. Current fitness functions offer probabilistic estimations of propagation~\cite{Fraser14:Mutation,Papadakis2013}. However, such estimations are generally too coarse-grained to accurately guide the search. 
\end{itemize}

This does not mean there is no way to effectively achieve such goals. Rather, we simply do not \textit{yet} know what fitness functions will be effective. There are many fitness functions available for use in search-based test generation. If we do not know of an effective fitness function that we can optimize to \textit{directly} achieve a goal, it may be possible to identify fitness functions that \textit{indirectly} achieve our goal. Careful selection of one or more of those functions could yield high goal attainment. For example, if optimizing the exception count fails to produce test suites that trigger exceptions, optimizing different functions (e.g., targeting both branch distance and exception count) might achieve that goal. 

\textbf{We simply need to identify that selection}. There are many combinations of fitness functions that could be selected, and the ``correct'' choices may be specific to the goal and system/class-under-test (SUT/CUT). In fact, the ``correct'' choices may even vary \textit{during} test case generation, as search-based processes evolve test suites over time in a stateful process. Therefore, \textbf{we seek a systematic method of automatically identifying and adapting the selection of fitness functions} that is appropriate for a variety of high-level testing goals. 

A class of search-based test generation approaches are known as \textit{hyperheuristic}, or self-adaptive, approaches~\cite{Jia15:HHICSE,Guizzo15:HMI}. These approaches incorporate a learning phase in order to automatically tune the search strategy. Hyperheuristic search has been used, for example, to change parameters of the search algorithm during evolution to improve solution quality~\cite{Jia15:HHICSE}. We propose a hyperheuristic search that strategically adjusts the chosen fitness functions throughout the generation process to maximize attainment of a desired goal.

Through the use of reinforcement learning~\cite{Sutton2018}, this approach is able to select the most appropriate fitness functions for a CUT and testing goal, and adjust that set as needed during generation. In this process, a measurement---representing the real goal of the search---is targeted as a high-level \textit{reward function}. A reinforcement learning agent selects fitness functions, and after evolving test suites using these functions for a set number of generations, the change in the reward score will be evaluated and the agent decides whether to continue using the set of functions known to best improve the reward (\textit{exploitation}) or to try different functions in order to refine expectations on the change in reward (\textit{exploration}). We refer to this hyperheuristic as \textbf{adaptive fitness function selection} (AFFS). 

We have implemented two reinforcement learning algorithms---Upper Confidence Bound (UCB) and Differential Semi-Gradient Sarsa (DSG-Sarsa)~\cite{Sutton2018}---in the EvoSuite unit test generation framework for Java~\cite{Rojas17:wholesuite}. We refer to the modified framework as \textbf{EvoSuiteFIT}. We have evaluated EvoSuiteFIT for each of the three goals listed above on a set of Java case examples in terms of (a) the ability to produce test suites that achieve the targeted goal and (b) the ability of the generated suites to detect real faults. In each case, we compare the two reinforcement learning approaches to three baselines: (a) current practice---a fitness function based on the goal that may not offer sufficient feedback, (b) a set of multiple fitness functions---the full set of functions that AFFS can choose among for that goal---that serves as a ``best guess'' a human might make at a combination of fitness functions that would produce effective test suites, and (c), a set of fitness functions randomly selected from the choices available to AFFS. We have found:
\begin{itemize}
\setlength{\itemsep}{1pt}
  \setlength{\parskip}{0pt}
  \setlength{\parsep}{0pt}
    \item Both EvoSuiteFIT techniques outperform all baselines with at least medium effect size for the goals of exception discovery and suite diversity---attaining improvements of up to 107.14\% in goal attainment. For the goal of Strong Mutation Coverage, no technique demonstrates significant improvements. When the search budget is a fixed number of generations rather than time, both EvoSuiteFIT techniques slightly outperform the baselines (up to 8.33\% improvement). However, the effect size is still negligible. 
    \item Both EvoSuiteFIT techniques detect faults missed by the other techniques for the exception discovery goal (up to 259.90\% improvement). UCB is able to detect more faults for the Strong Mutation goal (12.50\% improvement), and when the number of generations is fixed, both EvoSuiteFIT approaches outperform the baselines (up to 50.00\% improvement). Both techniques are outperformed by the random baseline for the diversity goal (34.74\% difference), but outperform the other baselines. 
    \item We find that AFFS is an appropriate technique to apply when an effective fitness function does not already exist for the targeted goal. However, AFFS requires a reward function that is fast to calculate, or requires additional time for test generation. Further, the effect of AFFS is limited by the span of fitness functions available to choose from. If none of the chosen functions correlate to the goal of interest, then improvements in goal attainment will be limited.
    \item Improvements in fault detection may arise because of higher attainment of goals thought to have a positive relationship with fault detection likelihood, optimizing multiple fitness functions---but avoiding needlessly complex and conflicting functions---and changing fitness functions as the suite evolves rather than applying all functions at once. However, higher goal attainment does not ensure fault detection.
    \item While reinforcement learning adds overhead to test generation, EvoSuiteFIT is often \textit{faster} than the default static configuration because the ability to avoid calculation of unhelpful fitness functions mitigates this overhead (up to 94.27\% faster than baselines). Further, feedback from effective fitness functions can help control computational costs. 
    \item The ability to adjust the fitness functions at regular intervals allows EvoSuiteFIT to make strategic choices that refine the test suite and allows us to attain a deeper understanding of the properties that link to goal attainment and how fitness functions can work together to imbue those properties. Fitness function combinations that are ineffective in a static context may be effective when used by AFFS to diversify a pre-evolved population of suites.
\end{itemize}

We have previously proposed adaptive fitness function selection, and demonstrated its potential for increasing the number of discovered exceptions~\cite{Gay20:RL}. We also have published a small pilot study for the Gson case examples and the diversity goal~\cite{Gay20:DivRL}. This publication extends both studies in significant ways:
\begin{itemize}
\setlength{\itemsep}{1pt}
  \setlength{\parskip}{0pt}
  \setlength{\parsep}{0pt}
    \item We perform more extensive experiments and analyses for the exception discovery goal, and perform the first full experiments for the diversity goal.
    \item We add a third testing goal---Strong Mutation Coverage.
    \item We have added a third baseline---random selection of fitness functions.
    \item We perform cross-goal analyses to better understand the capabilities of AFFS, leading to a richer discussion than in the previous studies.
\end{itemize}

Under the correct conditions, the use of AFFS allows EvoSuiteFIT to identify combinations of fitness functions effective at achieving our testing goals, and strategically vary that set of functions throughout the ongoing generation process. We hypothesize that other goals without known effective fitness function representations could also be maximized in a similar manner. We make EvoSuiteFIT\footnote{EvoSuiteFIT is available from \url{https://github.com/hukh/evosuite/tree/evosuitefit}.} and our empirical data\footnote{We make our research data available at \url{https://doi.org/10.5281/zenodo.4524786}.} available to others for use in research or practice. 
\section{Background}\label{sec:bg}

\subsection{Unit Testing}

Testing can be performed at various levels of granularity. In this research, we are focused on \textit{unit testing}~\cite{Pezze06:testing}. Unit testing is where the smallest segment of code that can be tested in isolation from the rest of the system---often a class~\cite{shamshiri15:generation}---is tested. Unit tests are written as executable code. We refer to a purposefully grouped set of test cases as a \textit{test suite}. When code changes, developers can re-execute the test suite to make sure the code works as expected after changing. Unit testing frameworks exist for many programming languages, such as JUnit for Java, and are integrated into most development environments. 

\begin{figure}[!t]
	\centering
	\begin{lstlisting}[language=Java,basicstyle={\scriptsize\ttfamily}]
	@Test
	public void testPrintMessage() {
	    String str = "Test Message";
	    TransformCase tCase = new TransformCase(str);
	    String upperCaseStr = str.toUpperCase();
	    assertEquals(upperCaseStr, tCase.getText());
	}
	\end{lstlisting}
	\caption{Example of a unit test case written using the JUnit notation for Java.}
	\label{fig:testcase}
\end{figure}

An example of a unit test, written in JUnit, is shown in Figure~\ref{fig:testcase}. A unit test consists of a \textit{test sequence (or procedure)}--a series of method calls to the CUT--with \textit{test input} provided to each method. Then, the test case will validate the output of the called methods and the class variables against a set of encoded expectations---the \textit{test oracle}---to determine whether the test passes or fails. In a unit test, the oracle is typically formulated as a series of assertions on the values of method output and class attributes~\cite{Harman13:oraclesurvey}. In the example in Figure~\ref{fig:testcase}, the \textit{test input} consists of passing a string to the constructor of the \texttt{TransformCase} class, then calling its \texttt{getText()} method. This method should transform the string to upper-case. To ensure this is the case, we use an assertion to check whether the output of the call is equal to an upper-case version of the provided string. 

\subsection{Search-Based Test Generation}

Automation has a critical role in controlling the cost of testing~\cite{Orso14:STR,Almasi17:IndustrialEval}. One particular task that has seen great attention is the selection of test input. Exhaustively applying all possible inputs is infeasible due to enormous number of possibilities. Therefore, \textit{which} input are tried becomes important. A promising method is \textit{search-based test input generation}.

Test input selection can naturally be seen as a search problem~\cite{Harman01:SBSE}. Out of all of the test cases that could be generated for a class, we want to select---systematically and at a reasonable cost---those that meet our goals~\cite{McMinn04:SBTesting,Ali10:SBST}. Given a testing goal and a scoring function denoting \textit{closeness to the attainment of that goal}---called a \textit{fitness function}---optimization algorithms can sample from a large and complex set of options as guided by a chosen strategy (the \textit{metaheuristic})~\cite{Bianchi09:Optimization}. 

Metaheuristics are often inspired by natural phenomena, such as swarm behavior~\cite{dorigo1997ant} or evolution~\cite{john1992adaptation}. While the particular details vary between algorithms, the general process employed by a metaheuristic is as follows: (1) One or more solutions are generated, (2), The solutions are scored according to the fitness function, and (3), this score is used to reformulate the solutions for the next round of evolution. This process continues over multiple generations, ultimately returning the best-seen solutions. The metaheuristic (genetic algorithm, simulated annealing, etc.) overcomes the shortcomings of a purely random selection when selecting test input by using a deliberate strategy to traverse the input space, gravitating towards ``good'' input and discarding ``bad'' input---as determined by the fitness function---through the incorporation of fitness feedback and mechanisms for manipulating a population of solutions. By determining how solutions are evolved and selected over time, the choice of metaheuristic impacts the quality and efficiency of the search process~\cite{Feldt15:GA}. 

In search-based test generation, the fitness functions capture testing objectives and guides the search. Through this guidance, the fitness function has a major impact on the quality of the solutions generated. Functions must be efficient to execute, as they will be calculated thousands of times over a search. Yet, they also must provide enough detail to differentiate candidate solutions and guide the selection of optimal candidates. Structural coverage of the source code is a common optimization target, as coverage criteria can be straightforwardly transformed into efficient, informative fitness functions~\cite{Arcuri13:Normalize}. Search-based generation often can achieve higher coverage than developer-created tests~\cite{Fraser13:AWT}. Due to the non-linear nature of software, resulting from branching control structures, a real-world program's search space is large and complex~\cite{Ali10:SBST}. Metaheuristic search---by strategically sampling from that space---can scale to larger problems than many other generation algorithms~\cite{Malburg11:SBDSE}. Such approaches have been applied to a wide variety of testing goals and scenarios~\cite{Ali10:SBST}.

A special class of search-based approaches are known as \textit{hyperheuristic}, or self-adaptive, approaches~\cite{Jia15:HHICSE,Gay20:RL,Gay20:DivRL}. These approaches incorporate a learning phase in order to automatically tune the search strategy towards particular problem instances~\cite{balera2019systematic}. Hyperheuristic search has been used, for example, to change parameters of the metaheuristic during evolution~\cite{Jia15:HHICSE}. 

Hyperheuristic search can, essentially, be thought of as ``using a heuristic to choose a heuristic.'' A hyperheuristic approach introduces an automated high-level search that can explore the lower-level space of options available to tune the metaheuristic algorithm, looking for the best options to solve the targeted problem. These options may include aspects of the metaheuristic such as population tuning mechanics (e.g., the crossover and mutation rates of a genetic algorithm) or, in this study, the choice of fitness functions. The metaheuristic operates directly on the problem space, attempting to optimize its own effectiveness using the options selected by the high-level hyperheuristic layer~\cite{drake2020recent,Kumari16:HHC}. 

Hyperhueristic approaches can be divided into two types---selection and generation. Selection-based approaches select low-level heuristics from a preexisting set. Generation-based approaches create new heuristics using the components of existing heuristics as building blocks~\cite{Burke2019Revisited}. Selection-based hyperheurstics are more common, especially in software testing research, as they are often easier to implement and are suited to a wider range of problems~\cite{balera2019systematic}. However, generation-based approaches may yield better solutions when applied successfully. In this research, we use a selection-based hyperheuristic approach, but may explore generation-based approaches in future work.

\subsection{Reinforcement Learning}

Reinforcement learning focuses on identifying an action that maximizes return, measured using a problem-specific numerical reward score. This return is gained after an agent interacts with the specified environment to reach the desired goal. To understand reinforcement learning, consider the $n$-armed bandit problem~\cite{Katehakis87:Bandit}. This problem describes a situation where you are repeatedly faced with a choice of $n$ different options. After each selection, you receive a reward chosen from a probability distribution dependent on the action selected. Reinforcement learning algorithms are designed to learn the optimal choice of action to maximize the reward earned~\cite{Sutton2018}.

Each action has an expected reward when it is selected. Over time, the reinforcement learning agent will try different actions and refine its estimations of their value. During each round, the agent will choose an action based on the expected reward of applying it in the current problem state. After applying the action, the agent will receive a reward value. The agent will update the expected reward for the chosen set using the new information---updating the \textit{policy} it uses to choose the next action.  

Reinforcement learning manages the trade-off between two concepts---exploration and exploitation---to maximize the reward. An agent must \textit{explore}---choosing different actions---until it reaches the point where it can \textit{exploit} that knowledge---favoring the actions known to provide a higher reward. At any time, there will be a portfolio with the greatest estimated value. If the algorithm selects that portfolio, it exploits its current knowledge to gain immediate reward. If instead, it chooses a portfolio with an unknown or potentially lower reward, it is exploring the option space to improve its estimate of a portfolio's value. Reinforcement learning is designed to effectively balance exploration and exploitation~\cite{Jia15:HHICSE,Sutton2018,Jia15:HHSBST}.

In this work, we consider two different types of reinforcement learning---tabular solution methods and approximation solution methods~\cite{Sutton2018}. Tabular methods are generally used in cases where states and action apace are small enough so they can be represented in table or array. For that, a method can find the exact solution for the given problem. However, finding an exact solution is not feasible when the state space is large or continuous. In this case, approximation methods attempt to find an approximate solution rather than a specific one by generalizing from previously encountered states~\cite{Sutton2018}.

\section{Technical Approach and Implementation}\label{sec:approach}

In theory, fitness functions should be selected to maximize attainment of the tester's overall goals. However, this is not always straightforward. In practice, many goals do not translate cleanly to effective fitness function representations---ones that offer detailed feedback to the search process to enable rapid optimization. 

Consider the three goals that we are focused on in this research: exception discovery, test suite diversity, and Strong Mutation Coverage. All three have existing fitness function representations---a simple count of exceptions thrown, the Levenshtein distance, and a probabilistic estimation of state propagation to output. All three of these fitness representations have weaknesses. Consider the fitness function for exception discovery. Counting the thrown exceptions meets the \textit{technical} requirement of a fitness function, in that it can distinguish a test suite that throws exceptions from one that does not. However, it offers no actionable feedback to the search. Finding new exceptions requires blind guessing. 

The other two goals are also difficult to optimize. The Levenshtein distance can effectively \textit{minimize} the distance between two strings, as the actions a test generation takes have a direct and learn-able impact on this score. It is less helpful when one wants to \textit{maximize} the distance---to make the test suites more diverse---and when it is not clear how to cause the most effective change in this score by manipulating method calls to the class-under-test. Similarly, Strong Mutation Coverage requires that a triggered fault propagate to an observable failure in the output. Offering feedback on the likelihood of propagation is a complex problem, and current approaches only provide course-grained estimations that insufficiently guide the search~\cite{Fraser14:Mutation,Papadakis2013}. 

All three of these goals contain \textit{elements that are either unknowable upfront, or are difficult to estimate}. Optimization of these functions \textit{does not map to the actions available} to the test case generator in a way that can be easily predicted, often requiring specific actions not suggested by fitness function feedback. Such properties are common when examining the goals a tester might have when creating test suites. In this research, our aim is not to find a better way to meet these three specific goals. Rather, our aim to develop a systematic approach capable of better meeting \textit{any goal} that does not already have an effective fitness function. 

Even if existing fitness functions are insufficient, \textit{such goals can still be met}. The existing fitness functions simply do not provide sufficient feedback. The problem to be solved is how \textit{to provide that feedback}. Search-based generation can simultaneously target multiple fitness functions~\cite{Gay14:coverage}. Each fitness function further shapes the test suite, imbuing it with additional properties. This offers an opportunity to provide that missing feedback. We can augment---or even replace---the existing fitness representations with fitness functions that better direct the search towards optimization of our core, high-level goal. 

We propose that careful selection---at different points in the generation process---of the set of fitness functions could result in test suites that better meet our goals than existing baselines. If this is true, identifying these fitness functions becomes a secondary search problem, tackled as an additional hyperheuristic optimization within the normal test generation process~\cite{Jia15:HHSBST}. We propose the use of reinforcement learning techniques to adapt the set of fitness functions over the generation process at regular intervals in service of matching the chosen CUT and a measurable testing goal. Given a measurable goal, each action---each choice of one or more fitness functions---has an expected reward when it is selected. \textit{If we use this function combination, we will increase attainment of our goal}. Because test generation is a stateful process---the population of test suites at round $N$ depends on the population from round $N-1$---reinforcement learning affords not just an opportunity to identify effective fitness functions, but to strategically adjust the functions based on the changing population of test suites. We refer to this process as \textbf{adaptive fitness function selection} (AFFS). 

In this work, we have implemented AFFS by extending the EvoSuite test generation framework~\cite{Rojas17:wholesuite} with two online reinforcement learning algorithms---Upper Confidence Bound (UCB) and Differential Semi-Gradient Sarsa (DSG-Sarsa)~\cite{Sutton2018}. EvoSuite is a search-based unit test generation framework for Java that uses a genetic algorithm to evolve test suites over a series of generations, forming new populations each generation by retaining, mutating, and combining the fittest solutions. It is actively maintained and has been successfully applied to a variety of projects~\cite{shamshiri15:generation}. In this study, we implemented AFFS in EvoSuite version 1.0.7. We call our approach \textbf{EvoSuiteFIT}.

In Sections~\ref{sec:ucb}-\ref{sec:sarsa}, we will explain the UCB and DSG-Sarsa algorithms. In Section~\ref{sec:evo}, we give an overview of the EvoSuite test generation framework. Finally, in Section~\ref{sec:imp}, we explain how AFFS is implemented into EvoSuite and present an overview of new fitness and reward functions implemented as part of our approach.

\subsection{Upper Confidence Bound (UCB) Algorithm}\label{sec:ucb}

\begin{algorithm}[!t]
\scriptsize
	\begin{algorithmic}[1]
	     \State \textbf{Initialization:}
	     \State max = 0
	     \For {a = 0 \ldots number of actions} \Comment{Initialize for all actions}
    	        \State numberTimesActionSelected$_a$ = 0 
	             \State sumReward$_a$ = 0
          \EndFor
         \State \textbf{Each time an action is selected:}  
	     \If{numberGenerationsElapsed $<$ numberActionsTried}
	        \State action = getAction(numberGenerationsElapsed) \Comment{Try all actions once before using RL}
	        \State numberActionsTried = numberActionsTried + 1
	     \Else
	         \For {a = 0 \ldots number of actions}
    	        \State upperBound = 0
	            
	            \If{numberTimesSelected$_{a}$  $>$ 0}
	                \State avgReward =($\frac{sumReward_{a}}{numSelection_{a}}$)
	                \State upperBound = (avgReward + c $\sqrt{\frac{\ln{(numberGenerationsElapsed)}}{numberTimesActionSelected_{a}}}$)
	            \Else
	                \State upperBound = doubleMaxValue
	            \EndIf
	            \If{upperBound $>$ max}
	                \State max = upperBound \Comment{Update estimation of best action.}
	                \State action = getAction(a) \Comment{Choose the action with the highest estimation.}
	            \EndIf
             \EndFor
	     \EndIf
	       \State numberTimesActionSelected$_{action}$ = numberTimesActionSelected$_{action}$ + 1
	     \State return action
	\end{algorithmic} 
	\caption{Overview of the UCB algorithm.}\label{alg:UCB_alg}
\end{algorithm}

In the $n$-armed bandit problem~\cite{Sutton2018}, an agent is presented with a machine with $n$ arms. Each time the agent chooses an arm, they will get a reward. Naturally, this agent will seek to identify the arms that give them the most reward. Even if the reward earned is non-deterministic, it is likely that certain arms will give more reward ``on average''. The problem, then, is to identify the arm that will give the greatest improvement in reward when chosen and to keep choosing that one until time runs out or the maximum reward is attained. This is challenging, of course, because one must decide to whether to exploit their current knowledge---choose the arm that you currently think is the best---or to explore---to refine your expected reward by trying a new or previously suboptimal option. Exploitation will lead to short-term improvement, but risks missing out on potentially greater gains in the long-term. However, too much exploration also risks resulting in a low reward by repeatedly trying poor options in the hope they improve. Approaches to the $n$-armed bandit problem seek to balance exploration and exploitation in an effective manner. 

The Upper Confidence Bound (UCB) algorithm is well-suited to addressing $n$-armed bandit problems~\cite{Sutton98:Reinforcement}. Each time a choice is made, UCB selects an action with a higher expected reward than the other possible actions. Each action returns a numerical value that is considered as the reward of taking that action. This means that a testing goal that is to be optimized using this approach requires the definition of a reward function representing the improvement attained in that goal by taking an action. In Section~\ref{sec:imp}, we discuss the specific reward functions used for each of our three goals. In contrast to fitness functions, these can be relatively simple functions. One could even use existing fitness functions and measure reward as the change in that score from the previous generation. 

Algorithm~\ref{alg:UCB_alg} outlines the UCB algorithm. For a selected action \(A\) at time step \(t\) (represented as \(A_t\)), the reward \(R_t\) represents the corresponding reward of taking action \(A_t\). In test generation, the time step is the current number of generations that have elapsed. Using this notation, the expected reward of action \(a\) is  \(q_*\doteq E[R_t|A_t=a]\). We apply the Upper Confidence Bound to select the action~\cite{Sutton98:Reinforcement}:

\begin{equation}
A_t \doteq max[ Q_t(a) + c \sqrt{\frac{ln(t)} {N_t(a)}}]
\end{equation}

\noindent where \(A_t\) represents the index of the combination that gives the highest expected reward. The \(c\) term represents the confidence level, determining the balance between exploration and exploitation in the algorithm. The value of \(c\) needs to be larger than 0. Otherwise, the algorithm will behave in a purely greedy manner. The confidence level is multiplied by the square root of the natural log of the time step divided by the number of times the action has been selected. \(Q_t(a)\) denotes the estimated value of choosing a combination of fitness functions (\(a\)), which can be calculated as:

\begin{equation}
Q_t(a) = \frac{1}{N_t} \sum_{i=1}^{t-1} R_i(a)
\end{equation}

This equation represents the total reward of a combination \textit{a} divided by the number of times that combination had been selected until the time \(t\). In this project, \(t\) denotes the number of generations of evolution that have occurred. In this algorithm, we first ensure that all actions are tried once in a random order (lines 8-10 in Algorithm~\ref{alg:UCB_alg}). This allows us to seed expected rewards of applying actions before using the standard selection procedure. We when proceed to apply the set of equations defined above to update our estimation of gain in reward and select the action with the highest estimation. 

Reinforcement learning approaches generally attempt to associate the reward of taking an action with a particular state. To control the size of the state space, we represent the state as a feature vector containing the current set of fitness functions, the current fitness value for that set of functions, the test suite size, and the coverage of the subgoals associated with the fitness functions.\footnote{For example, in Strong Mutation Coverage, this would be the percent of mutants detected through an observed difference in output.} 

UCB is an example of a tabular solution method, where it attempts to associate rewards with specific states. It logs those reward expectations in a table or list structure, and attempts to identify the exact reward that would be gained in that state. However, finding an exact solution is not feasible when the states space is large or continuous. This is a potential limitation of this approach during test generation, as the state space of even our limited representation is large, and our feature vector representation could potentially be met by a large number of actual test suites as is a summarization of facets of the suite. To address this potential limitation, we also implemented a second algorithm, DSG-Sarsa, which generalizes expectations from previously-encountered states~\cite{Sutton2018}.

\subsection{Differential Semi-Gradient Sarsa (DSG-Sarsa)}\label{sec:sarsa}

\begin{algorithm}[!t]
\scriptsize
	\begin{algorithmic}[1]
	      \State \textbf{After action A is taken, the test suite is in state S}
	      \If{numberGenerationsElapsed $<$ numberActionsSelected}
	        \State A$_{new}$ = getAction(numberGenerations) \Comment{Try all actions once before using RL}
	        \State numberActionsSelected = numberActionsSelected + 1
	     \Else
	        \State \textbf{Select action A$_{new}$ as a function of $q(S_0, . ,w)$ using $\epsilon-greedy$ policy}
	     \EndIf
	     \State \textbf{Observe S$_{new}$ and the reward after taking the action}
	     \State numTimesActionSelected$_{A_{new}}$ = numTimesActionSelected$_{A_{new}}$ + 1 
	     \State $\delta$ = reward + averageReward + q(S$_{new}$, A$_{new}$, w) - q(S, A, w) \Comment{Update the error function}
	     \State averageReward = averageReward + $\beta$ * $\delta$ \Comment{Update the average change in the reward score}
	     \State w = w + $\alpha$ * $\delta$ * q(S, A, w) \Comment{Update the weight vector}
	     \State A = A$_{new}$
	     \State S = S$_{new}$
	\end{algorithmic} 
	\caption{Overview of the DSG-Sarsa algorithm.}\label{alg:DSG_Sarsa_alg}
\end{algorithm}

Approximate solution methods generalize from previously encountered states~\cite{Sutton2018}. Therefore, approximate methods are appropriate for problems with a large or unconstrained state-space where finding exact solutions is not feasible with limited time~\cite{Busoniu11:Approx}. As test case generation has a potentially vast state space---even using a feature vector to summarize that state---we have explored using an approximate solution method, Differential Semi-Gradient Sarsa (DSG-Sarsa)~\cite{Sutton2018}.

DSG-Sarsa is semi-gradient, enabling continual and online learning. Relevant to our application domain, the algorithm is well-suited to problems in which there is no termination state. This is an ``on-policy'' method, which means that it tries to improve the policy that the agent has in place to make decisions. The agent leverages from past experiences to decide when to vary between exploitation and exploration~\cite{Sutton2018}. On-policy methods may be better suited to our application domain than off-policy methods. On-policy adjustment will allow more exploration than exploitation when necessary---this may be beneficial, given a large number of potential combinations of fitness functions that could be chosen.  
An overview of DSG-Sarsa is presented in Algorithm~\ref{alg:DSG_Sarsa_alg}. Each generation, an action---a choice of fitness functions---is applied, and the test suite evolves to a new state \(S\)`, with observed reward $R$. Again, we start by trying each action once in a random order to seed estimates (lines 2-4). Then, we choose a new action \(A\)`, using the formula:
\begin{equation}
\hat{q}(S,A,W) \doteq {W}^\top \cdot X(S,A) = \sum w{_i}x{_i}(S,A) 
\end{equation}
This action-value function is calculated by the inner product of weights and feature vectors. X(S,A) is the feature vector: \(X(s,A)=(x_1(S,A),x_2(S,A),\dots x_d(S,A))\). As noted previously, the feature vector describes the current state of a test suite as the current set of fitness functions, the current fitness value for that set of functions, the test suite size, and the coverage of function goals. 

\(W\) represents a weight vector, used to bias action selection~\cite{Sutton2018}. A weight is provided for each feature, and illustrates the importance of each feature in respect to its contribution to the action value. The weight for an action is updated each round using the semi-gradient with delta, controlled by the learning rate:
\begin{equation}
W_{t+1} \doteq W_t + \alpha\delta\nabla\hat{q}(S_t,A_t,W_t)
\end{equation}
To evaluate the chosen action, the algorithm calculates the error function ($\delta$), which represents the difference between the immediate reward $R$ and the average reward \(\bar{R_t}\) and the difference between the value of a target \(\hat{q}(S_{t+1},A_{t+1},W_t)\) and the value of the old estimate \(\hat{q}(S_t,A_t,W_t)\)~\cite{Sutton2018}. In each iteration, the current action---selection of fitness function---is used to generate a new state and reward. We use an action-value function to generate the action \(A\)'. In our case, we use \(\varepsilon-greedy\)~\cite{Sutton2018}. 
The reward return is calculated in terms of the difference between the current and the average reward. The corresponding value function that is used for this type of return is called a differential value function~\cite{Sutton2018}:
\begin{equation}
\delta_t = R_{t+1} - \bar{R}_t +\hat{q}(S_{t+1},A_{t+1},W_t) - \hat{q}(S_t,A_t,W_t)
\end{equation}
\(\bar{R}_t\) is the estimated average reward at time \(t\), calculated as:
\begin{equation}
\bar{R}_{t+1} = R_t + \beta\delta
\end{equation}
\(\beta\) is an algorithm parameter that represents the step size of updating the average reward. The notation $t$ represent the the time step (the number of generations).

By using the average reward, we consider the immediate reward as important as a delayed one. This means that we treat all fitness function combinations impartially without bias toward combinations that were selected first. Thus, there is no priority for the chosen combinations other than effectiveness. 

\subsection{EvoSuite Overview}\label{sec:evo}

 \begin{figure}[!t]
	\centering
	\includegraphics[width=4.75in]{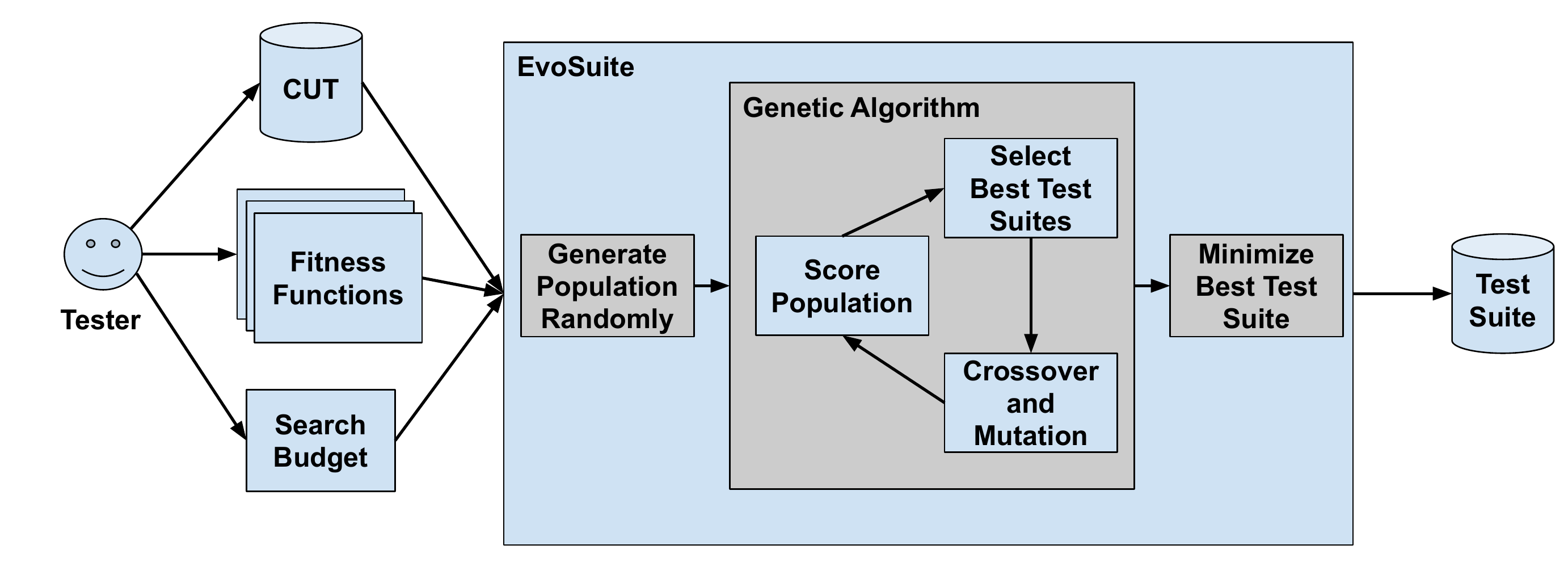}
	\caption{A simplified overview of EvoSuite's test generation process.}
	\label{fig:EvoSuite_process}
\end{figure}

We have implemented both reinforcement learning algorithms in the EvoSuite unit test generation framework~\cite{SBST17_competition,ssbse18_tutorial}. EvoSuite targets classes written in the Java language, and produces complete JUnit test cases that initialize the class-under-test, calls its methods with generated input, and applies generated assertions to check the results.\footnote{\normalsize Assertions are generated using the class-under-test, which means that generated assertions are not useful for fault detection in the tested code. Instead, assertions are used for regression testing scenarios or to check for differences between two versions of a class.} 

The general test generation process in EvoSuite is depicted in Figure~\ref{fig:EvoSuite_process}. 
EvoSuite takes, among other configuration options, a CUT, a set of chosen fitness functions, and a search budget---the time allocated to the test generation process. An initial population of test suites is generated randomly, then a metaheuristic algorithm evolves that population until the search budget is exhausted. In this research, we have integrated AFFS into the standard Genetic Algorithm (GA). 

Each generation, the GA evaluates the current population (a collection of test suites) using the chosen fitness functions. The score is calculated for each fitness function, the scores are normalized to a 0-1 scale, then the scores are summed into a single score. Lower scores are preferred. The standard GA in EvoSuite is not a true multi-objective approach, i.e., it does not try to balance each fitness function. Sufficient improvements to one of the chosen functions will result in a suite being favored, even if it attains worse scores in other functions than other test suites. 

Then, a new population is formed by retaining high-scoring solutions, mutating solutions, forming new solutions by combining elements of parent solutions (crossover), and generating a small number of new random solutions to maintain diversity. After the search budget has been exhausted, the best solution will go through a minimization process in which test cases that cover redundant goals are removed (using the goals set by the current fitness functions). For example, if one of the fitness functions represents the Branch Coverage, a test that does not cover additional goals not covered by already-selected tests will be removed. At the end, a small-but-effective test suite will be returned. EvoSuite supports a large number of fitness functions for test generation~\cite{Rojas15:Combining}. We make use of nine of these functions in our work:
\begin{itemize}
	\setlength{\itemsep}{1pt}
	\setlength{\parskip}{0pt}
	\setlength{\parsep}{0pt}  
	\item \textbf{Exception Count:} A count of the unique exceptions thrown by a test suite. Exceptions are tracked using the name of the Exception class and the method where the exception was thrown. In addition, exceptions are separated into those that are declared (in method signature), explicit (developer used a \texttt{throw} expression), and implicit (unplanned exceptions). 
	\item \textbf{Branch Coverage:} A test suite satisfies Branch Coverage if all control-flow branches are taken during test execution. For each program statement that can cause the execution path to diverge---such as \texttt{if} and \texttt{case} statements---test input should ensure that at all potential outcomes are covered at least once~\cite{Pezze06:testing}. To guide the search, the fitness function calculates the branch distance from the point where the execution path diverged from the targeted branch. If an undesired branch is taken, the function describes how ``close'' the targeted predicate is to be true, using a cost function based on the predicate formula~\cite{Arcuri13:Normalize}. 
	\item \textbf{Direct Branch Coverage:} Branch Coverage may be attained by calling a method directly or indirectly---i.e., a method call within a method that was directly called. Direct Branch Coverage requires each branch to be covered through a direct method call, while standard Branch Coverage allows indirect coverage. Each can detect faults missed by the other~\cite{Gay18:DBC}. 
	\item \textbf{Line Coverage:} A test suite satisfies Line Coverage if it executes each non-comment source code line at least once. To cover each line, EvoSuite tries to ensure that each basic code block is reached. For each conditional statement that is a control dependency of some other line in the code, the branch leading to the dependent code must be executed.
	\item \textbf{Method Coverage (MC):} Method Coverage requires that all the CUT's methods are executed at least once, through direct or indirect calls. 
	\item \textbf{Method Coverage (Top-Level, No Exception) (MNEC):} Generated test suites sometimes achieve high levels of Method Coverage by calling methods in an invalid state or with invalid parameters. MNEC requires that all methods be called directly and terminate without throwing an exception.
	\item \textbf{Output Coverage (OC):} Output Coverage rewards diversity in the method output by mapping return types to a list of abstract values~\cite{Alshahwan14:OutputCoverage}. A test suite satisfies Output Coverage if, for each public method in the CUT, at least one test yields a concrete return value characterized by each abstract value. For numeric data types, distance functions offer feedback using the difference between the chosen value and target abstract values.
	\item \textbf{Weak Mutation Coverage (WMC):} A test suite satisfies weak mutation coverage if, for each mutated statement, at least one test detects the mutation. The infection distance guides the search, a variant of branch distance tuned towards reaching and discovering mutated statements~\cite{Fraser14:Mutation}.
	\item \textbf{Strong Mutation Coverage (SMC):} Weak Mutation Coverage ensures that the mutated line of code is reached. However, it makes no guarantees that the infected program state is noticed by the tester. Strong Mutation Coverage adds an estimation of the likelihood of propagation, the propagation distance, by estimating the impact of corrupted state~\cite{Fraser14:Mutation}.
\end{itemize}
\noindent Rojas et al. provide more details on each of these fitness functions~\cite{Rojas15:Combining}. We additionally implemented a Test Suite Diversity fitness function based on the Levensthein distance, which we will discuss in Section~\ref{sec:div}

\subsection{Implementation of AFFS within EvoSuite}\label{sec:imp}

\begin{figure}[!t]
	\centering
	\includegraphics[width=4.75in]{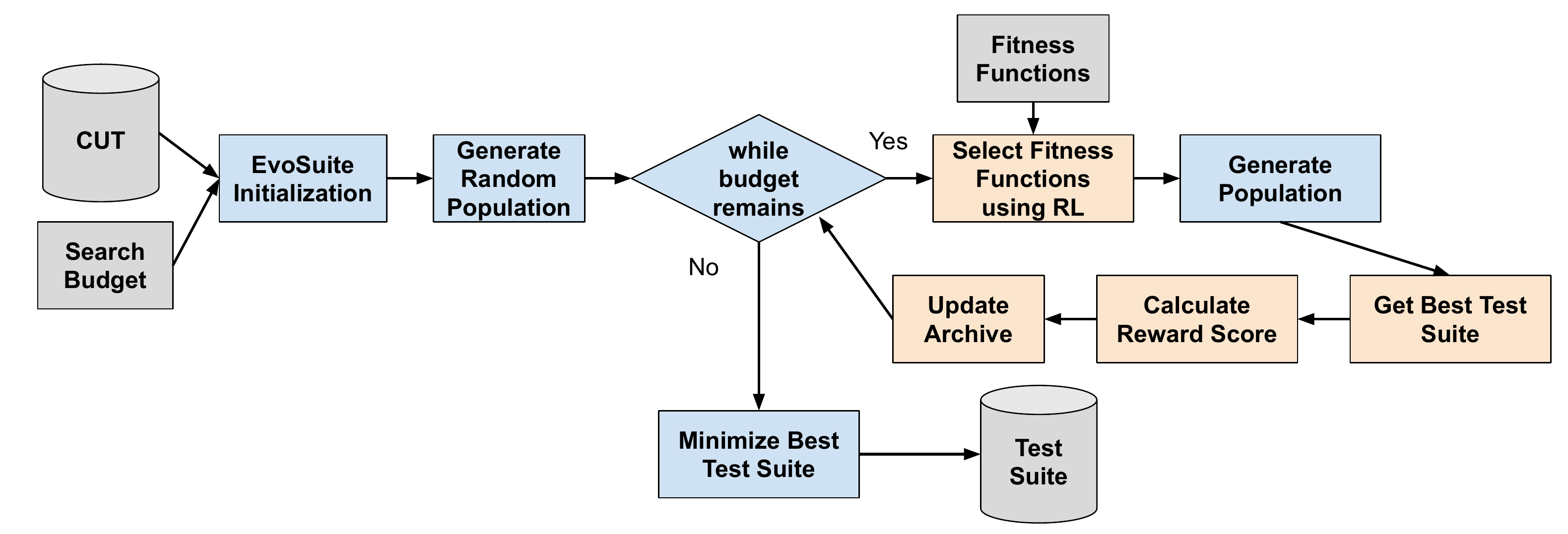} 
	\caption{The modified test generation process. Steps in orange are introduced or modified by AFFS.}
	\label{fig:process}
\end{figure}

\begin{algorithm}[!t]
\scriptsize
	\begin{algorithmic}[1]
	    \While{searchBudget $>$ 0}
			\State evolvePopulation() \Comment{Use fitness functions to evolve the population.}
			\State sortPopulation() \Comment{Sort by fitness score.}
			
			\If{ generation \% skipIter = 0 \textbf{and} approach $\neq$ NORL}
			    \Comment{Update reward estimate.}
			    \State bestSolution = GetBestIndividualFromPopulation()
		        \State reward = CalculateReward()
				\If{approach == DSGSARSA}
                     \State DSGSarsa(iterNum, bestSolution.coverage,
                            bestSolution.size(), reward, \mbox{bestSolution.fitness})
                \ElsIf{approach == UCB}
                    \State UCB(reward, action)
                \EndIf
                \State action = GetBestFFCombination() \Comment{Determine new fitness functions.}
                \State UpdateCurrentFitnessFunction() \Comment{Update fitness functions.}
                \State iterNum++
            \EndIf
            \State generation++
		\EndWhile
	\end{algorithmic} 
	\caption{Overview of the AFFS hyperheuristic. NORL = reinforcement learning is not used, skipIter = a user defined number of generations to improve the population before the fitness functions are changed (generally 3-5).}\label{alg:hyper}
\end{algorithm}

We have implemented both reinforcement learning algorithms in EvoSuiteFIT, and integrate their use into the standard GA. At a user-defined interval, the RL algorithm will choose a new set of one to four fitness functions. The specific sets of fitness functions are goal-dependent, and will be explained in the following subsections. The modified process is illustrated in Figure~\ref{fig:process}. Algorithm~\ref{alg:hyper} provides an overview of the reinforcement learning implementation in EvoSuiteFIT.

AFFS is an \textit{online} learning approach. The RL algorithm learns its policy during the test generation process, adapting to the CUT and the evolving state of the test suite. This stands in contrast to an offline process, which would attempt to apply a policy learned in an earlier process. We do not attempt to transfer learned policies to new classes in this work~\cite{Iqbal19:transfer}. The differences between classes may result in poor transfer success. However, this is a topic we will consider in future work. 

In the beginning, EvoSuiteFIT will make sure that all the actions have been tried once before it starts using the standard UCB or DSG-Sarsa selection mechanisms. This allows seeding of reward estimations. Before the initial selection occurs, the list of actions is randomized to avoid an ordering bias. This is important, as the population of test suites is shaped by the action used each generation. After this stage, every time the RL algorithm makes a selection, the set of chosen fitness functions will change unless the currently-selected combination is exploited. 

After changing the fitness functions, EvoSuiteFIT will proceed through the normal population evolution mechanisms, judging solutions using the new set of fitness functions (lines 2-3 in Algorithm~\ref{alg:hyper}). We use the reformulated population to calculate the reward---the gain in goal attainment from choosing an action (line 6 in Algorithm~\ref{alg:hyper}). Reward functions, too, are goal-specific and will be explained in the following subsections. Then, we use this reward to update the expectations of the RL algorithm. For UCB, we store the accumulated reward of each combination alongside the number of times each is selected \(N_t\), so we can calculate the average reward (line 10). Over time, the combination that gains the highest reward will be more likely to be selected again until reaching convergence. For DSG-Sarsa, after getting the reward, the new combination is selected using the learned policy. Based on the new and current combination, the new and current state, and the reward, the average reward and the weight of the state is updated (line 8). Then the current fitness function combination will change to the new one (lines 12-13).

After experimentation, we found that changing the fitness functions every three to five generations allows enough time to adequately adjust reward expectations. Fewer generations do not allow sufficient time for the chosen fitness function combination to reshape the test suite. This means that the GA will have a short time to reshape the population before reward is evaluated (line 4 in Algorithm~\ref{alg:hyper}).

In EvoSuiteFIT, test cases that cover a set of chosen goals can be retained in a test archive during the search and optimization process to prevent loss in coverage as the test suites are reshaped. Normally, this archive is based on the goals of the static set of fitness functions chosen when test generation starts. However, as we use RL to change the fitness functions, we have altered how the test archive is used. Instead, we use a set of goals associated with high-level testing goal. In the following subsections, we will discuss the goals used. After the search process completes, the archive is used to help produce the final test suite. This prevents the loss of test cases that may contribute to effectiveness due to changes in fitness functions. After generation concludes, the best solution is minimized with respect to this set of goals. The archive then is used to supplement this suite with coverage of any missing goals. 

In the following subsections, we will discuss specific adaptations made for the three high-level testing goals: exception discovery, test suite diversity, and Strong Mutation Coverage.

\subsubsection{Adaptations for Goal: Exception Discovery}

\noindent \textbf{Fitness Function Combinations:} EvoSuiteFIT chooses a combination of one to four of the following fitness functions: Exception Count, Branch Coverage, Direct Branch Coverage, Line Coverage, Method Coverage, MNEC, Output Coverage, and Weak Mutation Coverage. Initial experimentation revealed that effective combinations include the exception count, even though the count is rarely effective on its own. Therefore, we filtered the initial set of combinations down to all combinations of one to four fitness functions that include the exception count as one of the choices. EvoSuiteFIT can choose from 64 different sets of fitness functions.

\smallskip \noindent \textbf{Reward Function:} We measure reward as the sum of exceptions discovered during the entire generation process and the exceptions thrown by the current best test suite, encouraging discovery and retention of exceptions.

\smallskip \noindent \textbf{Goals Used for Minimization and Archiving:} We use the set of discovered unique exceptions as goals for minimization and archiving tests. A test that forces the CUT to throw a particular exception covers the ``goal'' for that exception. When the test suite is minimized, it is minimized to ensure that all discovered unique exceptions are covered. Tests detecting any exceptions no longer covered by that suite will be added from the archive, preventing loss of coverage.

\subsubsection{Adaptations for Goal: Test Suite Diversity}\label{sec:div}

\noindent\textbf{New Fitness Function:} EvoSuite does not already contain a fitness function intended to promote test suite diversity. Therefore, we have implemented a fitness function to measure test suite diversity based on the Levenshtein distance~\cite{Shahbazi15:Diversity}. The Levenshtein distance is the minimal cost of the sum of individual operations---insertions, deletions, and substitutions---needed to convert one string to another (i.e., one test to another). We compare the text of test cases within a test suite. 

The distance between two tests ($ta$ and $tb$) can be calculated as follows~\cite{Shahbazi15:Diversity}:
\begin{equation}
    lev_{ta,tb}(i,j)= 
    \begin{cases}
        max(i,j) & \text{if $min(i,j)==0$}\\
        min \begin{cases}
                lev_{ta,tb}(i-1,j)+1\\
                lev_{ta,tb}(i,j-1)+1 \\
                lev_{ta,tb}(i-1,j-1)+1_{(ta_{i} \neq tb_{j})}
            \end{cases} & otherwise
        \end{cases}
\end{equation}
\noindent where $i$ and $j$ are the letters of the strings representing $ta$ and $tb$. To calculate the diversity of a test suite ($TS$), we calculate the sum of the Levenshtein distance between each pair of test cases: 
\begin{equation}
div(TS) = \sum_{ta,tb}^{TS}{lev_{ta,tb}}
\end{equation}
\noindent To attain a normalized value between 0-1 for use in a multi-fitness function environment, we then calculate and attempt to minimize the final fitness as: 
\begin{equation}
\frac{1}{1+div(TS)}
\end{equation}

The fitness function calculation iterates through the test cases in a given test suite. Before calculating the distance, the variables and their values are extracted from the test cases. This includes extracting numeric primitive variables, null variables, strings, arrays, instance and class fields, methods, and constructor statements. Our analysis also includes partial assessment of aliasing. Consider the following fragment: \texttt{String x = "var"; String y = x; String z = y;}. Variables \texttt{x}, \texttt{y}, and \texttt{z} are different, but are initialized with the same value. These should not be considered diverse, so we statically trace the reference to the original value when possible to attain a more accurate estimation of diversity. The list of filtered statements is then used to calculate fitness. 

To calculate diversity, each pair of test cases is compared. From each pair of tests, each pair of statements is compared. The Levenshtein distance is calculated between each of these pairs and added to the diversity score, then it is returned to the core process. The Levenshtein distance calculation uses a classic matrix-based approach~\cite{Navarro01:strings} where the characters in the two strings are compared, and the final value stored in the matrix is returned. 

\smallskip\noindent\textbf{Fitness Function Combinations:} EvoSuiteFIT chooses a combination of one to four of the following fitness functions: Diversity, Exception Count, Branch Coverage, Direct Branch Coverage, Method Coverage, MNEC, Output Coverage, and Weak Mutation Coverage. To constrain the number of combinations, we use only the combinations that include the diversity score and remove a small number of semi-overlapping combinations (i.e., Branch and Direct Branch). Ultimately, EvoSuiteFIT can choose from 44 combinations of fitness functions.

\smallskip\noindent\textbf{Reward Function:} The change in the diversity fitness score is used as the reward function to identify the actions that best increase diversity.

\smallskip\noindent\textbf{Goals Used for Minimization and Archiving:} Unlike exception discovery and Strong Mutation Coverage, test suite diversity lacks a natural set of discrete goals. Test suites can be diverse in many different ways, and coverage lacks a direct analogue. To support the archiving and minimization process, we adapt the set of goals from Method Coverage. This means that suites are minimized using their coverage of the source code. This is a low-cost calculation that does not have a noticeable effect on overhead, while retaining diversity in the final suite.

\subsubsection{Adaptations for Goal: Strong Mutation Coverage}

\noindent\textbf{Fitness Function Combinations:} EvoSuiteFIT chooses a combination of one to three of the following fitness functions: Strong Mutation, Exception Count, Branch Coverage, MNEC, Output Coverage, and Weak Mutation Coverage. This provides EvoSuiteFIT with 31 combinations of fitness functions to choose from. This is a smaller pool of actions than was used for the other two goals. This is because the calculation of Strong Mutation Coverage requires more time than calculating other fitness functions. Attaining a clear estimation of the expected reward of choosing an action requires that each action be tried multiple times. If it is expensive to calculate fitness, however, the total number of generations that can be completed within a time period may be restricted. This reduces the time that can be spent exploring different actions. To compensate for this cost, we have reduced the number of possible actions by (1) limiting combinations to three fitness functions, and (2), removing potentially redundant fitness functions (Line Coverage, Direct Branch Coverage, Method Coverage). Unlike the other two goals, not all combinations include Strong Mutation Coverage. Instead, we conducted a small experiment and utilized the best combinations found in that experiment. 

\smallskip\noindent\textbf{Reward Function:} We use the mutation score as the reward function. This is the percentage of mutants detected: $\frac{Detected Mutants}{Total Number of Mutants} * 100$. The mutation score can be calculated using either Strong or Weak Mutation Coverage. The difference is that, in Strong Mutation Coverage, we require a noticeable difference in class output between the original and mutated version. In Weak Mutation Coverage, the mutated statement simply must be reached and the internal state of the execution must be corrupted at that point. 

Strong Mutation Coverage is much more expensive to calculate than Weak Mutation Coverage. To reduce the overhead that would occur when calculating Strong Mutation Coverage during reward estimation refinement, we iterate between Weak Mutation and Strong Mutation. The reward from choosing an action is the improvement in the mutation score.

\smallskip\noindent\textbf{Goals Used for Minimization and Archiving:} We use the set of goals calculated in order to attain the final Strong Mutation Coverage score. That is, each mutant that can be detected is a discrete goal. Suites are minimized in terms of coverage of these mutants and tests from the archive are added to the final suite to detect any mutants missed by the unaugmented suite.

\section{Related Work}\label{sec:related}

This section will provide an overview of related work on hyperheuristic search-based software testing approaches, as well as test generation research related to exception discovery, test suite diversity, and strong mutation coverage to give insight into past research in topics related to this study.

\subsection{Hyperheuristics in Search-Based Software Testing}

Hyperheuristic search has been employed in addressing multiple several search-based software engineering problems. Fitness function selection has been performed by hyperheuristic search in other domains, such as production scheduling~\cite{Crawford2013:Tuning,Ochoa09:HH}. However, our approach is the first automated technique for optimizing the set of fitness functions used during test generation. Related work, largely, uses the hyperheuristic to tune crossover and mutation operators used by an evolutionary algorithm. We briefly give an overview of this work below to illustrate how hyperheuristics have been used to improve other aspects of the search algorithm. 

Jia et al.~\cite{Jia15:HHICSE,Jia15:HHSBST} used reinforcement learning to tune the metaheuristic for Combinatorial Interaction Testing, using the Simulated Annealing algorithm in the outer layer and using an n-Armed Bandit approach for learning and choosing the best operator(s) (out of six) to tune the performance of the algorithm. Zamli et al. also used a hyper-heuristic approach for CIT~\cite{Zamli_Alkazemi_2016}, using Tabu search as a high-level hyperheuristic to select a low-level heuristic from four algorithms. Later, Zamli et al. used hyperheuristic search to learn optimal selection and acceptance mechanisms used by the metaheuristic in CIT~\cite{Zamli17:HHCIT}. Din et al. also applied hyperheuristic search to CIT~\cite{din8004298}, using parameter-free choice functions to rank low-level heuristics for selection. Din and Zamli use Exponential Monte Carlo with Counter (EMCO) as a hyperheuristic to select a low-level heuristic in CIT~\cite{Din_zamli_2018}. Ahmed et al.~\cite{Ahmed_2020} compare EMCO against an improved version using Q-learning, called Q-EMCO, to select the best operator based on historical information.

Guizzo et al. used a reinforcement learning-based hyper-heuristic search to tune the metaheuristic algorithm for optimizing the integration and test order problem~\cite{Guizzo15:HMI,Guizzo15:HHITO}. In later work, Guizzo et al. used hyperheuristic search to select an operator that can be executed by Multi-Objective Evolutionary Algorithms (MOEAs) to provide a solution for the ITO problem~\cite{Guizzo_Vergilio_2017}.  Guizzo et al. also applied a hyperheuristic to the NSGA-II MOEA to address ITO in Google Guava~\cite{Guizzo2017AHF}. Mariani et al. introduced an approach that depends on an offline hyperheuristic named GEMOITO to generate MOEAs to solve the ITO problem~\cite{Mariani_Guizzo_2016}. Guizzo et al. later used design patterns to improve the design of MOEA to reduce coupling and increase reusability of components~\cite{Guizzo2018APS}. They implemented the patterns into GEMOITO. They found that they were able to reuse MOEA components without decreasing the quality the algorithm results. 

Ferreira et al. proposed the use of hyperheuristic search in software product line (SPL) testing~\cite{Ferreira7895294}. Software Product Lines are sets of systems that share a common set of features that are customized for particular market segments or customers. In practice, all products cannot be tested. Therefore, search-based approaches can be used to select ``interesting'' ones to focus on. Building on earlier work~\cite{Ferreira7744315,STRICKLER20161232}, the authors proposed using a hyperheuristic MOEA to find a select product variants for testing. Their approach considers four objectives: the number of products, pairwise coverage, mutation score, and dissimilarity of products. Filho et al. also proposed a hyperheuristic that uses grammatical evolution to generate MOEAs for SPL testing~\cite{Filho3131152}. Their approach considers three factors---pairwise coverage, mutation score, and cost---and generates a MOEA using crossover and mutation operators tuned to the feature model being considered. Filho et al. extended this work~\cite{Filho3266275,Filho8477803} to Preference-Based Evolutionary Multi-objective Algorithms, which consider user preferences during the search. 

Kumari and Srinivas~\cite{Kumari16:HHC} used hyperheuristic search to tune software design---learning how to cluster classes for maximum cohesion and minimum coupling. This work applies reinforcement learning to select a low-level heuristic that will be used with an evolutionary algorithm to cluster software modules for further analysis. 

Moghadam et al.~\cite{Moghadam19:Perf} have proposed a framework that uses adaptive learning to generate test cases for stress testing. Bauersfeld et al.~\cite{bauersfeld2012reinforcement} introduced an automated testing approach for robustness testing of GUIs based on reinforcement learning. Building on their previous work~\cite{Bauersfeld6494948}, they introduce an approach to select input events for GUIs intended to improve coverage of deeply nested actions. They use Q-Learning to discover states and actions and learn the value function to maximize coverage of GUI actions. Grechanik proposed an adaptive, feedback-driven approach to generating input designed to highlight performance issues ~\cite{Grechanik2012}. Their technique, FOREPOST, initially generates test cases randomly, and the results are evaluated. Then, the results are feed to a machine learning classification algorithm, which will output a set of rules. These rules will be used in the next cycle as guidance to select input tests and generate test cases. This approach is not based on metaheuristic search, but still uses feedback to improve test case generation. 

\subsection{Crash and Exception Discovery}

Joffe el al.~\cite{joffe2019directing} use the results from an artificial neural network (ANN) classifier to construct a fitness function targeting crashes, which can be used in search-based test generation. They trained their ANN classifier on C programs to predict the likelihood of crashing, given a particular input. They modified American Fuzzy Lop---a search-based test generation tool---to consider the crash likelihood from the classifier. Romano et al.~\cite{Romano2011AnAF} focused on targeting null pointer exceptions, providing an approach that can identify code that can cause this exception by looking at execution paths. The approach generates a control flow graph, which is used to identify paths that could throw exceptions. Coverage of these paths is then targeted using search-based test generation. Although this approach is more likely to detect null pointer exceptions than a general test generation approach, coverage of these paths does not guarantee that a null pointer exception is triggered.

Due to inadequate detection of exceptions in automated test case generation, Goffi et al.~\cite{Goffi2931061,Blasi18:Spec} proposed the use of natural language processing to generate test oracles---assertions designed to assess the behavior of the system. Their approach extracts comments that are related to exceptional behaviors that can be thrown by a method or class. Then, these comments are translated into assertions, which are used in test cases to improve detection of faults. Extended work widens the range of behaviors that can be assessed by these oracles~\cite{Blasi18:Spec}. Their work, in contrast to ours, does not influence the selection of test inputs. Rather, it improves the likelihood of fault detection by existing inputs. Therefore, it could be combined with our approach, potentially improving fault detection further. 

\subsection{Test Suite Diversity}

Albunian investigated the impact of diversity on search-based test generation~\cite{Nasser66299}, proposing a phonetypic and genotypic representation to measure diversity. They studied the influence of five selection mechanisms and five fitness functions. Feldt et al.~\cite{Feldt7515474} proposed a new diversity fitness function based on normalized compression distance. Ma et al.~\cite{LinhaiMA2018162} proposed an adaptive approach that generated concurrent test cases targeting diversity metrics. They introduce two diversity metrics, static, which concerns diversity in structure, and dynamic, which is concerned with exposing untested thread schedules. Vogel et al.~\cite{vogel2019does} investigated using diversity metrics in search-based generation of test cases for Android mobile applications. They proposed an approach that diversifies the population at the initialization and selection steps, then preserves and improvse diversity during the search. All of these approaches are complementary to our proposed approach, and could potentially be used in combination with our approach to yield improved suite diversity.

\subsection{Strong Mutation Coverage}

Many approaches to test generation for mutation coverage aim at satisfying weak mutation coverage, where the impact of a fault does not need to propagate to the output. Strong mutation coverage, which requires that the program output differs from the unmutated (correct) program, is harder to satisfy. Fraser el al.~\cite{Fraser14:Mutation} proposed a fitness function representation for strong mutation that is implemented in EvoSuite. This function estimates propagation of change using an impact measurement, which measures the difference between control flow and data that results from running the tests on an original program and mutants. We use this fitness function in our work, and attempt to use hyperheuristic search to further improve optimization of this function. Souza el al.~\cite{Souza2016} proposed an automated test generation approach for strong mutation using Hill Climbing, a simple local search algorithm. The proposed fitness function uses three metrics, called the Reach Distance, Mutation Distance, and the Impact Distance. These metrics are used to guide the search toward satisfying three goals; reaching the mutant, changing the program state, and propagating the state change to the program output. Papadakkis and Malevris~\cite{Papadakis2013} proposed using alternating variable method---a search algorithm---to generate tests to optimize a fitness function based on strong mutation. The proposed fitness function is composed of four parts. The first are the approach level and the branch distance, used in branch coverage to measure distance of the execution path from a targeted statement. They measure distance from covering the mutated line of code. The third is the mutation distance, which assesses how close program state is to being corrupted. Finally, the impact distance approximates the likelihood of the mutant impacting the output by quantifying how much of an effect the mutation had on the program state when exposed. 

Like with suite diversity, all of these fitness function representations are compatible with our approach, and could potentially be used within reward functions targeted by the hyperheuristic search. We used the strong mutation function proposed by Fraser et al.~\cite{Fraser14:Mutation}, as it was already implemented in EvoSuite. However, any of the other functions could have been implemented instead, and could be considered in future work.

In the domain of policy testing, Xu et al.~\cite{Dianxiang3395599} proposed using strong mutation to generate XACML policy tests automatically. Their approach is based on three constraints: reachability, necessity, and propagation. These constraints are used to capture the differences between mutants and original policies in terms of the responses to access requests. Harman et al. proposed an approach that aims to achieve strong coverage of first and higher-order mutants~\cite{Mark2025144}. Mutants that alter one line are ``first-order'' mutants, while higher-order mutants change multiple lines. Most mutation approaches are based on first-order mutants. Their approach, called SHOM, is a hybrid of dynamic symbolic execution (DSE) and search-based test generation aimed at overcoming limtiations of earlier work with regard to higher-order mutants. The approach includes applying three transformations to the program that reduce constraint and path analysis effort without impacting the semantics of programs under test. 
\section{Methodology}\label{sec:method}

To better understand the effectiveness and applicability of adaptive fitness function selection, we have assessed EvoSuiteFIT using case examples from the Defects4J fault benchmark~\cite{Just14:Defects4J} for each of our goals---exception discovery, test suite diversity, and Strong Mutation Coverage. We will address the following research questions:
\begin{enumerate}
\setlength{\itemsep}{1pt}
  \setlength{\parskip}{0pt}
  \setlength{\parsep}{0pt}  
  \item For each goal, is either EvoSuiteFIT approach more effective than test generation using static fitness function choices at attaining that goal?
  \item For each goal, is either EvoSuiteFIT approach more effective than test generation using static fitness function choices in terms of attained fault detection?
  \item What impact does the computational overhead from reinforcement learning have on the test generation process?
  \item Are there observations that can be discerned in the combinations of fitness functions chosen by either EvoSuiteFIT approach that help explain the success (or lack of success) of an approach for a goal?
  \end{enumerate}
  
The first two questions provide us with an understanding of the effectiveness of EvoSuiteFIT compared to baseline approaches representing current practice. We hypothesize that adaptive fitness function selection is capable of increasing our attainment of difficult-to-optimize goals. We must evaluate whether that is true. 

Increased goal attainment does not \textit{necessarily} suggest higher likelihood of fault detection. However, each of the three goals we are maximizing are thought to be indicators of fault detection. That is, if the number of exceptions, suite diversity, or Strong Mutation coverage are increased, it is theorized that the likelihood of fault detection will rise as well. If EvoSuiteFIT is able to improve goal attainment, the number of faults detected may increase as well. Note, however, that we are asking a broader question than whether increased goal attainment leads to increased likelihood of fault detection. We are asking if any element of the AFFS process increases that likelihood. AFFS is a complex process, and other factors---like varying the fitness functions over time---could also impact fault detection. 

The third question will address the consequences of using reinforcement learning during the test generation process. This question will focus on the computational overhead of reinforcement learning. Test generation uses a time budget. Additional overhead from reinforcement learning may impact the number of generations of evolution the population of test suites goes through during that time---potentially negating the benefits of using reinforcement learning in the first place. At the same time, it is also expensive to calculate certain fitness functions or large sets of functions, and reinforcement learning may be able to avoid such functions. Therefore, we must examine the relationship between reinforcement learning and the cost of computing each generation of evolution. Finally, to better understand AFFS, we will also examine trends in the fitness functions choices. We will also identify and discuss limitations of the current implementation. 

In order to investigate these questions, we have performed the following experiment for each of the three goals:
\begin{enumerate}
\setlength{\itemsep}{1pt}
  \setlength{\parskip}{0pt}
  \setlength{\parsep}{0pt}
\item{\textbf{Collected Case Examples:} We have used a collection of case examples, from the Defects4J fault benchmark, as test generation targets (Section~\ref{sec:d4j}).}
\item{\textbf{Generated Test Suites:}  We target the classes affected by each fault for test generation. For each class, we generate 10 suites per approach. Approaches include the two reinforcement learning algorithms---UCB and DSG-Sarsa---and three baselines---an existing fitness function for that goal (current practice), a combination of all fitness functions that AFFS can chose from (a ``best guess''), and random selection from the choices available to AFFS. A search budget of 10 minutes is used per suite (Section~\ref{sec:testgen}).}
\item{\textbf{Removed Non-Compiling and Flaky Tests:} Any tests that do not compile, or that return inconsistent results, are removed (Section~\ref{sec:testgen}).}
\item{\textbf{Assessed Effectiveness:} We measure goal attainment for each test suite, the number of faults detected by each approach, the likelihood of fault detection for each fault and approach, the number of generations of evolution that occur during the generation process, and other data that can be used to analyze the behavior of both AFFS and traditional test generation (Section~\ref{sec:data}).}
\end{enumerate}
\noindent We use the gathered data to analyze the performance of AFFS for each individual goal, as well as to analyze the general behavior of AFFS across all goals.

\subsection{Case Examples}\label{sec:d4j}

Defects4J is a benchmark of real faults extracted from Java projects~\cite{Just14:Defects4J}.\footnote{\normalsize Available from \url{http://defects4j.org}} For each fault, Defects4J provides access to the faulty and fixed versions of the code, developer-written test cases that expose the fault, and a list of classes and lines of code modified by the patch that fixes the fault. Defects4J provides test execution, generation, code coverage, and mutation analysis capabilities.

Each fault is required to meet three properties. First, a pair of code versions must exist that differ only by the minimum changes required to address the fault. The ``fixed'' version must be explicitly labeled as a fix to an issue, and changes imposed by the fix must be to source code, not to other project artifacts such as the build system. Second, the fault must be reproducible---at least one test must pass on the fixed version and fail on the faulty version. Third, the fix must be isolated from unrelated code changes such as refactoring.

Our first goal, exception discovery, was assessed using Defects4J 1.4, which consists of 395 faults from six projects: Chart (26 faults), Closure (133 faults), Lang (65 faults), Math (106 faults), Mockito (38 faults), and Time (27 faults). Nine of the faults were excluded from our analysis---Closure faults 38, 44, 47, and 51, Math faults 13, 31, and 59, Mockito fault 6, and Time fault 21---as no technique caused exceptions to be thrown. 

The other goals, suite diversity and Strong Mutation Coverage, were assessed later using Defects4J 2.0. The experiments for exception discovery were not repeated due to experiment cost. However, as we already accounted for differences between Java 7 and 8---the primary semantic difference between Defects4J 1.4 and 2.0---results would not differ between versions of the benchmark. To compare results between the three high-level goals, we focus on the same projects. In both the diversity and Strong Mutation Experiments, we use the following 434 faults: Chart (26 faults), Closure (174 faults), Lang (64 faults), Math (106 faults), Mockito (38 faults), and Time (26 faults). In addition, for the diversity goal, we also use the Gson project (18 faults)---which was initially assessed in a pilot study~\cite{Gay20:DivRL}---bringing the total case examples for the diversity experiment to 452. 

\subsection{Test Suite Generation}\label{sec:testgen}

For all three goals, and for each bug-affected class from each case example used from Defects4J, we have generated test suites using UCB and DSG-Sarsa. In addition, we generate tests for 2-3 baseline approaches representing current practice:
\begin{itemize}
\setlength{\itemsep}{1pt}
  \setlength{\parskip}{0pt}
  \setlength{\parsep}{0pt}  
  \item \textbf{Current Practice}: We use the existing fitness function representation of that goal.\footnote{All three functions are explained in more detail in Section~\ref{sec:approach}.} This would be the likely starting point for a tester interested in these goals, and thus, represent current practice. These are, as follows: 
      \begin{itemize}
      \setlength{\itemsep}{1pt}
      \setlength{\parskip}{0pt}
      \setlength{\parsep}{0pt}   
        \item \textbf{Exception Count}: A count of the number of unique exceptions thrown by a test suite. 
         \item \textbf{Strong Mutation Coverage}: The existing fitness function in EvoSuite for measuring Strong Mutation Coverage, based on an estimated propagation of corrupted state~\cite{Fraser14:Mutation}.
        \item \textbf{Diversity Score (Levenshtein Distance)}: A new fitness function based on the textual changes required to transform one test case into another. 
      \end{itemize}
  \item \textbf{Combination of all Functions (``Default Approach'')}: A combination of all of the individual fitness functions used in each experiment is used as a baseline as this combination attains reasonable fulfillment of each individual function, and in theory, will produce multifaceted test suites effective at fault-finding~\cite{Rojas15:Combining}. This configuration represents a ``best guess'' at what would produce effective test suites, and would be considered a reasonable approach in the absence of a known, informative fitness function or ``best'' combination.
  \item \textbf{Random Selection of Functions}: The final baseline is a random selection of fitness functions, chosen from the combinations available to AFFS. For each fault, we make a random selection and use that selection for all trials for that fault. We employ this baseline for the exception and diversity goals, but omit it for the Strong Mutation goal in order to control experiment costs, and due to limited value from adding this baseline for that goal (as we will discuss further in Section~\ref{sec:sm_results}). 
  \end{itemize}
  
Test suites are generated that target the classes reported as relevant to the fault by Defects4J. Tests are generated using the fixed version of the CUT and applied to the faulty version in order to eliminate the oracle problem. In practice, this translates to a regression testing scenario, where tests are generated using a version of the system understood to be ``correct'' in order to guard against future issues~\cite{shamshiri15:generation}. Tests that fail on the faulty version, then, detect behavioral differences between the two versions.\footnote{This is identical practice to other studies using EvoSuite and Defects4J, e.g.~\cite{shamshiri15:generation,Gay18:fitness}.}

To perform a fair comparison between approaches, each is allocated a ten minute search budget for test generation. In past work, 10 minutes was used as the maximum generation time and represented a point of ``diminishing returns'' for detection of the faults in Defects4J~\cite{Gay18:fitness}. 

To control experiment cost, we deactivated assertion filtering---all possible regression assertions are included. All other settings were kept at their default values. As results may vary, we performed 10 trials for each fault and search budget. For the Exception experiment, this resulted in the generation of 19,750 test suites (ten trials, five approaches, 395 faults), representing over 3,291 hours of computation time. For the Diversity experiment, this resulted in the generation of 22,600 test suites (ten trials, five approaches, 452 faults), representing over 3,766 hours of computation time. Finally, in the Strong Mutation experiment, this resulted in the generation of 17,360 test suites (ten trials, four approaches, 434 faults), representing over 2,893 hours of computation time. We performed experiments on Amazon EC2 infrastructure, where all VMs shared an identical hardware and software configuration (t2.large instances, with two CPU threads and 8GB of RAM, running Amazon Linux).

Generation tools may generate flaky (unstable) tests~\cite{shamshiri15:generation}. For example, a test case that makes assertions about the system time will only pass during generation. We automatically remove flaky tests. First, all non-compiling test suites are removed. Then, each remaining test suite is executed on the fixed version five times. If the test results are inconsistent, the test case is removed. This process is repeated until all tests pass five times in a row. On average, less than one percent of tests tends to be removed from each suite.

\subsection{Data Collection}\label{sec:data}

In order to address our research questions, we collect the following data for each test suite, based on the goal of the experiment:
\begin{itemize}
\setlength{\itemsep}{1pt}
  \setlength{\parskip}{0pt}
  \setlength{\parsep}{0pt}
  \item \textbf{Exception Experiment:}
      \begin{itemize}
      \setlength{\itemsep}{1pt}
      \setlength{\parskip}{0pt}
      \setlength{\parsep}{0pt}
      \item \textbf{Number of Unique Exceptions Discovered During Generation}
      \item \textbf{Number of Unique Exceptions Thrown by the Final Test Suite}: Tests that trigger an exception can be lost during the generation process. We calculate this number by monitoring test suite execution.
      \end{itemize}
  \item \textbf{Strong Mutation Experiment:}
      \begin{itemize}
      \setlength{\itemsep}{1pt}
      \setlength{\parskip}{0pt}
      \setlength{\parsep}{0pt}
      \item \textbf{Number of Mutants}: The number of mutants inserted into the CUT. 
      \item \textbf{Strong Mutation Coverage}: Percentage of mutants detected, meeting the conditions of Strong Mutation.
      \end{itemize}
 \item \textbf{Diversity Experiment:} 
      \begin{itemize}
      \setlength{\itemsep}{1pt}
      \setlength{\parskip}{0pt}
      \setlength{\parsep}{0pt}
      \item \textbf{Diversity Score}: The diversity score (based on the Levenshtein Distance) for the final test suite. 
      \end{itemize}
  \item \textbf{All Experiments:}
      \begin{itemize}
      \setlength{\itemsep}{1pt}
      \setlength{\parskip}{0pt}
      \setlength{\parsep}{0pt}
      \item \textbf{Number of Faults Detected}
      \item \textbf{Number of Generations of Evolution}: The amount of time that it takes to complete one generation of evolution is not static, and each approach may complete a different number of generations during the test generation process based on the time needed to calculate each employed fitness function. Reinforcement learning will add additional overhead to this process, further decreasing the number of completed generations. We collect the number of generations to assess the impact of fitness function choice and RL overhead.
      \item \textbf{Decisions Made by EvoSuiteFIT}: The reinforcement learning algorithms reformulate the fitness function combination in use at regular intervals. Each time a combination is selected, we log the decision made. This can assist in understanding how the reinforcement learning algorithms function, and how they make decisions in service of goal attainment.
      \end{itemize}
  \end{itemize}

\section{Results and Discussion}\label{sec:results}

We are interested in understanding the effectiveness of EvoSuiteFIT in terms of attainment of our high-level goals---exception discovery, test suite diversity, and Strong Mutation Coverage---and in terms of detection of faults. We are also interested in the impact of the overhead of reinforcement learning on the generation process, how the approaches makes their fitness function selections, and the limitations of adaptive fitness function selection. The following subsections outline and discuss our observations.


\subsection{Goal: Exception Discovery}

\subsubsection{Ability to Discover Exceptions}

Our first question asks whether AFFS can be used to more effectively meet our goal of generating tests that trigger more unique exceptions than baseline static fitness function configurations. We do not know a priori which exceptions can be thrown by a CUT. However, we know that the number of possible exceptions varies from class to class. Therefore, it is not fair to compare raw counts of exceptions between each case example. If we discover thirty exceptions when testing one class, and five when testing another, we should not compare five to thirty. Instead, we \textit{normalize} exception counts between 0-1 for each class-under-test, using the formula:
\begin{equation}
\scriptsize
    (\frac{\text{Number of Exceptions Observed In This Trial For CUT}}{\text{Maximum Number of Observed Exceptions In Any Trial For CUT}})
\end{equation}
This normalization allows fair comparison between case examples.

\begin{table}[!t]
\centering
\scriptsize
\caption{Median count of exceptions discovered for each technique. Counts are normalized between 0-1 for each fault to allow comparison across case examples. Higher scores are better. The highest median is \textbf{bolded}.}
\label{table:resultsThrown}
\begin{tabular}{l|rrrrr}
\hline 
\textbf{System} & \textbf{DSG-Sarsa} & \textbf{UCB} & \textbf{Exception Count} & \textbf{Default} & \textbf{Random} \\ \hline
Chart & \textbf{0.83} & 0.82 & 0.33 & 0.63 & \textbf{0.83} \\ 
Closure & \textbf{0.83} & 0.82 & 0.30 & 0.55 & 0.59 \\ 
Lang & \textbf{0.92} & 0.91 & 0.61 & 0.83 & \textbf{0.92}  \\ 
Math & \textbf{0.89} & \textbf{0.89} & 0.54 & 0.67 & 0.67  \\ 
Mockito & 0.83 & \textbf{0.86} & 0.50 & 0.50 & 0.50  \\ 
Time & \textbf{0.87} & 0.83 & 0.39 & 0.63 & 0.71  \\ \hline
\textbf{Overall} & \textbf{0.87} & 0.86 & 0.42 & 0.63 & 0.68 \\ \hline
\end{tabular}
\end{table}

\begin{figure}[!t]
\centering
\includegraphics[width=3.5in]{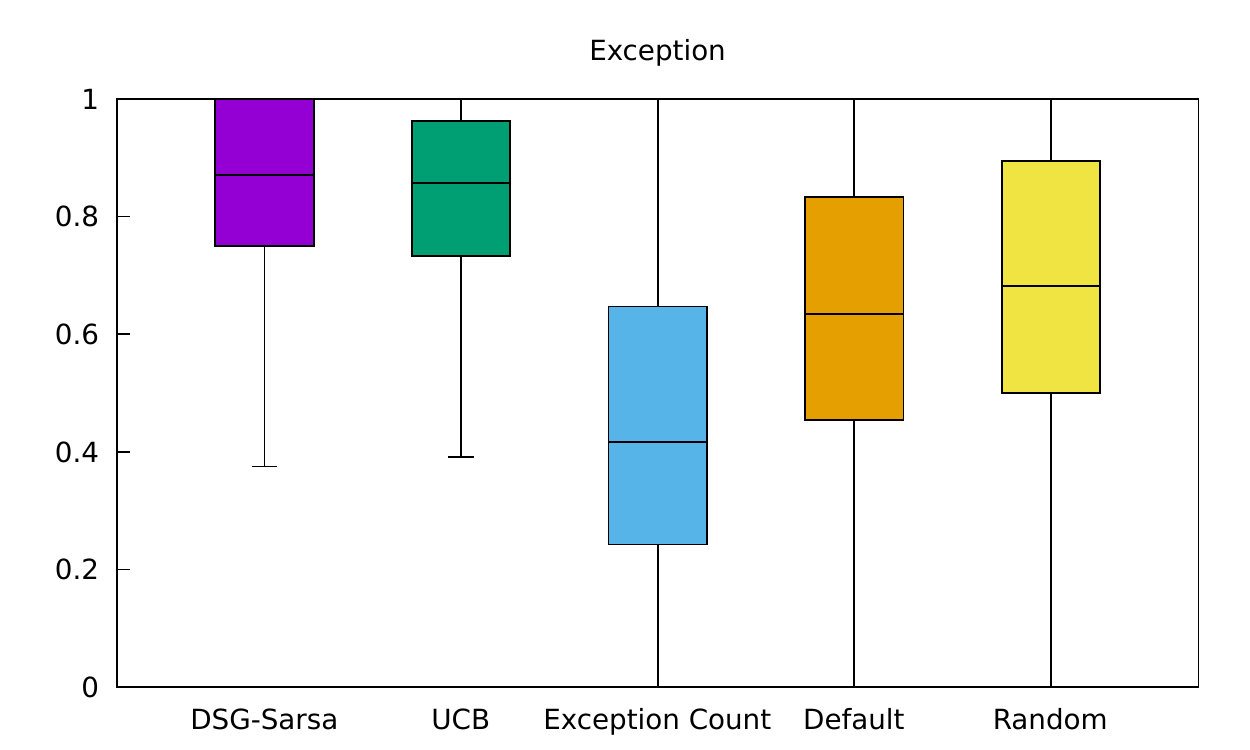} 
  \caption{Unique exceptions discovered by each technique. Counts are normalized between 0-1 for each fault to allow comparison across case examples.}
  \label{fig:boxDisc}
\end{figure}

The median count of unique exceptions discovered for each technique is listed in Table~\ref{table:resultsThrown} for each project and overall. Boxplots of the exceptions discovered are shown in Figure~\ref{fig:boxDisc}. \textbf{Higher scores are better.} Overall, both AFFS techniques have a higher median performance in both measurements than all three baselines. In particular, both approaches outperform current practices---attaining up to a 176.67\% improvement in median exceptions discovered over the basic exception count and up to a 72.00\% improvement over the eight-function default configuration. Overall, DSG-Sarsa attains a 107.14\% improvement in median exception discovery over the simple exception count and 38.10\% over the default combination. EvoSuiteFIT also tends to retain all discovered exceptions, while the default configuration may discard a small number of exception-triggering tests if offered improvements in the other fitness functions. 

On a per-system basis, the third baseline---a random selection---ties in median performance with DSG-Sarsa for Chart and Lang. However, it is outperformed by both AFFS approaches for the other systems. Overall, DSG-Sarsa outperforms the random baseline by 27.94\% in median performance, and UCB outperforms it by 26.47\%. Notably, the random baseline outperforms both of the other baselines---using the existing fitness function and combining several fitness functions. Testers would be better served by choosing a small random set of fitness functions than by blindly combining all options available. We will discuss the primary reasons for this shortly. 

Figure~\ref{fig:boxDisc} also shows that both techniques not only offer a higher median than the baselines, but also have a narrower interquartile spread, showing relatively consistent performance. UCB yields more consistent performance, as shown by the decreased variance. However, DSG-Sarsa has a slightly higher median performance and third quartile. 

We perform statistical analysis to assess our observations. First, we are interested in establishing whether there are differences in performance between the different AFFS techniques and the baselines. If so, we are then interested in examining the magnitude of the differences. For each pair of
techniques and baselines, we formulate hypothesis and null hypotheses:
\begin{itemize}
\setlength{\itemsep}{1pt}
  \setlength{\parskip}{0pt}
  \setlength{\parsep}{0pt}  
\item{$H$: Generated test suites have different distributions of exception discovery results depending on the technique used to generate the suite.}
\item{$H0$: Observations of exception discovery for all techniques are drawn
from the same distribution.}
\end{itemize}

\begin{table}[!t]
\centering
\scriptsize
\caption{Results of Vargha-Delaney A Measure for exceptions discovered. Large positive effect sizes are \textbf{bolded}. Medium positive effect sizes are \textit{italicized}.}
\label{table:resultsVDThrown}
\begin{tabular}{l|rrrrr}
\hline
 & \textbf{DSG-Sarsa} & \textbf{UCB} & \textbf{Exception} & \textbf{Default} & \textbf{Random} \\ \hline
 \textbf{DSG-Sarsa} & -    & 0.52 & \textbf{0.86} & \textbf{0.74} & \textit{0.69} \\ 
 \textbf{UCB} & 0.48 & -    & \textbf{0.85} & \textit{0.73} & \textit{0.67} \\ 
 \textbf{Exception} & 0.14 & 0.15 & - & 0.33 & 0.30 \\ 
 \textbf{Default} & 0.26 & 0.27 & \textit{0.67}  & - & 0.45 \\ 
 \textbf{Random} & 0.32 & 0.33 & \textit{0.70} & 0.55 & - \\ \hline
\end{tabular} 
\end{table}

Our observations are drawn from an unknown distribution.
To evaluate the null hypothesis without any assumptions on distribution, we use the Friedman test,
a non-parametric test for determining whether there are any statistically significant differences between the distributions of three or more paired groups. For each fault, we have a set of paired observations based on the exception discovery performance of the suites generated for that fault by each technique. We apply the test with $\alpha = 0.05$. This test yielded a p-value $<$ 0.001, indicating differences in performance between the techniques. 

Therefore, we then used the Vargha-Delaney A measure to assess effect size~\cite{Vargha00:Measure}. The results for exception discovery are listed in Table~\ref{table:resultsVDThrown}, with large effect sizes in bold ($E \geq 0.80$) and medium effect sizes in italics ($0.80 > E \geq 0.70$). DSG-Sarsa outperforms the default and exception count baselines with large effect size, and the random baseline with medium effect size. UCB outperforms the exception count baseline with a large effect size, and outperforms the default and random baselines with a medium effect size. DSG-Sarsa also outperforms UCB, but with a negligible effect size. 

\begin{center}
\begin{framed}
Both EvoSuiteFIT techniques discover more exceptions than the baseline techniques with significance. DSG-Sarsa outperforms the exception count and default baselines with large effect size (107.14\%, 38.10\% improvement in median) and the random baseline with medium effect (27.94\%).
\end{framed}
\end{center}

\subsubsection{Fault Detection Effectiveness}

\begin{table}[!t]
\centering
\scriptsize
\caption{Percentage of faults detected by each approach for the exception goal. The best approach is \textbf{bolded}.}
\label{table:resultsFaults}
\begin{tabular}{l|rrrrr}
\hline
\textbf{System} & \textbf{DSG-Sarsa} & \textbf{UCB} & \textbf{Exception Count} & \textbf{Default} & \textbf{Random} \\ \hline
Chart & 80.77\%  & \textbf{84.62\%} & 38.46\% & 65.39\% & 69.23\% \\ 
Closure & 6.77\%  & 6.02\% & 3.76\% & 15.04\% & \textbf{15.79\%} \\ 
Lang & 58.46\%  & \textbf{64.62\%}  & 16.92\% & 52.31\% & 53.85\% \\ 
Math & \textbf{68.87\%}  & \textbf{68.87\%}  & 12.26\%  & 57.55\% & 41.51\% \\ 
Mockito & 13.16\%  & \textbf{15.79\%}  & 7.89\% & 13.14\% & 7.89\% \\ 
Time & 59.25\% & \textbf{66.67\%} & 18.52\% & 51.85\% & 55.56\% \\ \hline
\textbf{Overall} & 41.01\% & \textbf{42.78\%}  & 11.90\% & 38.23\% & 34.43\% \\ \hline
\end{tabular}
\end{table}

In theory, forcing the class-under-test to throw exceptions will help developers discover faults in the system. Therefore, our second research question revolves around the ability of the generated test suites to trigger and detect failures. Table~\ref{table:resultsFaults} lists the percentage of faults detected by each technique. We can see that both EvoSuiteFIT techniques generate suites that are able to detect faults that are missed by suites generated using the baselines. UCB, in particular, detects the most faults---4.32\% more than DSG-Sarsa, 11.90\% more than default, 24.25\% more than the random baseline, and 259.50\% more than the exception count. 

The default and random baselines outperform EvoSuiteFIT for one project---Closure. It is likely that triggering these faults requires incorrect output, rather than an exception. The baselines are outperformed on all other systems.

DSG-Sarsa yielded slightly better performance at goal attainment. However, UCB detected more faults. The difference between the two may come down to how fitness functions are chosen. The reinforcement learning strategy, by impacting how and which fitness functions are selected, will impact how input is selected. Differences in how UCB and DSG-Sarsa make selections will influence the resulting likelihood of fault detection. 

As an initial assessment of the connection between goal attainment and fault detection, we calculated the point-biserial correlation coefficient between the normalized goal attainment and whether the fault was detected (the dichotomous variable). This calculation yielded a coefficient of only 0.22, indicating only a weak correlation between fault detection and goal attainment. This suggests that, while improving the ability of the suite to throw exceptions has a positive relationship with detection of the specific faults used in this study, the relationship is far from the only factor influencing fault detection. Rather, factors such as the fitness functions applied in service of improving the goal may also increase the likelihood of selecting input that triggers a particular fault. Further analysis is required to understand the full impact that reinforcement learning strategy can have on fault detection capability. Still, the broad hypothesis that triggering exceptions can aid fault discovery may have merit.

\begin{center}
\begin{framed}
For the exception discovery goal, both EvoSuiteFIT techniques detect faults missed by the other techniques. UCB detects up to 259.90\% more faults than the baselines.
\end{framed}
\end{center}

\subsubsection{Impact of Reinforcement Learning Overhead}

\begin{table}[!t]
\centering
\scriptsize
\caption{Median time per generation (in seconds) for exception discovery goal. The lowest median is \textbf{bolded}.}
\label{table:resultsGens}
\begin{tabular}{l|rrrr}
\hline
 & \textbf{DSG-Sarsa} & \textbf{UCB} & \textbf{Random} & \textbf{Default}\\ \hline
Chart & \textbf{0.24} & 0.26 & 1.38 & 3.29 \\ 
Closure & \textbf{0.32} & 0.49 & 0.45 & 5.71 \\ 
Lang & \textbf{0.30} & 0.44 & 2.02 & 4.38 \\ 
Math & \textbf{0.14} & 0.22 & 2.64 & 3.03 \\ 
Mockito & \textbf{0.03} & \textbf{0.03} & 0.05 & 0.08 \\ 
Time & 0.33 & 0.43 & 2.67 & 3.72 \\ \hline
\textbf{Overall} & \textbf{0.22 }& 0.31 & 0.91 & 3.84 \\ \hline
\end{tabular}
\end{table}

Search-based test generation approaches are generally benchmarked using a fixed time budget~\cite{shamshiri15:generation}. During this period, the amount of work completed by each algorithm may not be equal. The number of generations of evolution will largely depend on total cost to calculate fitness. The addition of reinforcement learning will further impact this cost due to reward score calculation and action selection mechanisms. We are interested in understanding whether the cost of reinforcement learning has more of an effect than the cost of fitness calculation, and the further impact of being able to change the set of fitness functions on this cost.

Table~\ref{table:resultsGens} lists the median time per generation for DSG-Sarsa, UCB, and the default and random baselines. An issue in the version of EvoSuite deployed prevented us from collecting accurate generation times for the exception count alone, but---as it is an extremely simple count that does not require sophisticated instrumentation---it can be assumed that its computation is far less expensive than any other option. 

From Table~\ref{table:resultsGens}, we can see that the median time per generation tends to increase as additional fitness functions are added to the calculation, with the time for the default combination often being many times higher than either EvoSuiteFIT approach. While reinforcement learning may add to the cost of generation, its overhead is less than that required to compute a large number of fitness functions. 

Most of the potential users of a test generation framework would, rightfully, not be interested in tinkering with fitness functions until they stumbled on the right approach. In the absence of perfect knowledge, using the ``default''---all eight fitness functions---is a reasonable idea. It is also an expensive option. Reinforcement learning can reduce the time required to generate effective test cases. 

Both AFFS approaches are similar in speed to the random selection, if not faster. Figure~\ref{fig:choices} helps explain why. In Figure~\ref{fig:choices}, we show the ten actions chosen most often by DSG-Sarsa for each system. While DSG-Sarsa can choose combinations of up to four fitness functions, it rarely does so in practice. Often, either the simple exception count is used, or the combination of the exception count and method (no exception) coverage. The latter is a count of methods called without throwing an exception, which can be calculated efficiently. Because EvoSuiteFIT can strategically change its fitness function selection, overhead added by reinforcement learning is mitigated by the gain in speed from the ability to avoid calculating unhelpful fitness functions. 

The need to calculate these fitness scores helps explain the difference in performance. The default baseline requires more time per generation than AFFS techniques. In turn, this means that AFFS techniques are able to refine test suites further during the time allocated to the search than the default baseline, or even the random baseline. Reinforcement learning adds overhead to the generation process, but the ability to vary the generation strategy can mitigate the impact of that overhead.

\begin{center}
\begin{framed}
The ability to avoid unhelpful fitness functions mitigates reinforcement learning overhead. Both AFFS approaches are able to complete more generations of evolution during than the default and random baselines, with DSG-Sarsa being 94.27\% and 75.82\% faster on average.
\end{framed}
\end{center}

\subsubsection{Actions Selected by AFFS}

\begin{figure}[!t]
\centering
   \begin{subfigure}[t]{0.5\textwidth}
        \centering
        \includegraphics[width=2.37in]{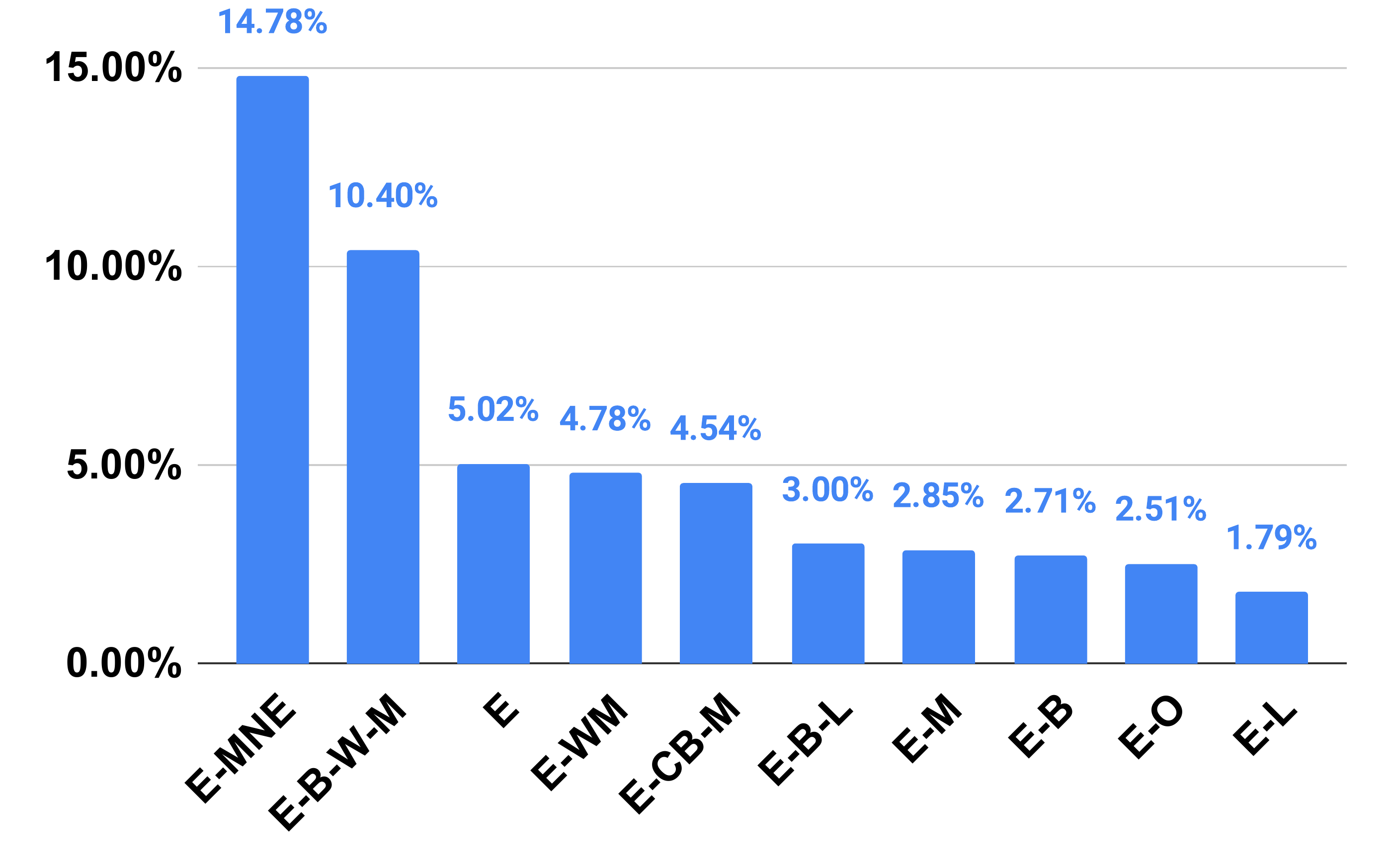} 
        \caption{Chart}
    \end{subfigure}%
    \begin{subfigure}[t]{0.5\textwidth}
        \centering
        \includegraphics[width=2.37in]{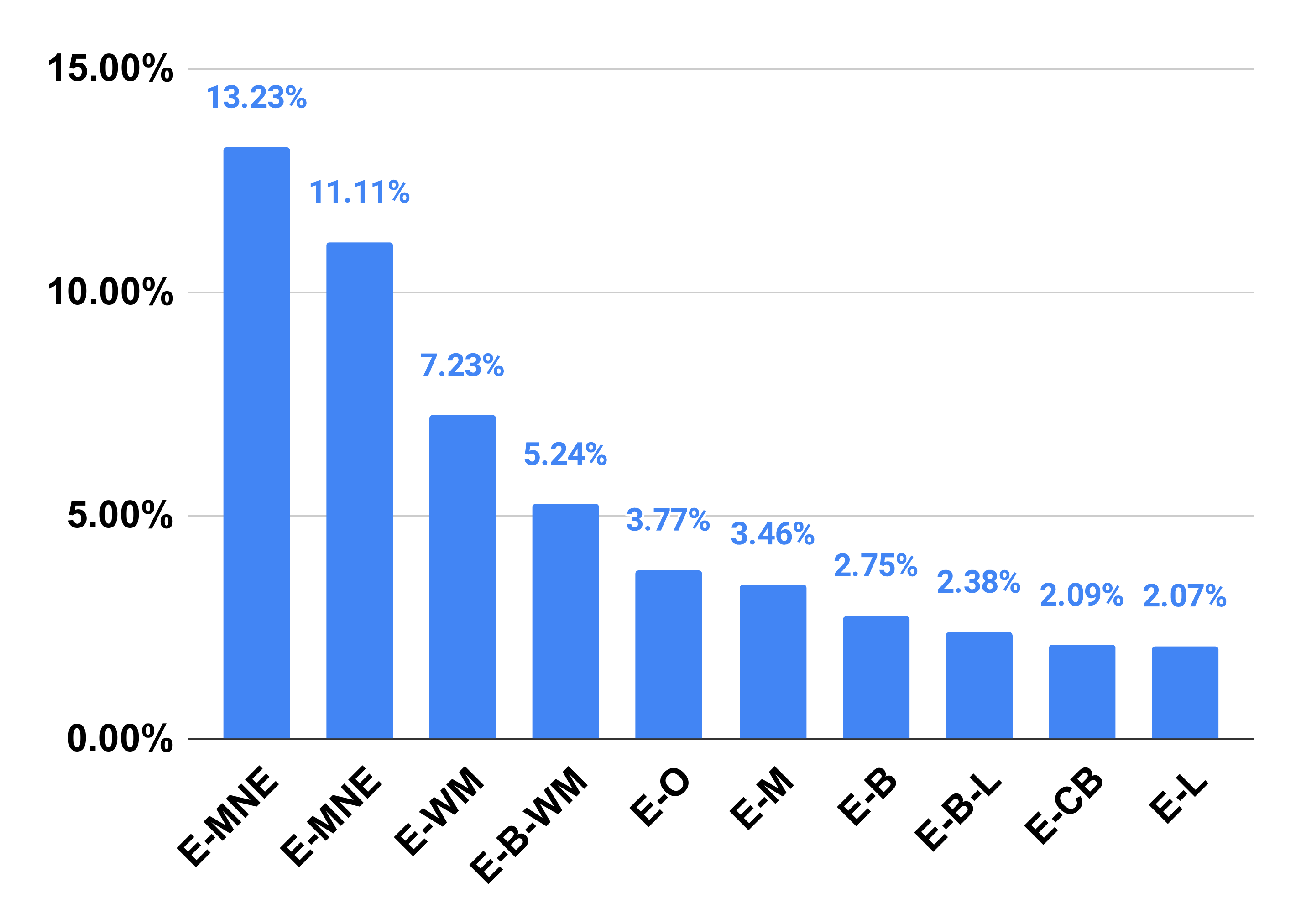} 
        \caption{Closure}
    \end{subfigure}%
    
    \begin{subfigure}[t]{0.5\textwidth}
        \centering
        \includegraphics[width=2.37in]{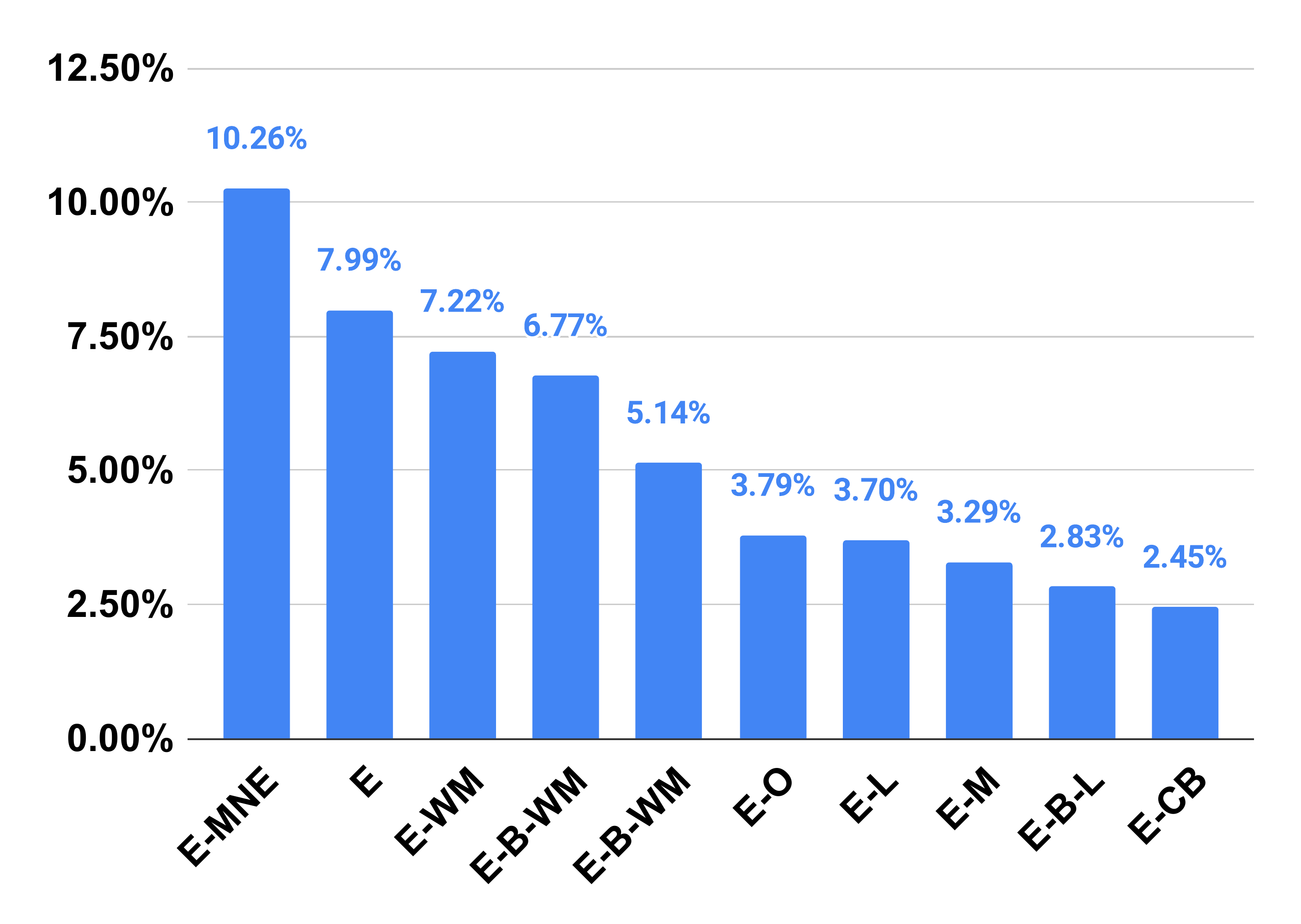} 
        \caption{Lang}
    \end{subfigure}%
    \begin{subfigure}[t]{0.5\textwidth}
        \centering
        \includegraphics[width=2.37in]{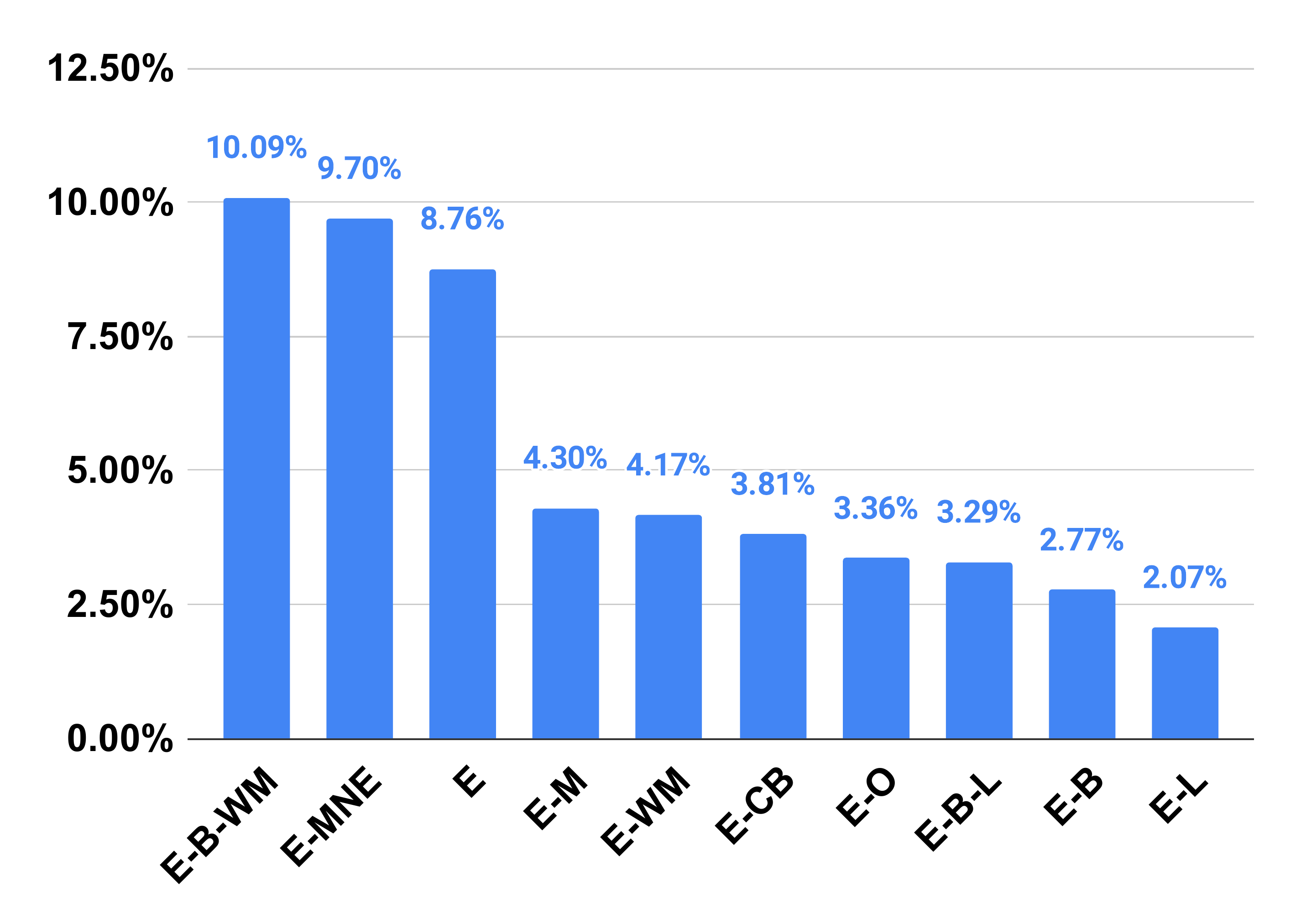} 
        \caption{Math}
    \end{subfigure}%
    
    \begin{subfigure}[t]{0.5\textwidth}
        \centering
        \includegraphics[width=2.37in]{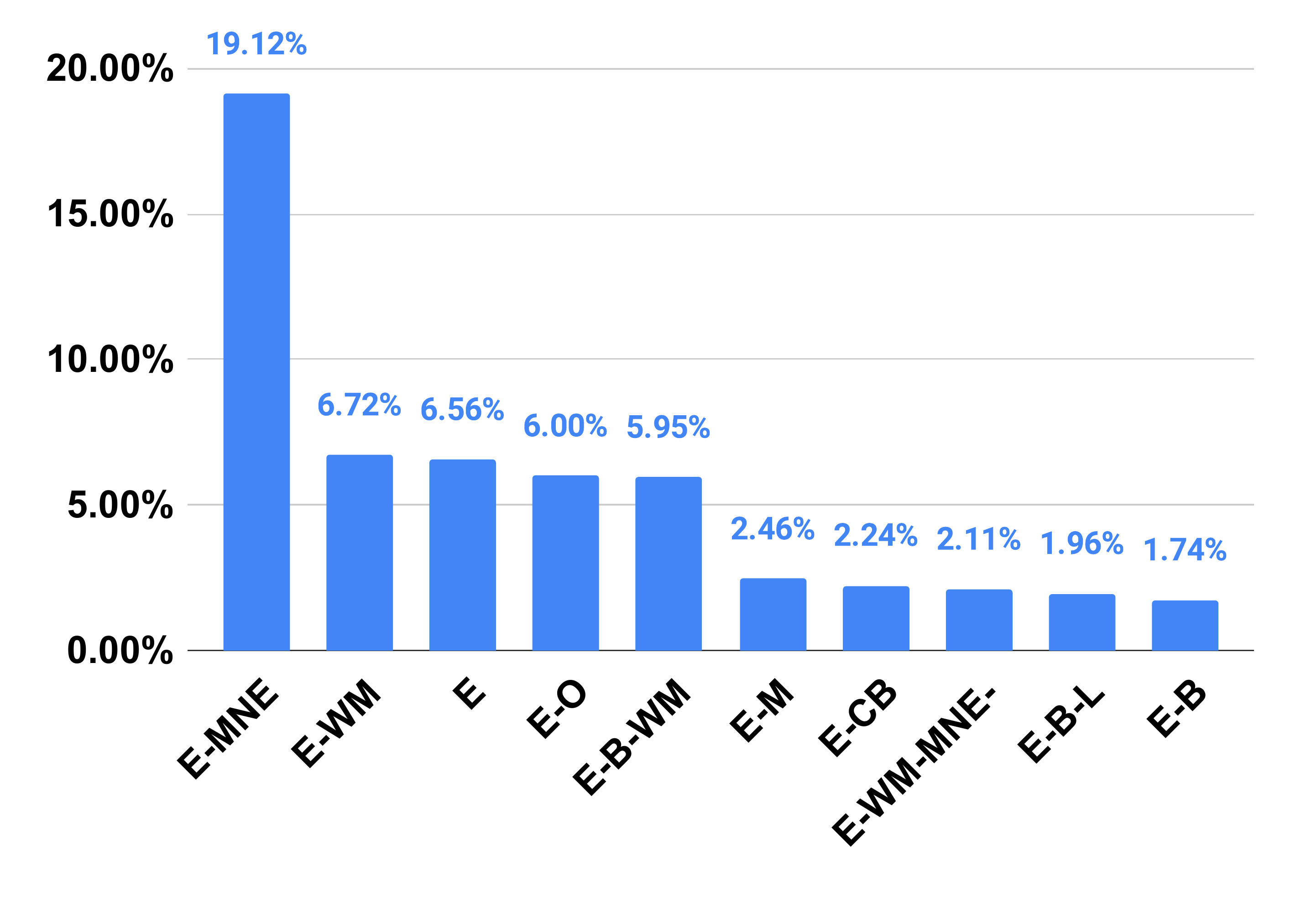} 
        \caption{Mockito}
    \end{subfigure}%
    \begin{subfigure}[t]{0.5\textwidth}
        \centering
        \includegraphics[width=2.37in]{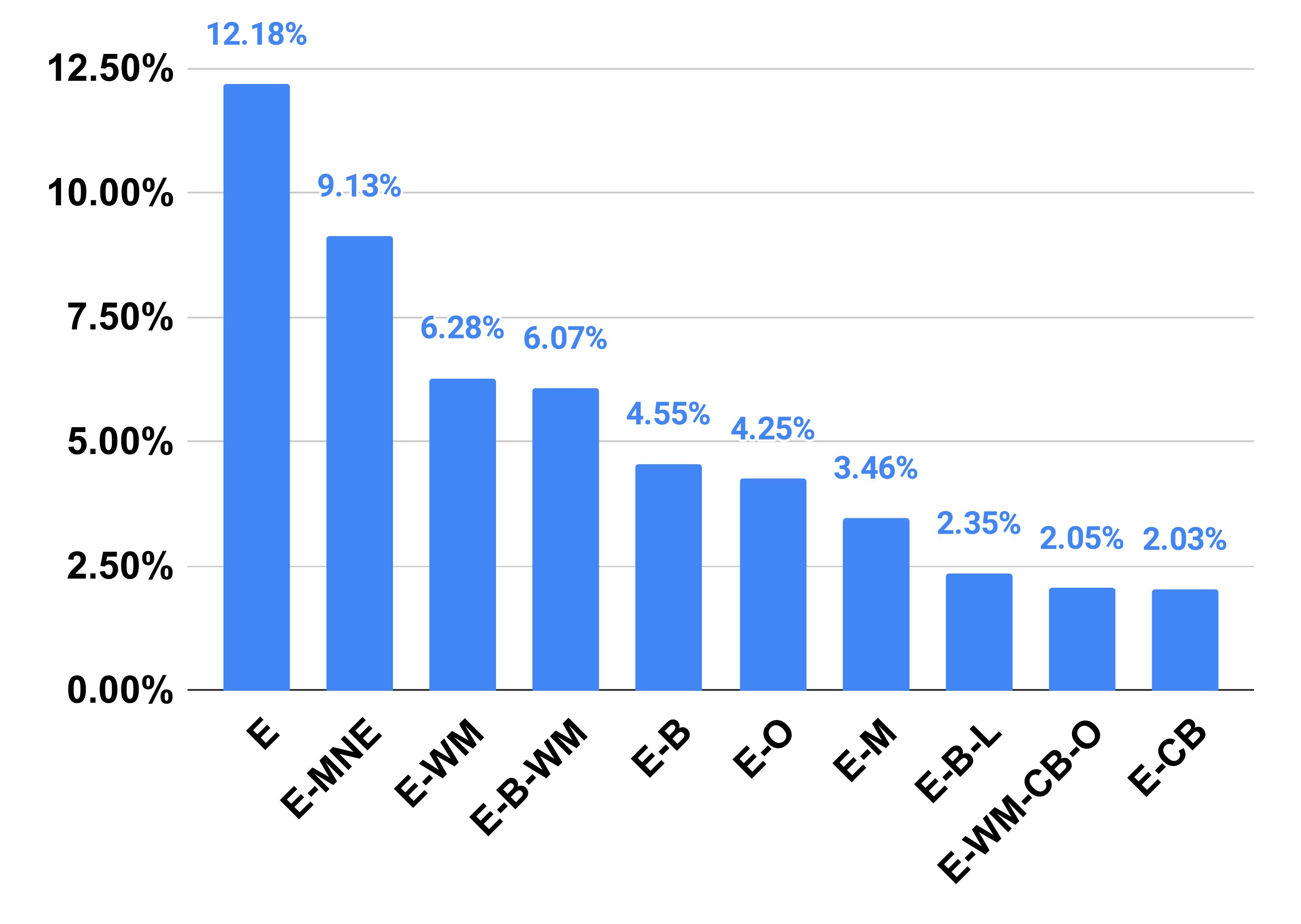} 
        \caption{Time}
    \end{subfigure}%
\caption{Top ten function combinations chosen by \textbf{DSG-Sarsa} for each system for the exception discovery goal. E = Exception Count, B = Branch Coverage, CB = Direct Branch Coverage, L = Line Coverage, O = Output Coverage, M = Method Coverage, MNE = Method (No Exception), WM = Weak Mutation Coverage}
  \label{fig:choices}
\end{figure}

\begin{figure}[!t]
\centering
   \begin{subfigure}[t]{0.5\textwidth}
        \centering
        \includegraphics[width=2.37in]{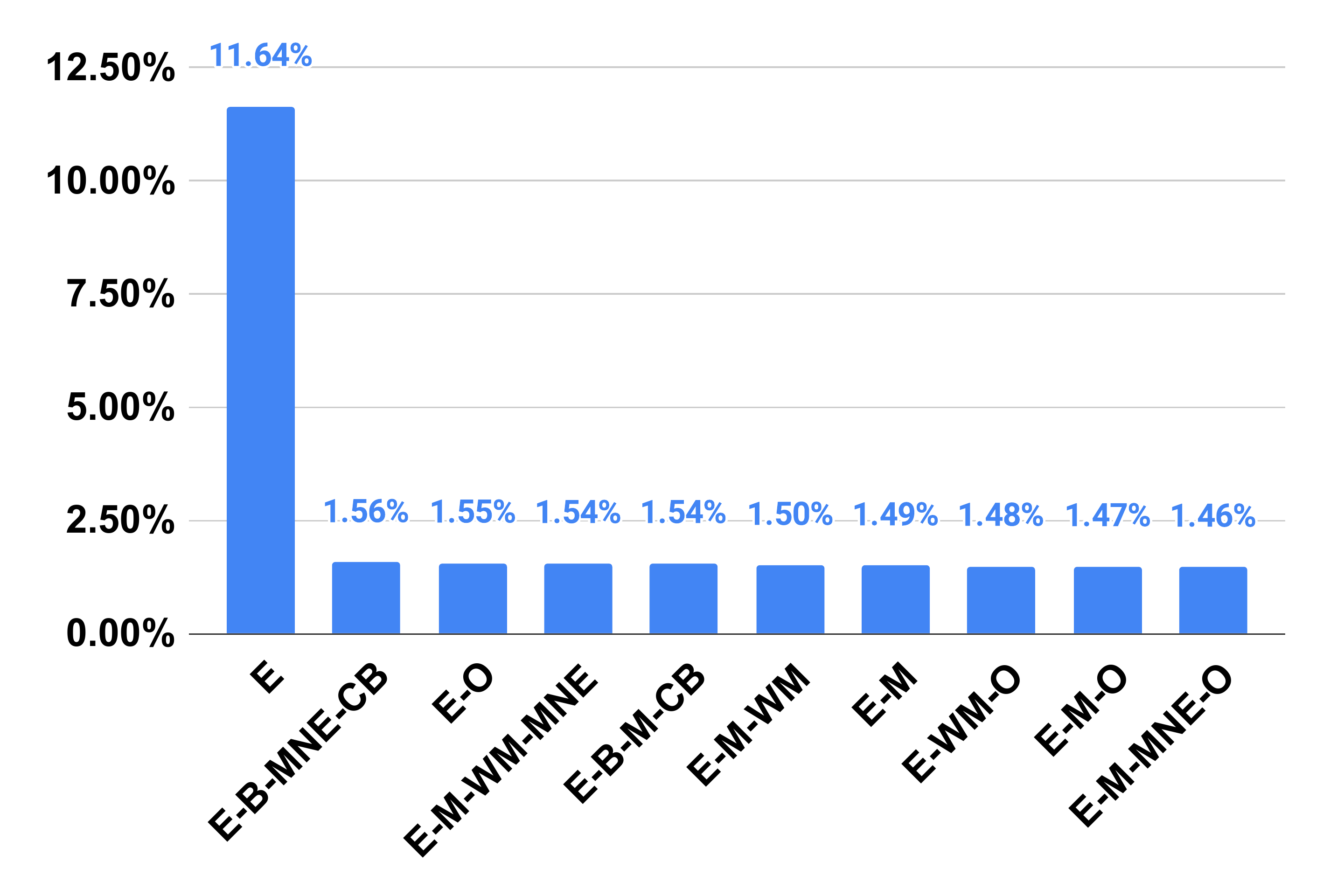} 
        \caption{Chart}
    \end{subfigure}%
    \begin{subfigure}[t]{0.5\textwidth}
        \centering
        \includegraphics[width=2.37in]{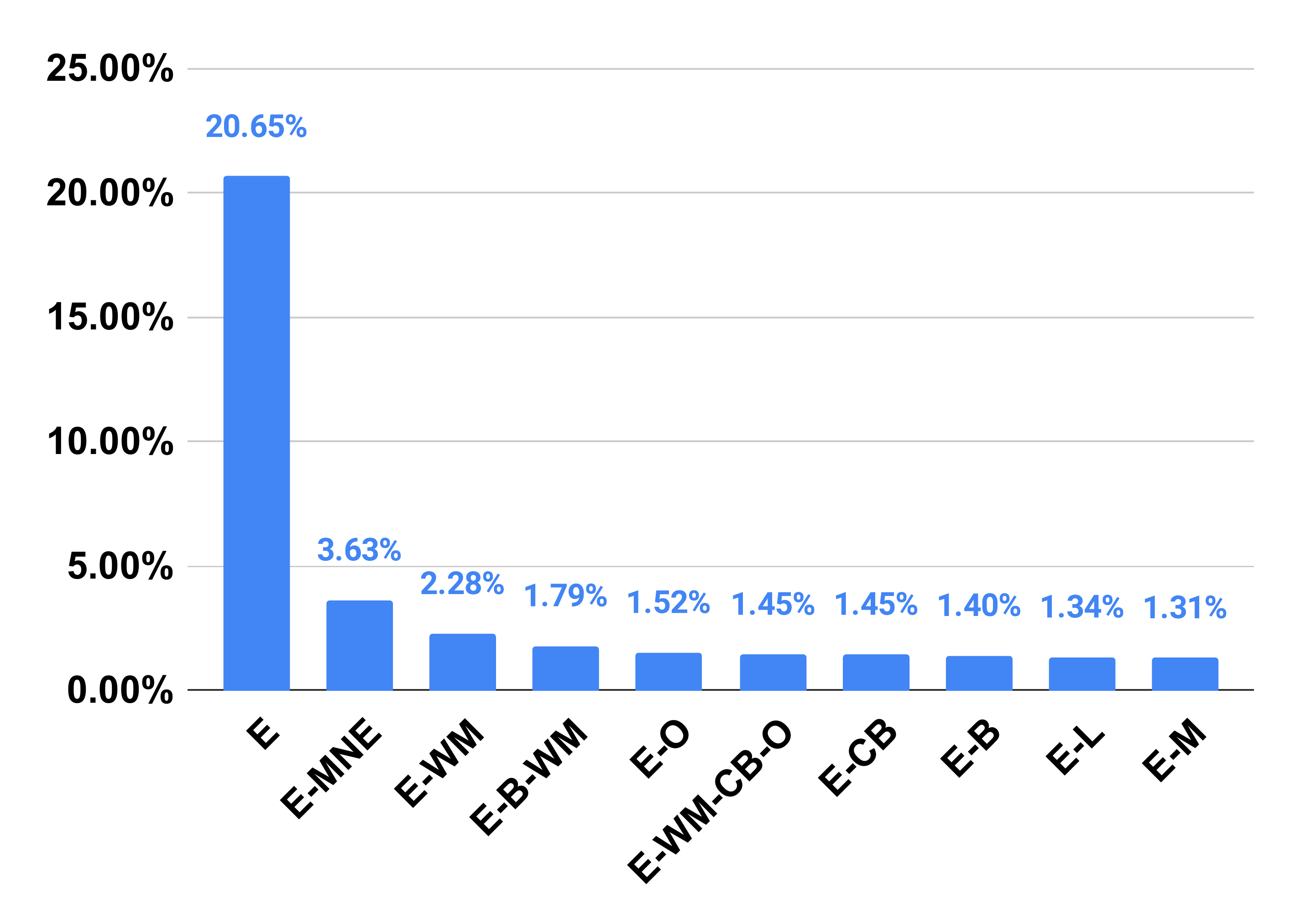} 
        \caption{Closure}
    \end{subfigure}%
    
    \begin{subfigure}[t]{0.5\textwidth}
        \centering
        \includegraphics[width=2.37in]{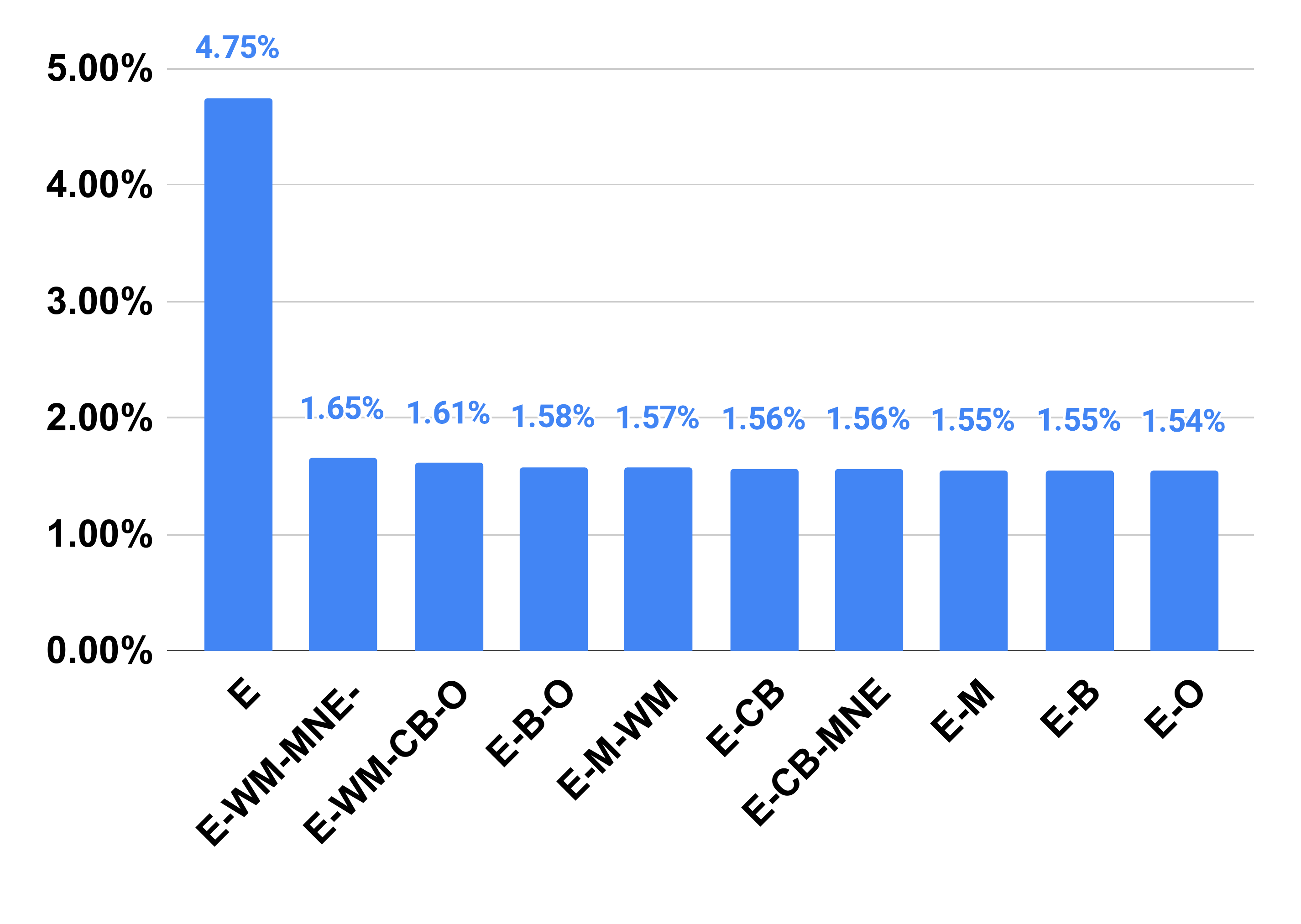} 
        \caption{Lang}
    \end{subfigure}%
    \begin{subfigure}[t]{0.5\textwidth}
        \centering
        \includegraphics[width=2.37in]{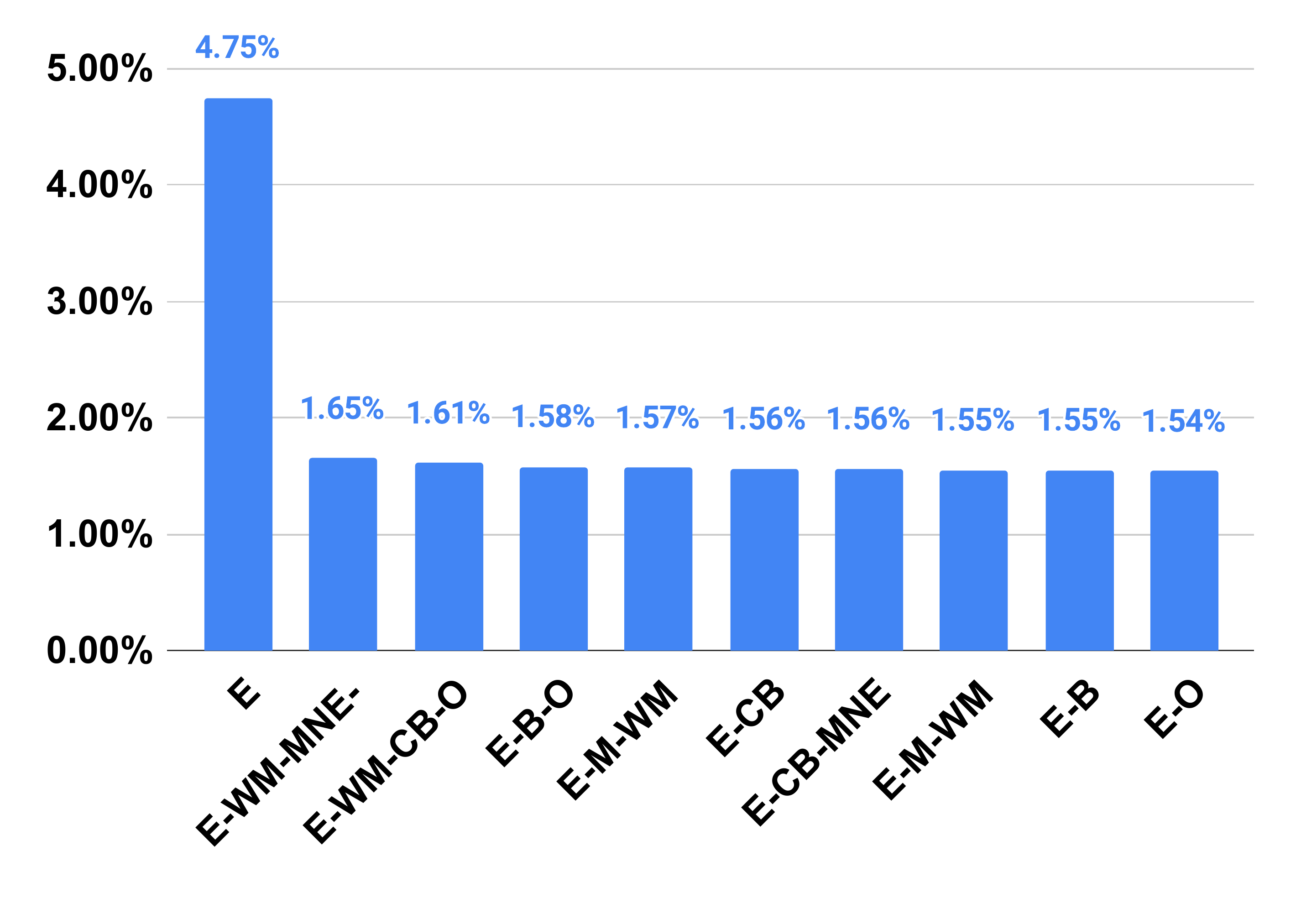} 
        \caption{Math}
    \end{subfigure}%
    
    \begin{subfigure}[t]{0.5\textwidth}
        \centering
        \includegraphics[width=2.37in]{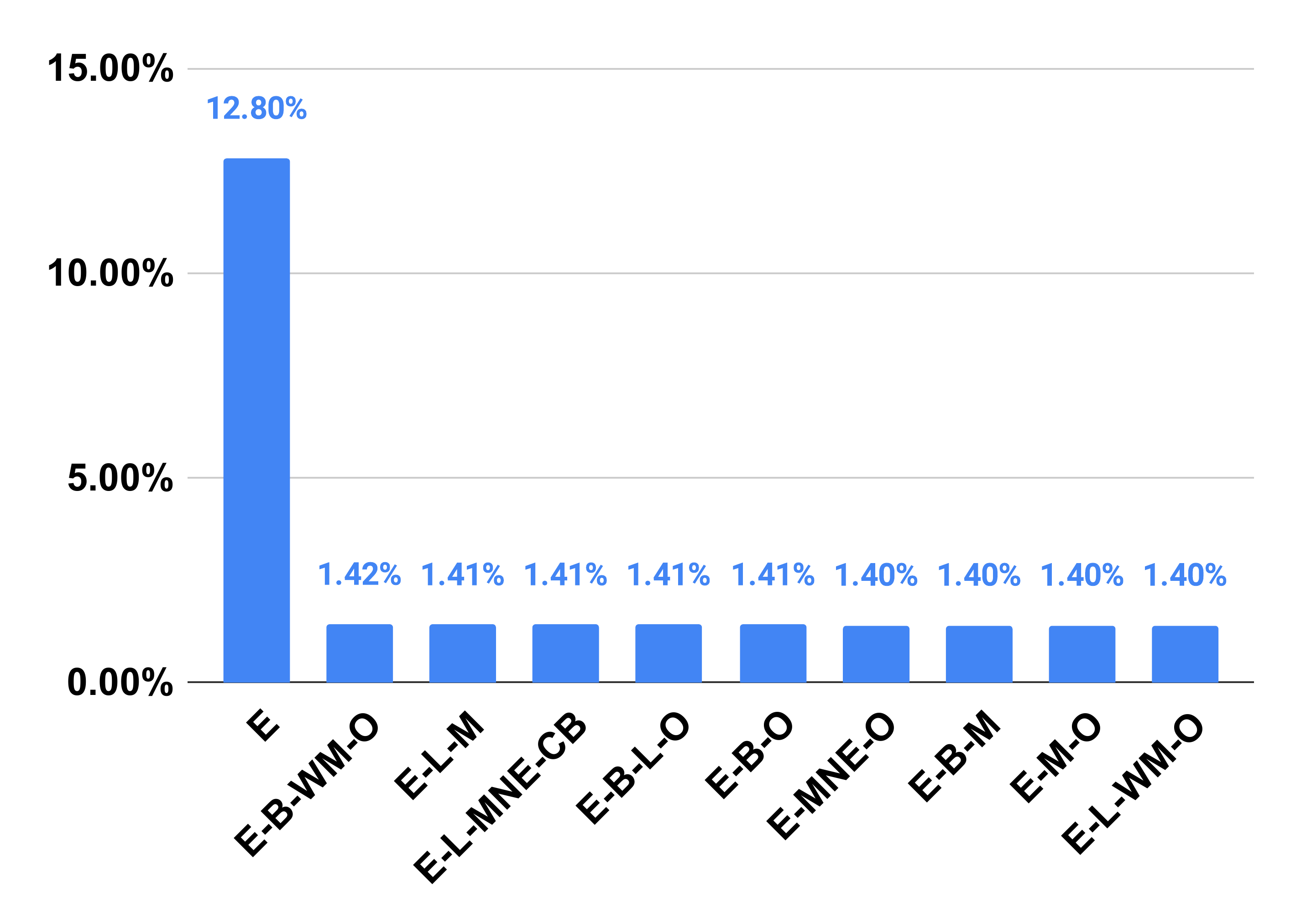} 
        \caption{Mockito}
    \end{subfigure}%
    \begin{subfigure}[t]{0.5\textwidth}
        \centering
        \includegraphics[width=2.37in]{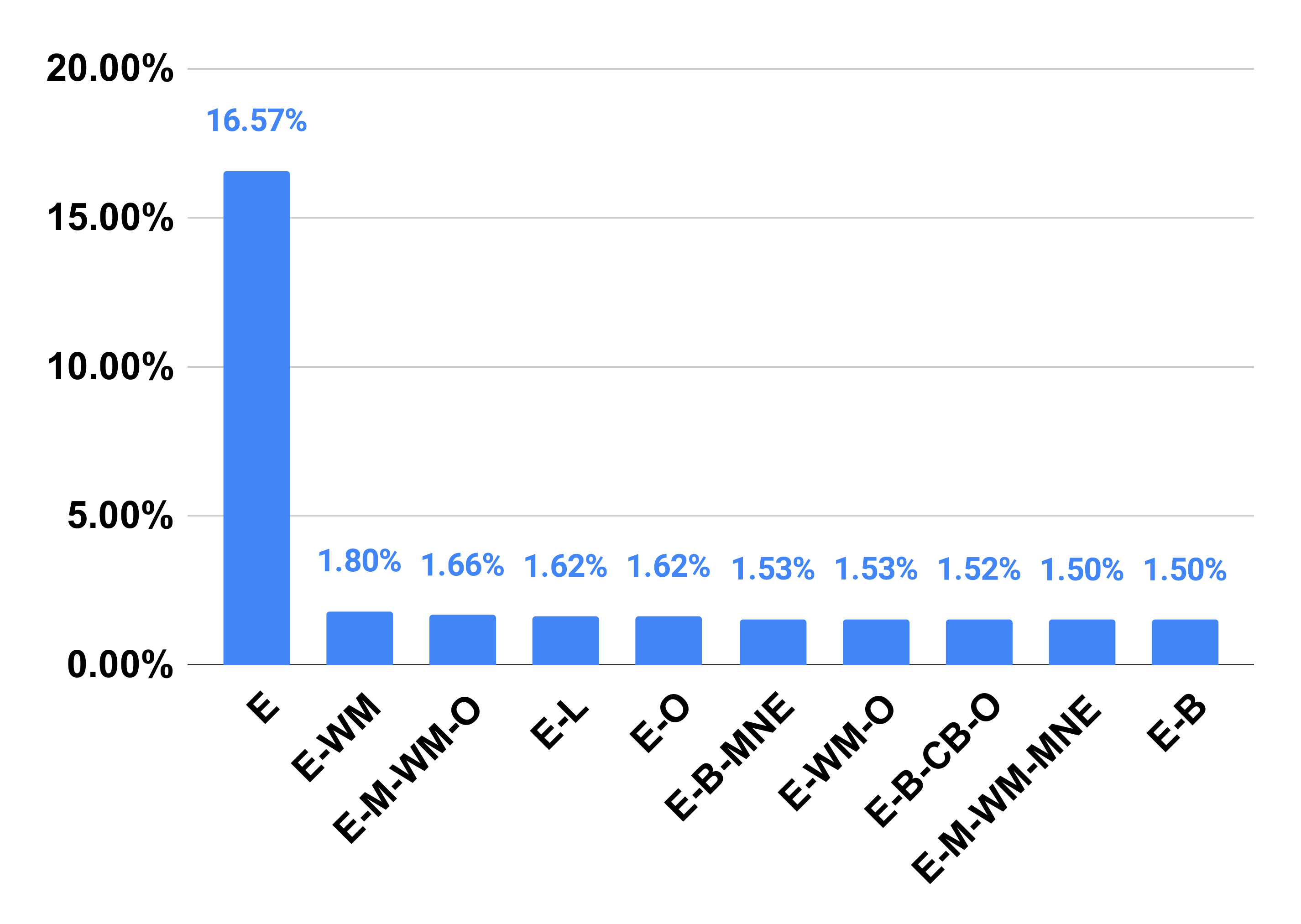} 
        \caption{Time}
    \end{subfigure}%
    
\caption{Top ten function combinations chosen by \textbf{UCB} for the exception discovery goal.  E = Exception Count, B = Branch Coverage, CB = Direct Branch Coverage, L = Line Coverage, O = Output Coverage, M = Method Coverage, MNE = Method (No Exception), WM = Weak Mutation Coverage Coverage}
  \label{fig:choices_ucb}
\end{figure}

For the exception discovery goal, EvoSuiteFIT is able to freely alternate between 64 combinations of fitness functions. To help understand why reinforcement learning is effective, we should examine the actions chosen by AFFS techniques. In Figure~\ref{fig:choices}, we display the ten actions chosen most often by DSG-Sarsa for each system, and in Figure~\ref{fig:choices_ucb}, we do the same for UCB. 

From Figure~\ref{fig:choices}-\ref{fig:choices_ucb}, we can see differences between projects in terms of which choices are made and how often choices are made. For example, DSG-Sarsa frequently used a combination of exception count, Direct Branch Coverage, Weak Mutation Coverage, and Output Coverage for the Time system, but not for others. Although the ordering differs, however, there are also a lot of commonalities in the choices. 

For the most part, the combinations favored by DSG-Sarsa are simple---pairing exception count with one additional fitness function. It is reasonable that simple combinations would be used frequently. Larger combinations introduce a risk of conflicting goals, and are harder to maximize. Simple combinations offer enough feedback to increase the exception count, without adding noise to the search. 

UCB chooses complex sets of actions, combinations of 3-4 fitness functions, somewhat more often than DSG-Sarsa. However, it does not necessarily do so significantly more often than it chooses simple combinations. The most noticeable factor about UCB, as seen in Figure~\ref{fig:choices_ucb}, is that it heavily favors the simple exception count---applying it far more often than it does any other action. 

Many of the fitness function combinations chosen heavily by DSG-Sarsa or UCB would yield poor results when used on their own, as static fitness functions for suite generation. Both often use the pure exception count, when this yields poor results when used as the sole fitness function. Similarly, we know from past unpublished experiments that the EX-MNE combination produces poor results when used as a static choice, yet DSG-Sarsa applies it heavily. 

The EX-MNE combination appears to be contradictory at first. MNE rewards calling each method of the CUT and it executing without throwing an exception. However, it is possible to attain high fitness in both functions at the same time, as each test case can call multiple methods (and there are multiple test cases in a suite). A test case might call method \texttt{X()} twice. If it executed once without an exception and threw an exception in the second call, it would increase fitness for both functions. 

It is important to remember that test generation is a stateful process. Each round of the generation process builds on the results of previous rounds. There are times where the choices that DSG-Sarsa makes are relevant given the state of generation, even if those choices yield poor results when used in a static context. For example, if a suite \textit{already} has achieved a high level of code coverage, it would make sense to switch to pure use of the exception count to further tune the population of test suites. Similarly, the exception and MNE combination makes sense as a strategic choice because it adds a light feedback mechanism to the exception count. When the combination is employed, new exceptions may be discovered, but the simple count of methods called might prevent loss of code coverage as other fitness functions are explored. The EX-MNE combination may be ineffective in a static context, as it does not offer enough feedback to fully explore the code structure. However, it can be very effective if chosen at the right stage of the generation process, as part of an adaptive process.

AFFS may use combinations early on that---for example---rapidly advance coverage of the source code. Combinations involving Branch Coverage could be used for early gain, then a lightweight combination of exception count and MNE could further sculpt the test suite in a way that allows discovery of additional exceptions. Combinations like the exception count and Output Coverage would potentially be very useful in this same situation to diversify input selections after the suite has already evolved to achieve high code coverage.

\begin{center}
\begin{framed}
The ability to adjust the fitness functions at regular intervals allows EvoSuiteFIT to make strategic choices that refine the test suite. Fitness function combinations that are ineffective in a static context may be effective when used by AFFS to diversify a pre-evolved population of suites.
\end{framed}
\end{center}


\subsection{Goal: Test Suite Diversity}

\subsubsection{Ability to Improve Suite Diversity}

\begin{table}[!t]
\centering
\scriptsize
\caption{Median diversity fitness score of the produced test suite. Score is between 0-1, with \textbf{lower scores being better}. The lowest median is \textbf{bolded}.}
\label{table:diversity_results}
\begin{tabular}{l|rrrrr}
\hline
 & \textbf{DSG-Sarsa} & \textbf{UCB} & \textbf{Default} & \textbf{Diversity Score} & \textbf{Random} \\ \hline
Chart & 7.45E-07 & \textbf{4.20E-07} & 1.11E-06 & 4.40E-06 & 1.12E-06\\ 
Closure & \textbf{1.49E-06} & 1.59E-06 & 1.82E-06 & 5.64E-06 & 2.40E-06 \\
Gson & 1.43E-06 & \textbf{9.51E-07} & 1.86E-06 & 3.46E-06  & 2.28E-06 \\ 
Lang & 5.21E-07 & \textbf{3.05E-07} & 1.11E-06 & 3.85E-06 & 1.32E-06\\ 
Math & 1.32E-06 & \textbf{1.06E-06} & 1.54E-06 & 4.02E-06 & 1.88E-06 \\ 
Mockito & \textbf{2.32E-06}  & 2.35E-06 & 3.69E-06 & 5.20E-06 & 4.30E-06 \\ 
Time & 6.58E-07 & \textbf{4.39E-07} & 9.74E-07& 3.51E-06 & 9.00E-07 \\ \hline
\textbf{Overall} & 1.32E-06 & \textbf{1.02E-06} & 1.73E-06 & 4.30E-06 & 1.87E-06 \\ \hline
\end{tabular}
\end{table}

\begin{figure}[!t]
\centering
\includegraphics[width=3.5in]{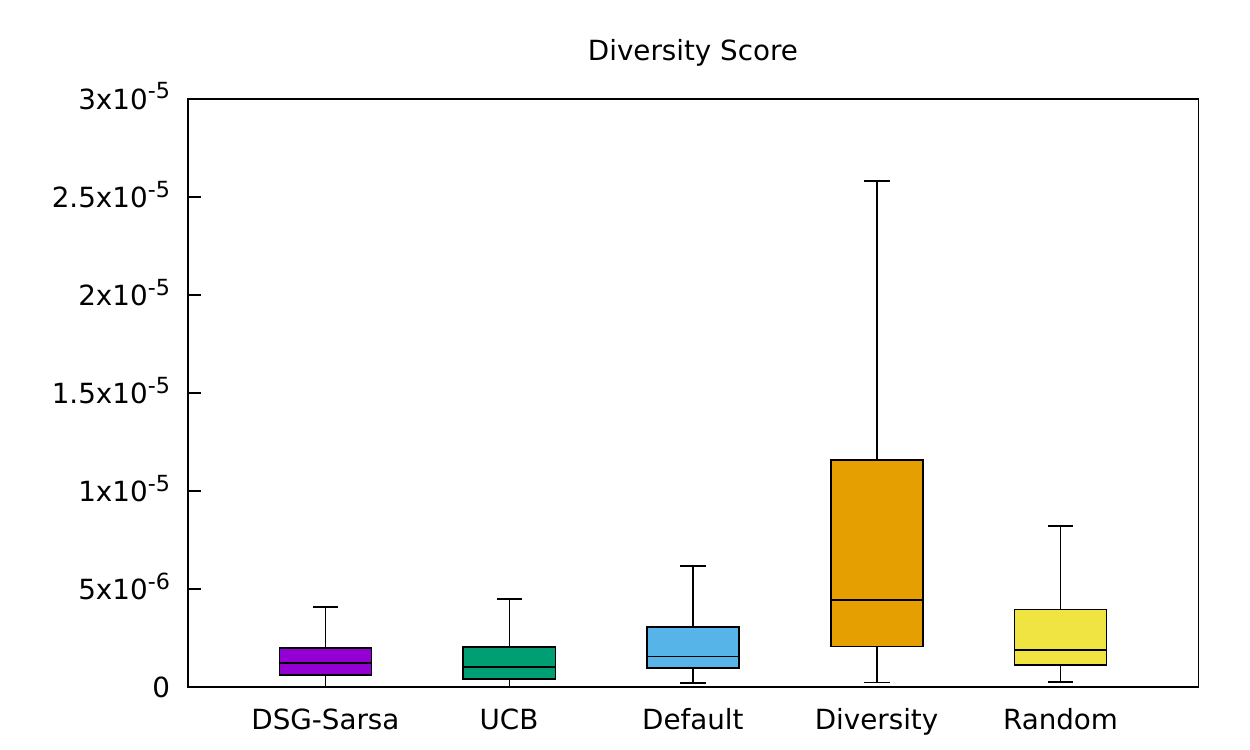}
  \caption{Diversity fitness scores of the produced test suites. Score is between 0-1, with lower scores being better.}
  \label{fig:boxDiversityScore}
\end{figure}

Again, our first question concerns the ability of AFFS to meet our goal of diverse test suites. We assess this by examining the diversity fitness score. In this case, scores range between 0-1, and \textbf{lower scores} indicate higher levels of diversity. In Table~\ref{table:diversity_results}, we indicate the median diversity score for each technique for each project and overall. In Figure~\ref{fig:boxDiversityScore}, we show boxplots for each technique.

From Table~\ref{table:diversity_results} and Figure~\ref{fig:boxDiversityScore}, we see that both AFFS techniques outperform all three baselines. Diversity-alone serves as a poor fitness target, confirming our initial concerns. This fitness function---while representing a valid high-level goal---offers insufficient feedback to achieve that goal. This can be seen in the worse median score and the wide variance in Figure~\ref{fig:boxDiversityScore}. 

The default baseline attains better results than diversity-alone or the random baseline. However, both AFFS techniques outperform it. Overall, the best technique, UCB, attains 76.27\% better median performance than diversity alone, 45.45\% better than the random baseline, 41.04\% better than the default combination, and 22.72\% better than DSG-Sarsa. From Figure~\ref{fig:boxDiversityScore}, we also see that DSG-Sarsa also shows less variance in its results than UCB. UCB attains \textit{better} results, but DSG-Sarsa is more consistent. The state approximation performed by DSG-Sarsa may result in less variance in performance. 

\begin{table}[!t]
\centering
\scriptsize
\caption{Results of Vargha-Delaney A Measure for diversity score. Large positive effect sizes are \textbf{bolded}. Medium effect sizes are \textit{italicized}.}
\label{table:resultsVDDiv}
\begin{tabular}{l|rrrrr}
\hline
 & \textbf{DSG-Sarsa} & \textbf{UCB} & \textbf{Default} & \textbf{Diversity Score} & \textbf{Random} \\ \hline
 \textbf{DSG-Sarsa}      & -    & 0.47 & \textit{0.63} & \textbf{0.87} & \textit{0.68} \\ 
 \textbf{UCB}            & 0.53 & -    & \textit{0.66} & \textbf{0.86} & \textit{0.71} \\ 
 \textbf{Default}    & 0.36 & 0.34 & -    &  \textbf{0.78} &  0.56 \\ 
 \textbf{Diversity Score} & 0.13 & 0.13 & 0.23 & - & 0.28  \\ 
 \textbf{Random} & 0.32 & 0.29 & 0.44 & \textit{0.72} & - \\ \hline
\end{tabular} 
\end{table}

We again perform statistical analysis to assess our observations. We formulate hypothesis and null hypothesis:
\begin{itemize}
\setlength{\itemsep}{1pt}
  \setlength{\parskip}{0pt}
  \setlength{\parsep}{0pt}  
\item{$H$: Generated test suites have different distributions of diversity score results depending on the technique used to generate the suite.}
\item{$H0$: Observations of diversity score for all techniques are drawn from the same distribution.}
\end{itemize}

The Friedman test confirms, with p-value $<$ 0.001, that there are significant differences between the distributions of the AFFS approaches and baselines. The results for the Vargha-Delaney A measure are listed in Table~\ref{table:resultsVDDiv}, with large effect sizes in bold and medium effect sizes in italics. The results of this test further confirm our observations. Both techniques outperform the diversity score baseline with large effect size and the other baselines with medium effect size. The default combination outperforms diversity-only with medium effective size, and the random baseline outperforms the diversity-only baseline with medium effect size. 

\begin{center}
\begin{framed}
Both EvoSuiteFIT techniques produce more diverse test suites than static baselines with significance. UCB outperforms the diversity score with large effect size (76.27\% improvement in median) and the default and random baselines with medium effect (41.04\%, 45.45\%).
\end{framed}
\end{center}

\subsubsection{Fault Detection Effectiveness}

\begin{table}[!t]
\centering
\scriptsize
\caption{Percentage of faults detected by each approach for the diversity goal. The best approach is \textbf{bolded}.}
\label{table:diverstity_faultDetectionCount}
\begin{tabular}{l|rrrrr}
\hline
 & \textbf{DSG-Sarsa} & \textbf{UCB} & \textbf{Default} & \textbf{Diversity Score} & \textbf{Random} \\ \hline
Chart & 34.61\% & 42.31\% & 34.61\% & 26.92\% & \textbf{53.85\%} \\ 
Closure & 10.80\% & 8.52\% & 7.34\% & 5.68\% & \textbf{12.50\%} \\ 
Gson & \textbf{22.00\%} & 16.67\% & 16.67\% & 11.11\% & 16.67\%  \\ 
Lang & 26.15\% & 21.54\% & 24.61\% & 18.46\% & \textbf{32.31\%} \\ 
Math & 31.13\% & 30.19\% & 30.19\% & 21.69\% & \textbf{46.22\%} \\ 
Mockito & 5.26\%  & 5.26\% & 5.26\% & 5.26\% & \textbf{10.53\%} \\ 
Time & 29.63\% & 29.63\% & 29.63\% & 22.22\% & \textbf{40.74\%} \\ \hline
\textbf{Overall} & 20.18\% & 18.64\% & 18.20\% & 13.60\% & \textbf{27.19\%} \\ \hline
\end{tabular}
\vspace{0pt}
\end{table}

Proponents of test suite diversity have noted a positive relationship between diversity and the likelihood of fault detection. Logically, test suites that \textit{apply a larger variety of stimuli} to the CUT should be more likely to detect faults just by virtue of not performing the same actions over and over again. AFFS does increase suite diversity. Therefore, we also are curious about whether it increases the potential for fault detection. Table~\ref{table:diverstity_faultDetectionCount} lists the percentage of faults detected by each approach. 

Overall, the AFFS approaches detect more faults than the diversity score and default baselines. DSG-Sarsa detects 8.26\% more faults than UCB, 10.88\% more than the default combination, and 48.38\% more more than optimizing for diversity alone. However, the random baseline detects significantly more faults than either the other baselines or the AFFS techniques. The random baseline detects 34.74\% more faults than DSG-Sarsa and 45.87\% more than UCB.

We again calculated the point-biserial correlation coefficient between diversity and fault detection. The calculated coefficient was 0.01---indicating a practically non-existent relationship between goal attainment and fault detection in this experiment. This should not be interpreted as a conclusion that improved diversity will not improve the likelihood of fault detection \textit{in general}. However, in this specific experiment, the diversity of a test suite was not a significant factor in whether faults were detected. Highly-specific input is needed to trigger many of the faults in Defects4J, and improving suite diversity does little to locate those faults. 

This can also be seen by comparing the percentage of faults detected between this experiment and the exception discovery experiment. Many more faults were detected in that experiment, suggesting that other fitness functions may offer better guidance for locating the specific input that is needed to trigger those faults. Again, we stress that this does not mean that test suite diversity is in unimportant goal, or that it is not helpful in general. However, it may be less helpful for the specific examples in Defects4J than other fitness functions.

\begin{figure}[!t]
\centering
\includegraphics[width=3in]{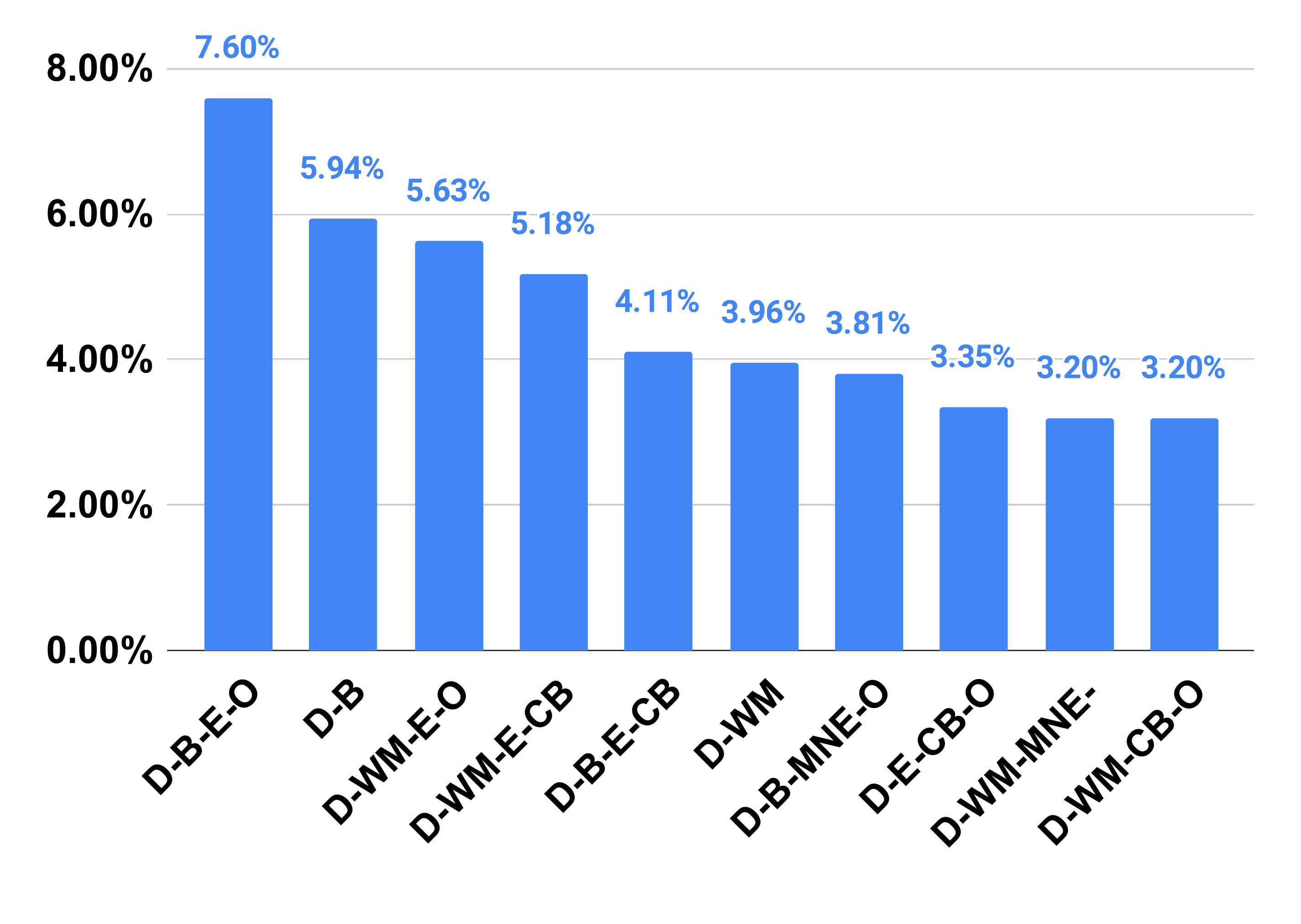}
  \caption{The ten fitness function combinations chosen most often when generating tests for the random baseline when a fault was detected.}
  \label{fig:randomDIVchoices}
\end{figure}

The fitness function combinations that could be selected for the random baseline are the same that AFFS techniques can choose from when attempting to improve diversity. The AFFS techniques attain higher diversity, but the functions that best improve diversity may differ from those that are most likely to result in detection of these specific faults. In Figure~\ref{fig:SarsaDivchoices}, we show the ten fitness function combinations chosen most often by DSG-Sarsa for each system, and in Figure~\ref{fig:UCBDIVchoices}, we do the same for UCB. To contrast, we show the ten most-selected fitness function combinations for the random baseline when a fault was detected in Figure~\ref{fig:randomDIVchoices}. We see little overlap between these figures, further indicating that there is little connection between improved diversity and the likelihood of fault detection for these case examples. 

DSG-Sarsa detects more faults than UCB, even though UCB attains higher diversity. Given the observations above, the difference is likely to be due to a combination of the stochastic nature of search-based generation and differences in the decision making processes for the two algorithms. DSG-Sarsa is more likely to choose fitness functions that are better for detecting the studied faults than UCB. 

\begin{center}
\begin{framed}
The random baseline detects 34.74\% more faults than DSG-Sarsa and 45.87\% more than UCB. Improved diversity does not lead to improved likelihood of fault detection for these case examples.
\end{framed}
\end{center}

\subsubsection{Impact of Reinforcement Learning Overhead}

\begin{table}[!t]
\centering
\scriptsize
\caption{Median time per generation (in seconds) for the goal of test suite diversity. The lowest median is \textbf{bolded}.}
\label{table:diverstity_resultstimeGens}
\begin{tabular}{l|rrrrr}
\hline
 & \textbf{DSG-Sarsa} & \textbf{UCB} & \textbf{Default} & \textbf{Diversity Score} & \textbf{Random} \\ \hline
Chart & 5.26 & \textbf{3.39} & 8.38 & 6.93 & 7.73 \\ 
Closure & 8.63 & \textbf{6.44} & 12.58 & 12.37 & 11.84 \\ 
Gson & 4.09 & \textbf{2.95} & 4.70 & 4.64 & 4.33 \\ 
Lang & 3.65 & \textbf{2.63} & 7.09 & 4.60 & 6.27 \\ 
Math & 4.09 & \textbf{3.05} & 5.68 & 4.15 & 4.64 \\ 
Mockito & 5.98 & \textbf{4.16} & 6.22 & 5.96 & 5.90 \\ 
Time & 3.63 & \textbf{2.77} & 6.32 & 5.29 & 5.28 \\ \hline
\textbf{Overall} & 4.09 & \textbf{3.05} & 8.39 & 6.81 & 7.56 \\ \hline
\end{tabular}
\end{table}

In Table~\ref{table:diverstity_resultstimeGens}, we display the median time per generation for each approach. This, again, allows us to compare the overhead introduced by reinforcement learning to the cost of calculating fitness. Immediately, we see that AFFS is faster than the default and random baselines. While reinforcement learning introduces overhead---including the calculation of diversity as part of the reward score---\textit{this cost is less than naively calculating unnecessary fitness functions}.

A surprising result, however, was that \textit{AFFS is faster than targeting diversity alone}. Intuitively, calculating multiple fitness functions should be more computationally expensive than calculating one fitness function. However, the cost of computing the Levenshtein distance is based on the quantity of text being compared. If a test suite is larger---containing a greater number of tests, longer tests with more interactions with the CUT, or both---then fitness computation will be more expensive. In inspecting changes in the size of test suites throughout the generation process, we found that the test suites evolved targeting diversity alone were significantly larger than those being evolved by DSG-Sarsa, with the latter being 41\% smaller on average in the studied examples. 

In the absence of feedback from additional fitness functions, optimizing the diversity fitness function alone led to unconstrained growth in test suites. Creating longer tests is one \textit{potential} path to improving diversity, but not a guaranteed one---it could still result in similar test cases. Ultimately, the diversity fitness function was not only limited in its ability to suggest means of improving fitness, but actually detrimental to goal attainment by limiting the number of generations that could be completed during the search budget. AFFS was able to both improve diversity and control the growth of test suites, in turn controlling the cost of fitness calculation as well.  
\begin{center}
\begin{framed}
Both AFFS approaches complete more generations of evolution during the search than the baselines, with UCB being 63.65\% faster than the default baseline, 59.66\% faster than random, and 55.22\% faster than diversity alone. By incorporating additional feedback, AFFS controls the cost of the diversity calculation by preventing unconstrained test suite growth.
\end{framed}
\end{center}

\subsubsection{Actions Selected by AFFS}

\begin{figure}[!t]
\centering
   \begin{subfigure}[t]{0.49\textwidth}
        \centering
        \includegraphics[width=2.35in]{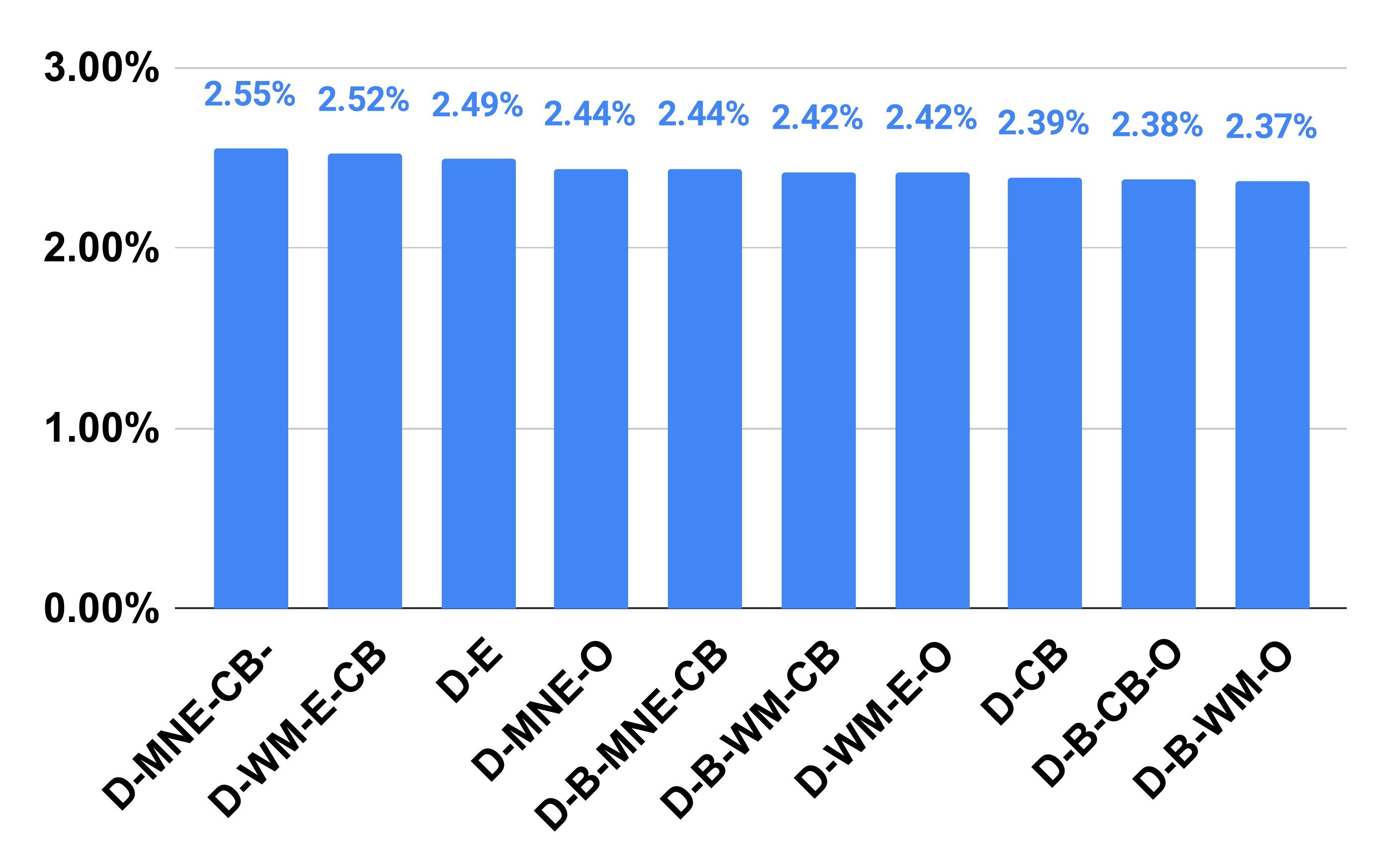} \vspace{-10pt}
        \caption{Chart}
    \end{subfigure}%
    \begin{subfigure}[t]{0.49\textwidth}
        \centering
        \includegraphics[width=2.35in]{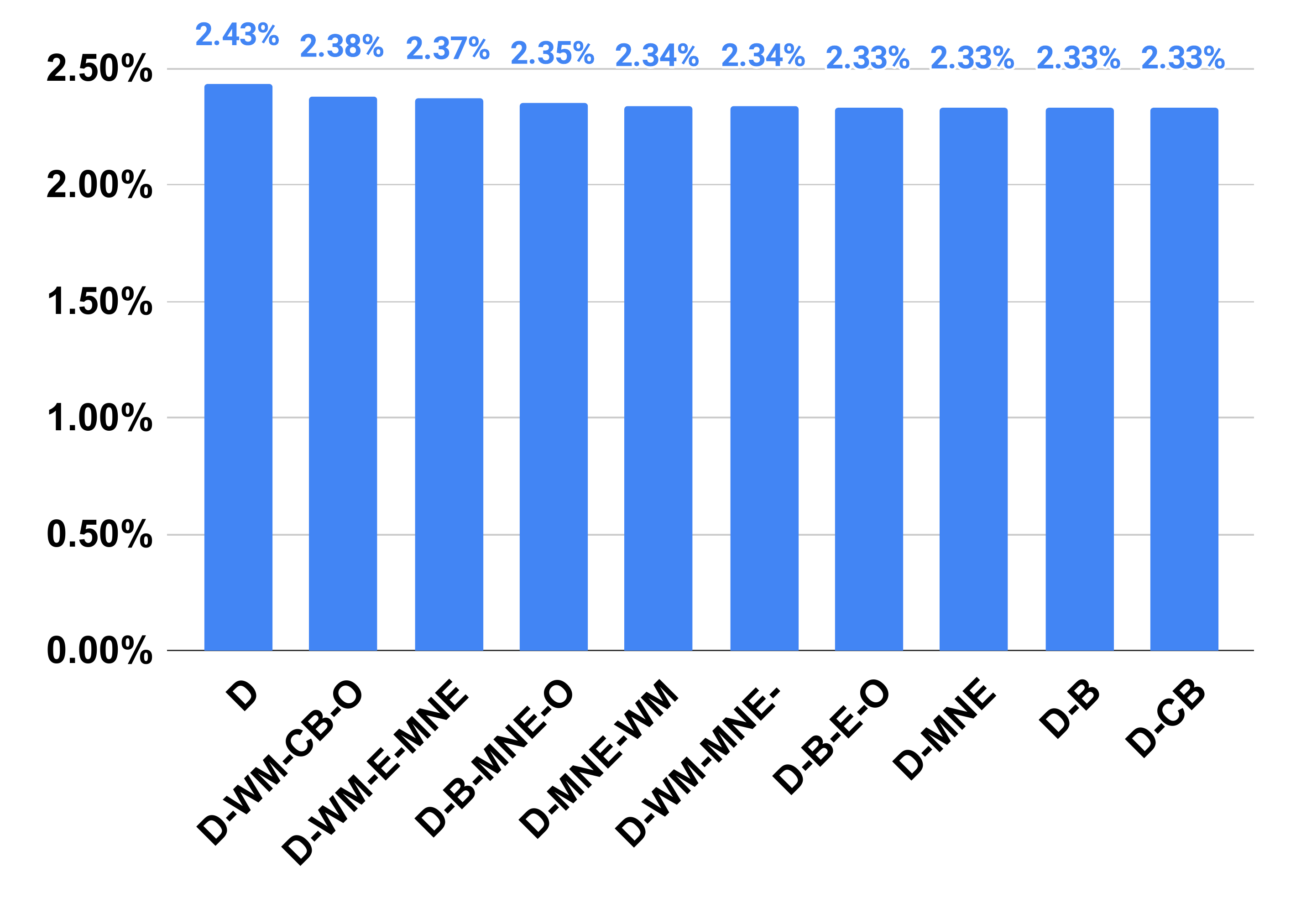} \vspace{-10pt}
        \caption{Closure}
    \end{subfigure}%
    
    \begin{subfigure}[t]{0.49\textwidth}
        \centering
        \includegraphics[width=2.35in]{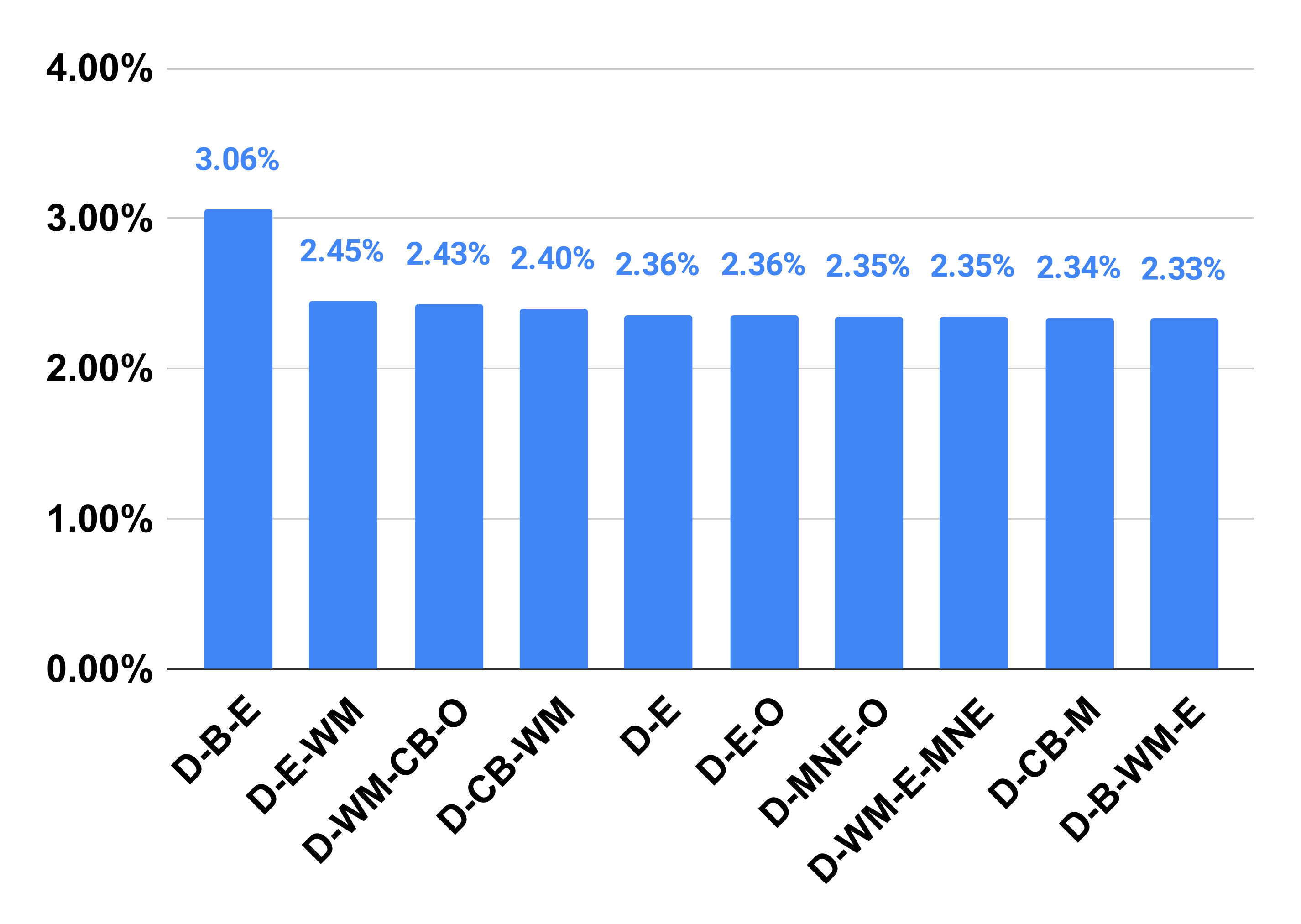}\vspace{-10pt}
        \caption{Gson}
    \end{subfigure}%
    \begin{subfigure}[t]{0.49\textwidth}
        \centering
        \includegraphics[width=2.35in]{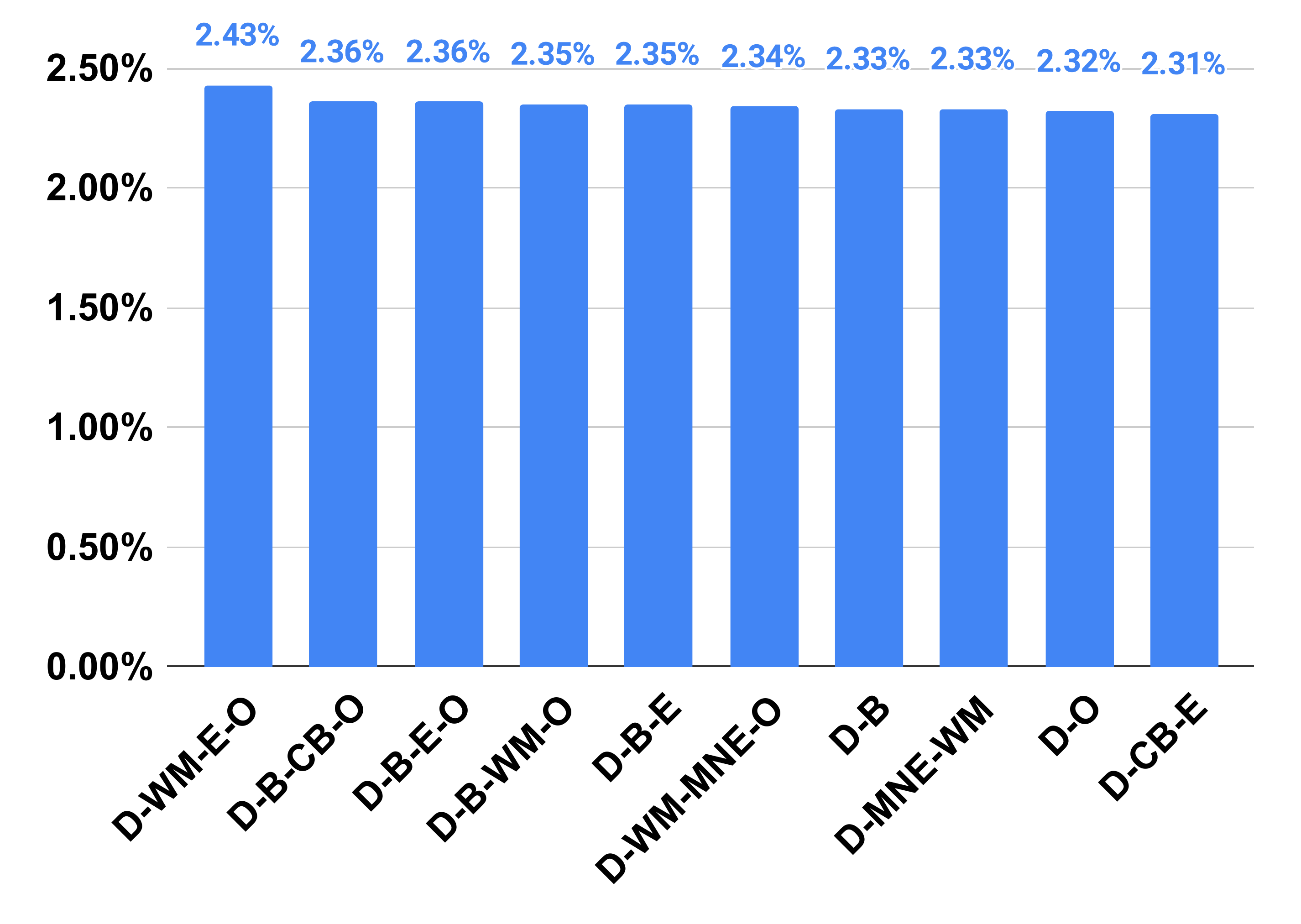} \vspace{-10pt}
        \caption{Lang}
    \end{subfigure}%
    
    \begin{subfigure}[t]{0.49\textwidth}
        \centering
        \includegraphics[width=2.35in]{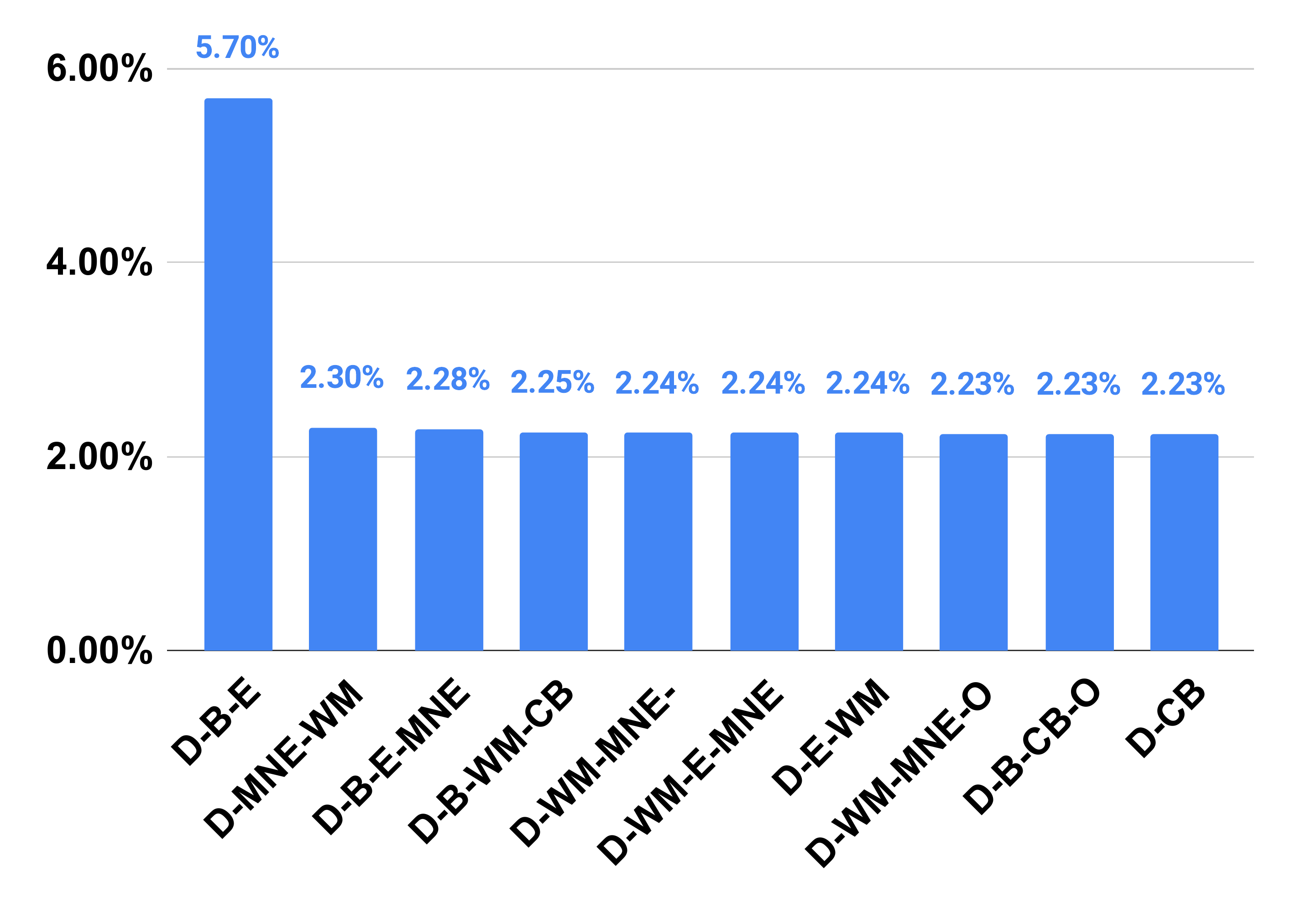}\vspace{-10pt}
        \caption{Math}
    \end{subfigure}%
    \begin{subfigure}[t]{0.49\textwidth}
        \centering
        \includegraphics[width=2.35in]{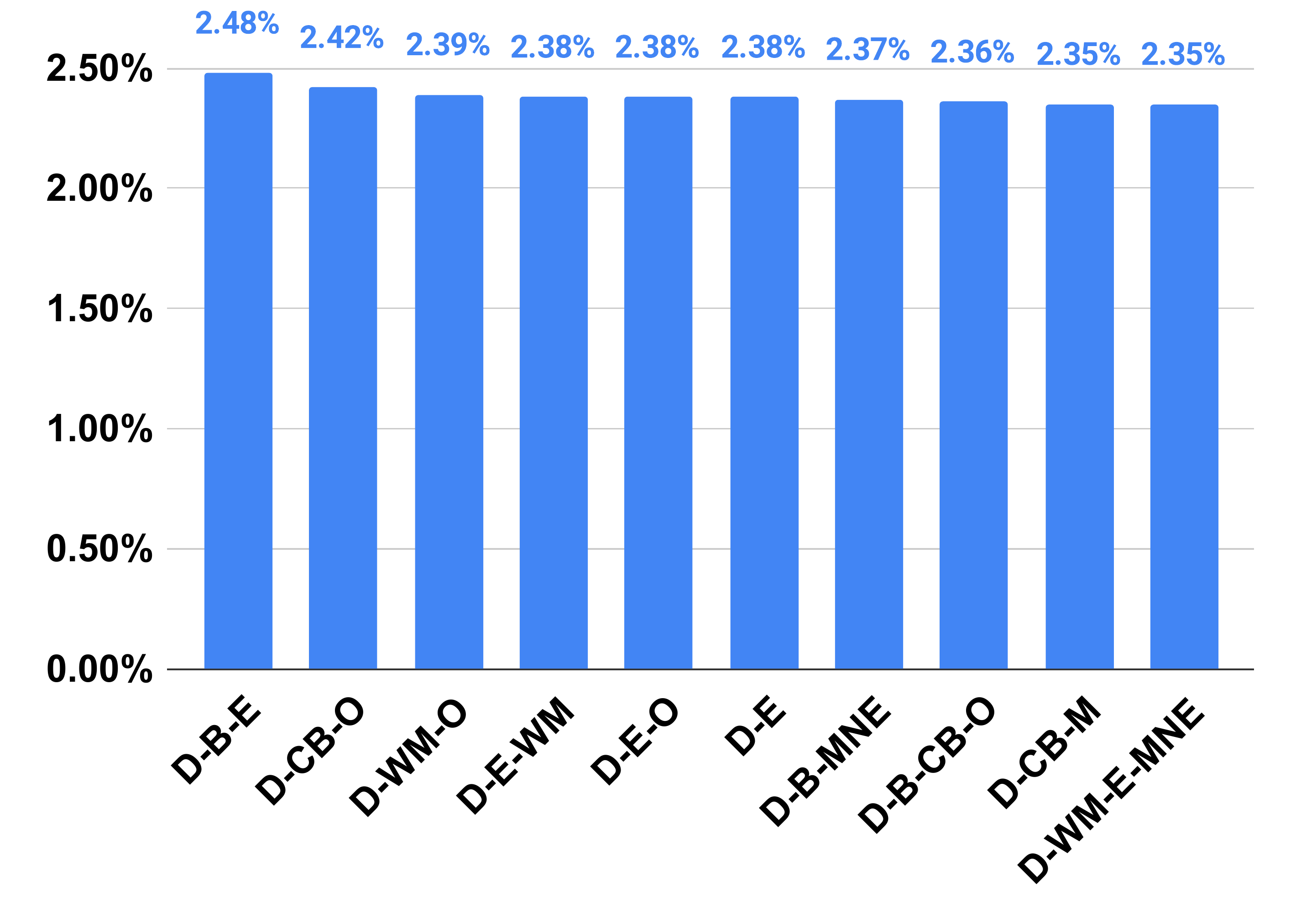}\vspace{-10pt}
        \caption{Mockito}
    \end{subfigure}%
    
    \begin{subfigure}[t]{0.49\textwidth}
        \centering
        \includegraphics[width=2.35in]{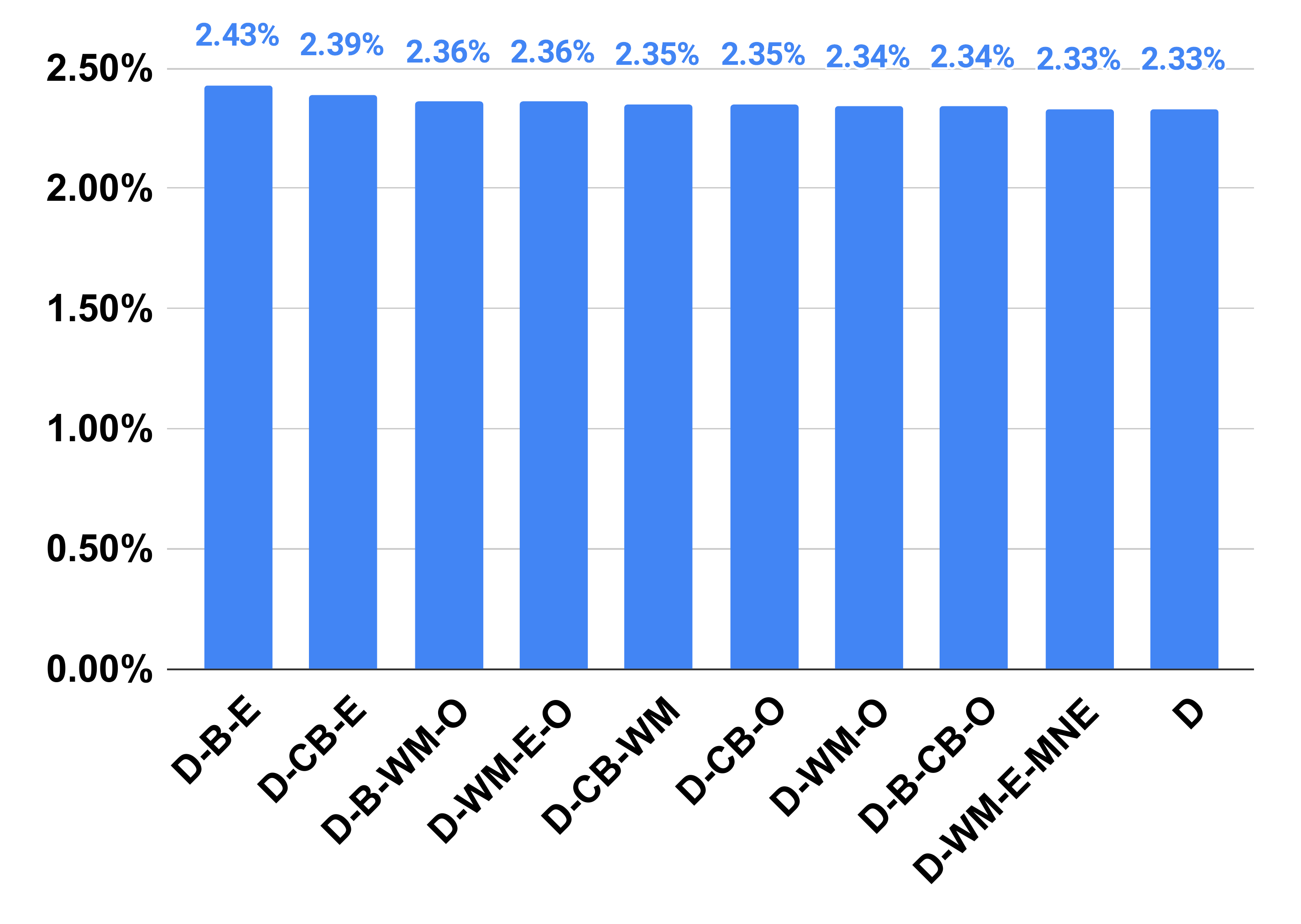}\vspace{-10pt}
        \caption{Time}
    \end{subfigure}%
  \caption{Top ten function combinations chosen by \textbf{DSG-Sarsa} for diversity. D = Diversity Score, B = Branch, CB = Direct Branch, O = Output, M = Method, MNE = Method (No Exception), WM = Weak Mutation}
  \label{fig:SarsaDivchoices} \vspace{-40pt}
\end{figure}

\begin{figure}[!t]
\centering
   \begin{subfigure}[t]{0.5\textwidth}
        \centering
        \includegraphics[width=2.35in]{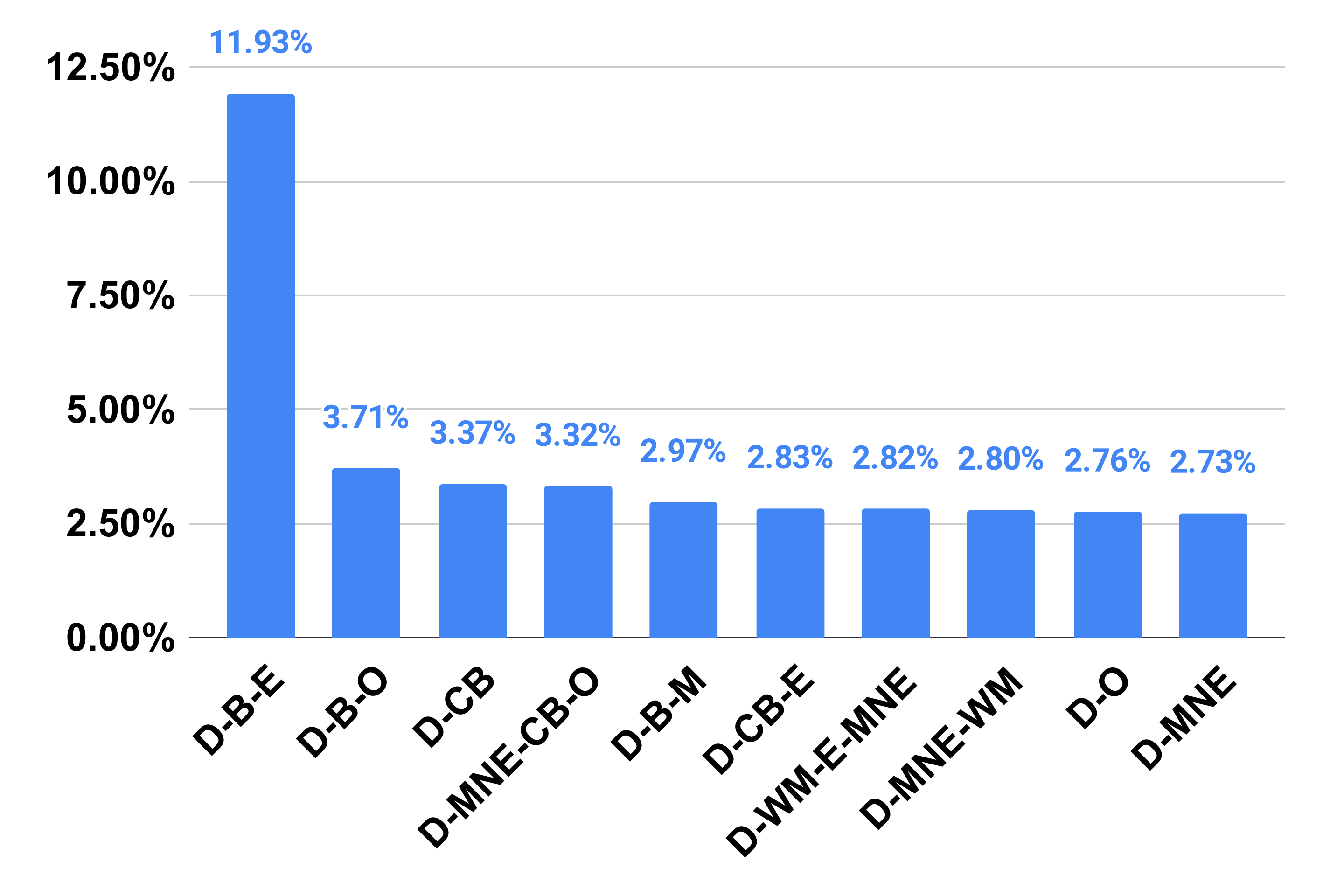} \vspace{-10pt}
        \caption{Chart}
    \end{subfigure}%
    \begin{subfigure}[t]{0.5\textwidth}
        \centering
        \includegraphics[width=2.35in]{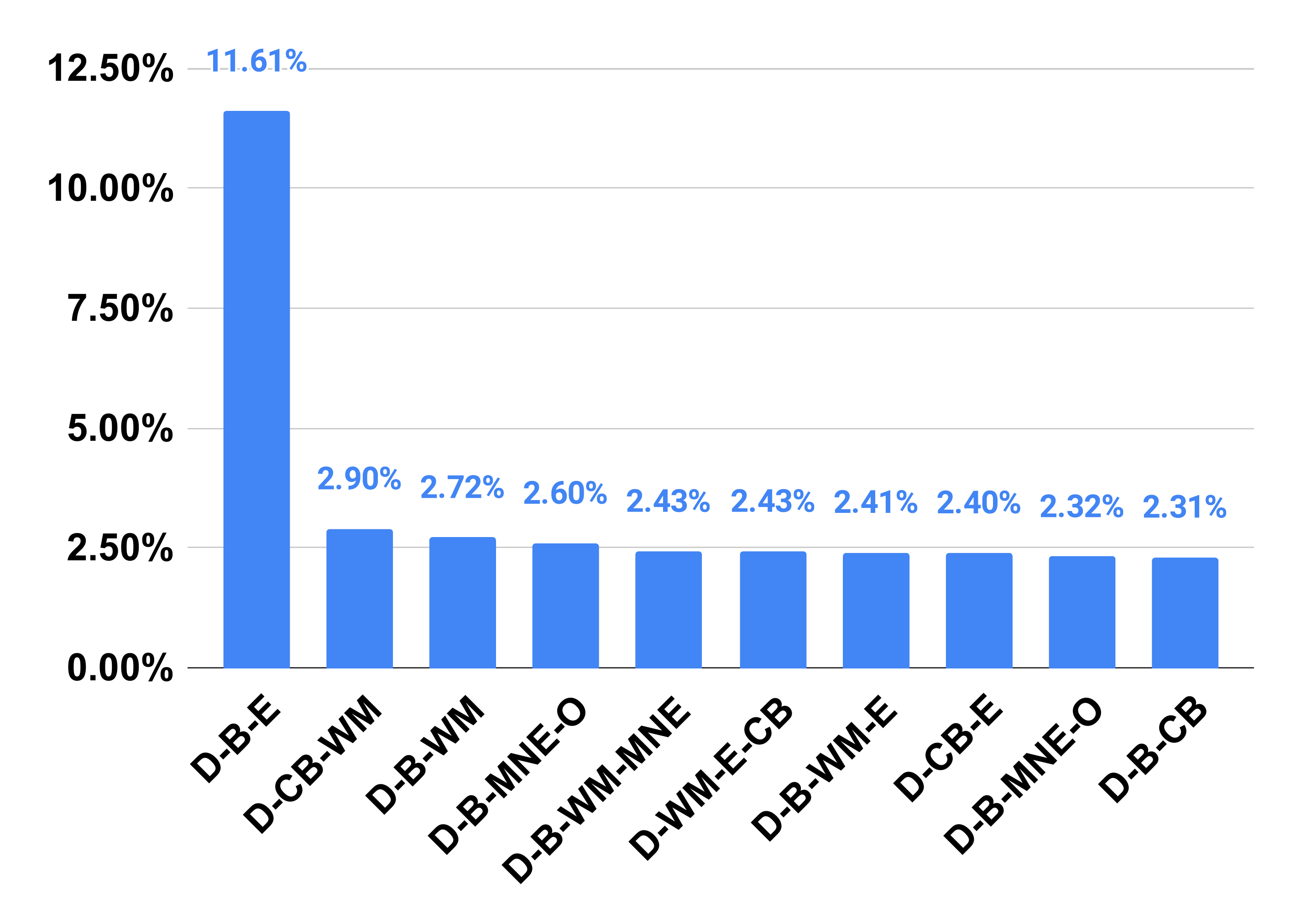} \vspace{-10pt}
        \caption{Closure}
    \end{subfigure}%
    
    \begin{subfigure}[t]{0.5\textwidth}
        \centering
        \includegraphics[width=2.35in]{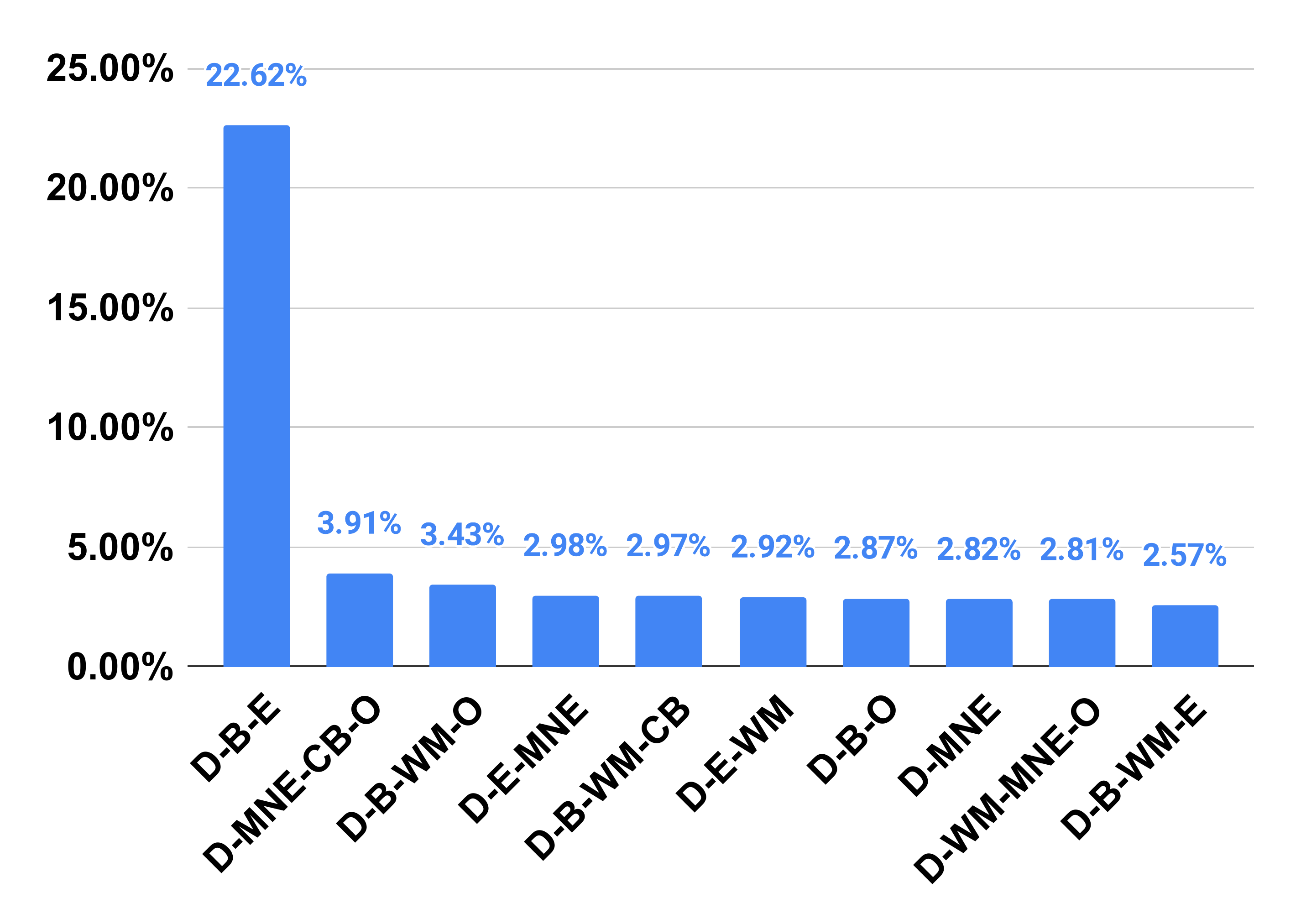} \vspace{-10pt}
        \caption{Gson}
    \end{subfigure}%
    \begin{subfigure}[t]{0.5\textwidth}
        \centering
        \includegraphics[width=2.35in]{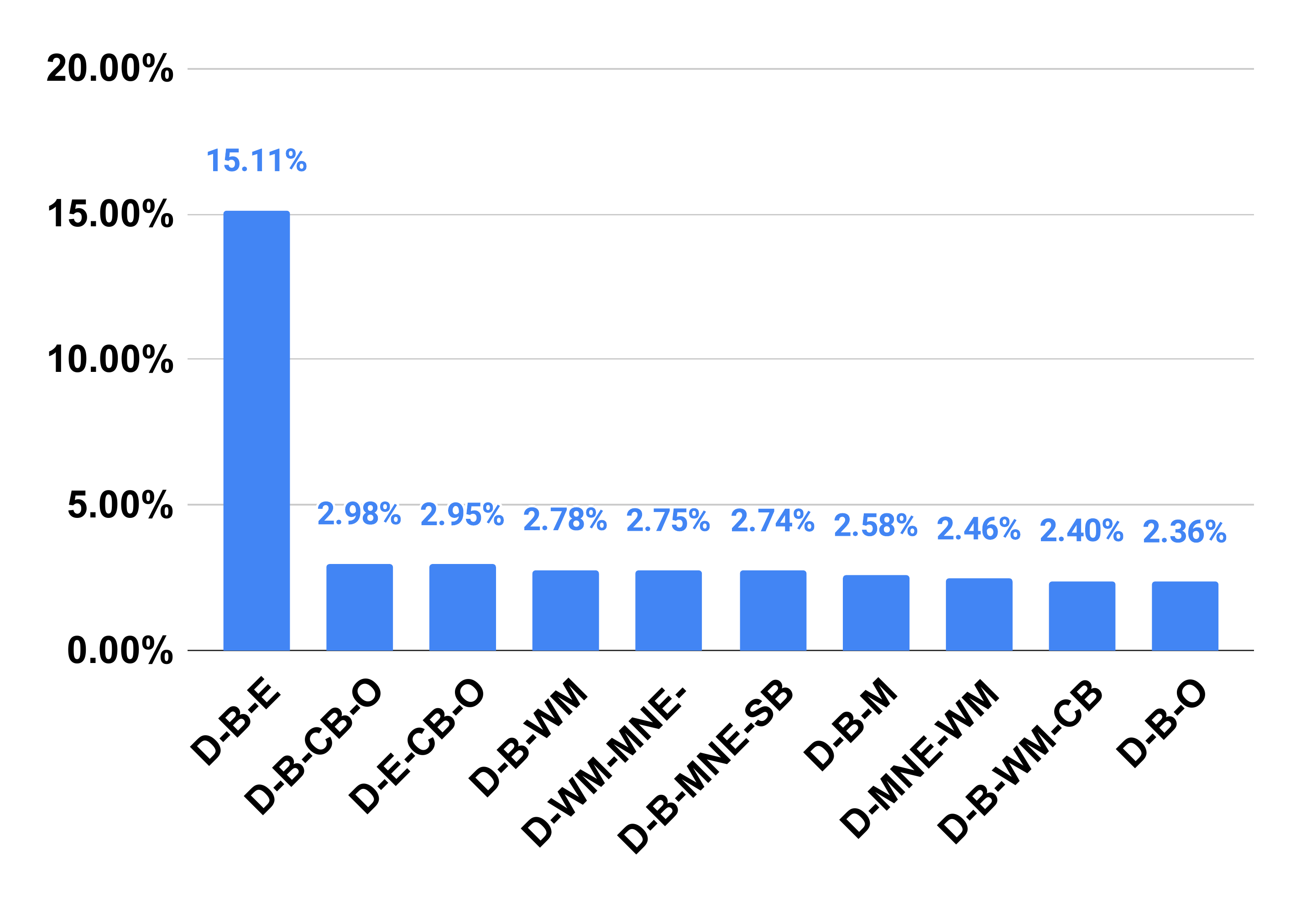} \vspace{-10pt}
        \caption{Lang}
    \end{subfigure}%
    
    \begin{subfigure}[t]{0.5\textwidth}
        \centering
        \includegraphics[width=2.35in]{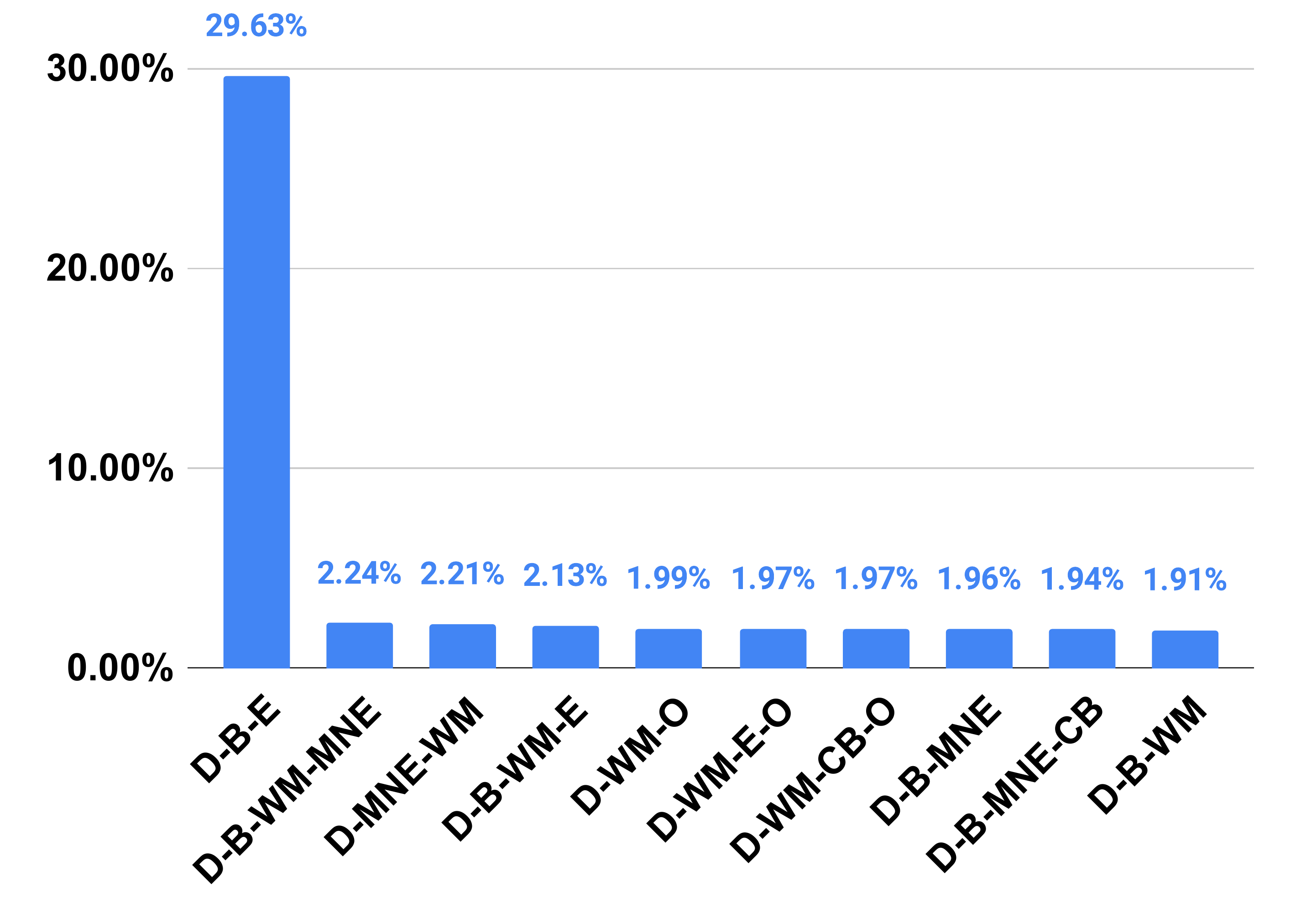} \vspace{-10pt}
        \caption{Math}
    \end{subfigure}%
    \begin{subfigure}[t]{0.5\textwidth}
        \centering
        \includegraphics[width=2.35in]{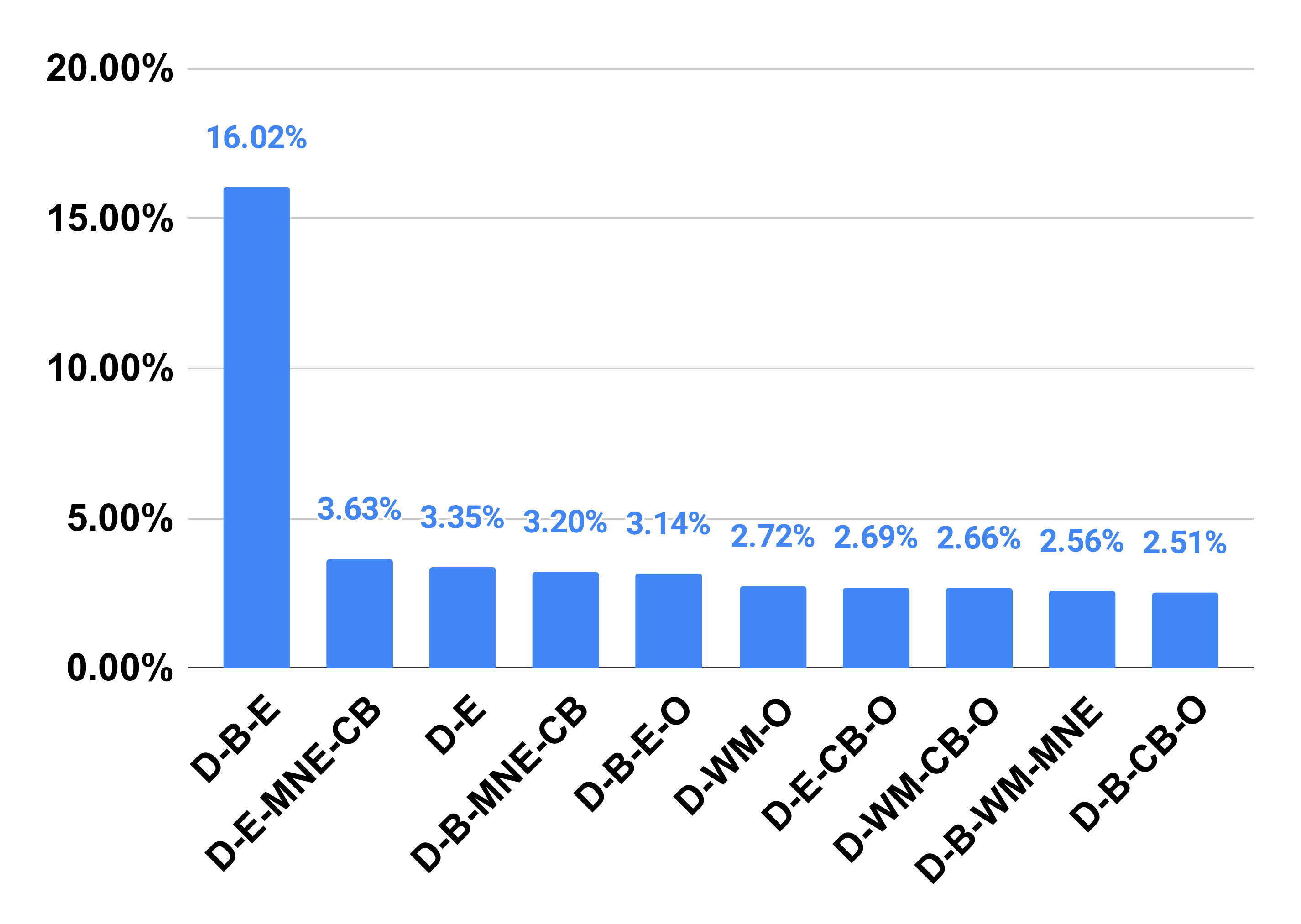} \vspace{-10pt}
        \caption{Mockito}
    \end{subfigure}%
    
    \begin{subfigure}[t]{0.5\textwidth}
        \centering
        \includegraphics[width=2.35in]{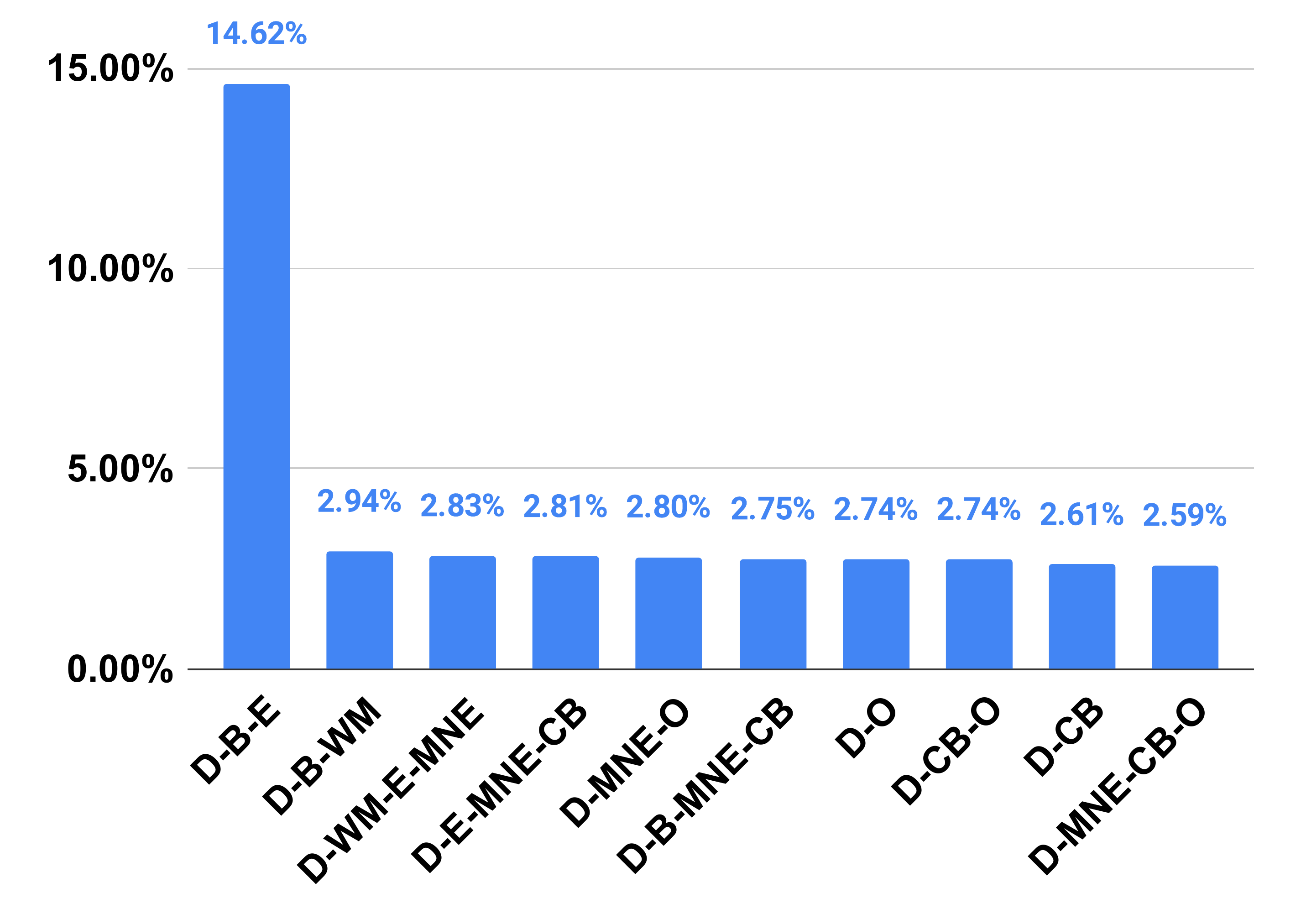} \vspace{-10pt}
        \caption{Time}
    \end{subfigure}%
  \caption{Top ten fitness function combinations chosen by \textbf{UCB} for the diversity goal. D = Diversity Score, B = Branch, CB = Direct Branch, O = Output, M = Method, MNE = Method (No Exception), WM = Weak Mutation}
  \label{fig:UCBDIVchoices} \vspace{-10pt}
\end{figure}

In Figure~\ref{fig:SarsaDivchoices}, we display the ten fitness function combinations chosen most often by DSG-Sarsa for the diversity goal, and in Figure~\ref{fig:UCBDIVchoices}, we do the same for UCB. Examining these choices may offer insight into the results attained by AFFS.

The first observation we make is that the combination of diversity score, exception count, and Branch Coverage is chosen the most often. It is the top choice made by DSG-Sarsa for four of the seven systems, and the top choice for UCB for all projects. This combination provides several key ingredients for attaining diversity. Branch Coverage encourages exploration of the structure of the CUT, building strong test suites that the other functions can tune. The exception count imbues the suite with a wider range of input choices. Finally, the diversity score encourages further input exploration. 

Each function is insufficient on its own. The diversity score needs external feedback to drive diversity. Branch Coverage and the exception count both offer this. Branch Coverage alone will only result in as much diversity as is required to cover more of the code. The other functions force diversification of the input choices. The exception count could be a great driver of diversity, but needs Branch Coverage to aid code exploration. Together, these three functions offer each other feedback, resulting in more diversity than could be attained individually. 

Unlike the exception discovery goal, both UCB and DSG-Sarsa favor complex combinations of three-four fitness functions. For the goal of diversity, this makes some sense. We seek test suites that \textit{try a lot of different things}. Even if poor coverage is attained of some of the fitness functions in a combination, and even if conflicts exist, more functions could drive the generation process towards attempting to satisfy a huge variety of goals.

\begin{center}
\begin{framed}
The combination of Branch Coverage, exception count, and diversity score seems effective at improving test suite diversity. These functions (and other combinations) act in concert, providing feedback to the other functions.
\end{framed}
\end{center}

We again see that UCB tends to exploit one combination above all others, while DSG-Sarsa will spend more time exploring different options. As UCB attains better results, it may be that heavier exploitation is a good idea for this goal. A greater tendency towards exploitation may enable better goal attainment, as less time is spent trying potentially weak function combinations.

Like we saw with the exception discovery goal, certain selections that would not work well in a static context may be useful to refine pre-evolved suites. We see this with DSG-Sarsa and the diversity score. Optimizing the diversity score in a static context yields poor results, but is used quite often by DSG-Sarsa to refine test suites that have been shaped by other function combinations. This allows diversification of test suites that have already been built up to do things like explore the code base.


\subsection{Goal: Strong Mutation Coverage}\label{sec:sm_results}

\subsubsection{Ability to Improve Coverage and Impact of Overhead}

\begin{table}[!t]
\centering
\scriptsize
\caption{Percentage of Strong Mutation Coverage attained when all approaches execute for 10 minutes. \textbf{Higher values are better}. The highest median is \textbf{bolded}.}
\label{table:SM_coverage}
\begin{tabular}{l|rrrr}
\hline
 & \textbf{DSG-Sarsa} & \textbf{UCB} & \textbf{Default} & \textbf{Strong Mutation} \\ \hline
Chart   & 47.00 &	47.00 & 52.00 & \textbf{54.00} \\ 
Closure & 16.00 &	16.00 & 18.00 & \textbf{19.00}  \\ 
Lang    & 61.00 &	60.00 & 62.00 & \textbf{63.00}  \\ 
Math    & 73.00 &	\textbf{74.00} & 73.00 & 73.00  \\ 
Mockito & \textbf{12.00} &	8.50 & 11.00 & 10.00   \\ 
Time    & 67.00 & 66.00 & 67.00 & \textbf{68.00}  \\ \hline
\textbf{Overall} & 38.00 & 39.00 & \textbf{40.00} & \textbf{40.00} \\ \hline
\end{tabular}
\end{table}

\begin{figure}[!t]
\centering
 \begin{subfigure}{3in}
        \centering
\includegraphics[width=3in]{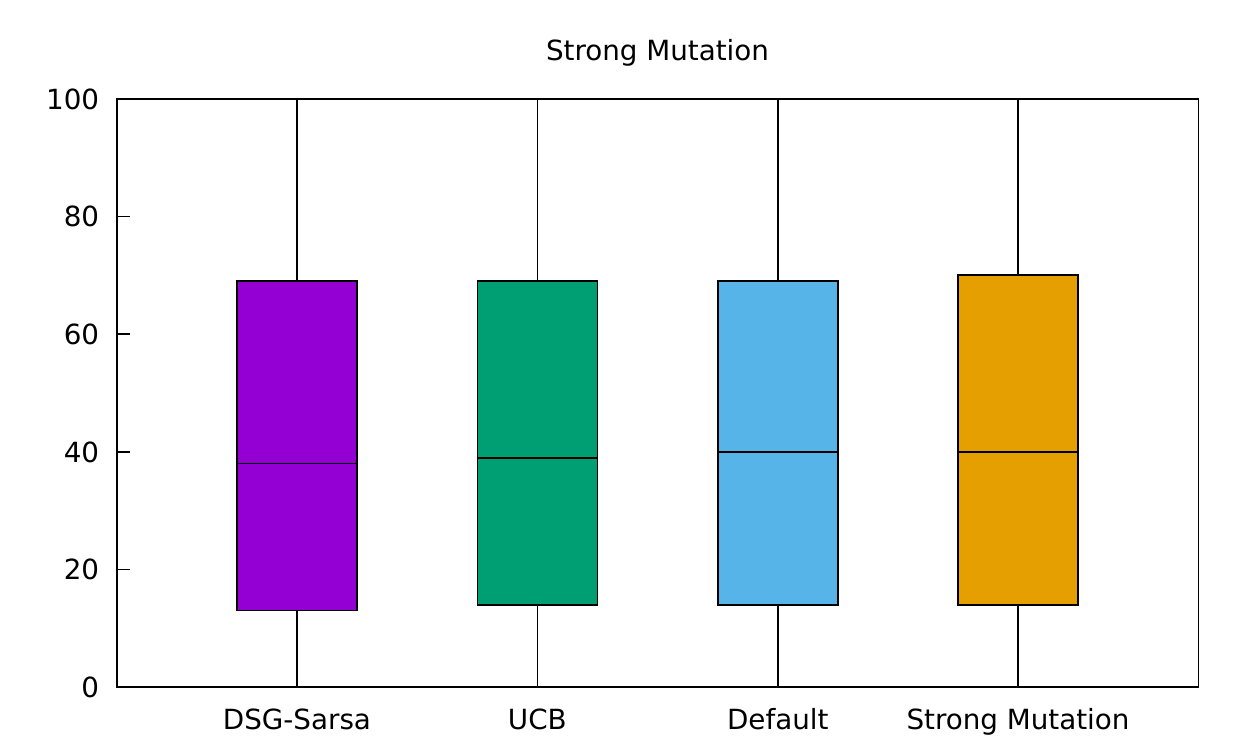}
    \end{subfigure}
    
 \begin{subfigure}{3in}
        \centering
\includegraphics[width=3in]{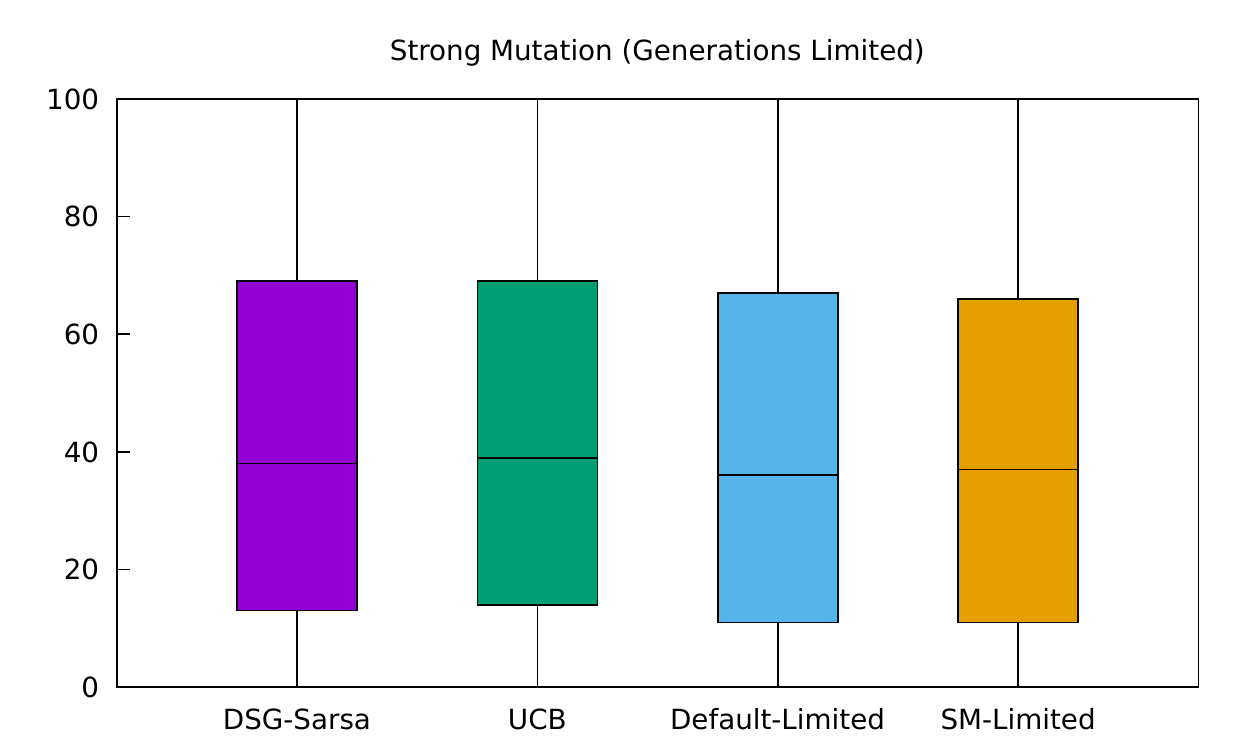}
    \end{subfigure}
  \caption{Strong Mutation Coverage attained by final test suites when (top) all approaches run for 10 minutes, and (bottom), when all approaches are fixed to the number of generations of evolution completed by DSG-Sarsa in 10 minutes.}
  \label{fig:SM_coverage} 
\end{figure}

We assess attainment of our third goal using the attained percentage of Strong Mutation Coverage. In Table~\ref{table:SM_coverage}, we note the median Strong Mutation Coverage for each AFFS technique and baseline configuration\footnote{We omit the random baseline for this experiment to (a) control the cost of running experiments, and (b), because of the similarity of results for the AFFS approaches and the other two baselines. It is likely that the random baseline would attain similar results as well.} for each system and overall, bolding the approach with the highest median result. In the top diagram in Figure~\ref{fig:SM_coverage}, we show a boxplot of the Strong Mutation results for all approaches. 

From Table~\ref{table:SM_coverage}, we see that, for four of the six projects, \textit{both AFFS approaches and the default configuration attain worse median results than simply optimizing Strong Mutation Coverage directly}. From Figure~\ref{fig:SM_coverage}, we can see that all four approaches yield very similar boxplots, with Strong Mutation having a slightly higher third-quartile than the other approaches. Overall, optimizing Strong Mutation alone or targeting the default configuration yields a median improvement of 2.56\% over UCB and 5.26\% over DSG-Sarsa. 

\begin{table}[!t]
\centering
\scriptsize
\caption{Results of Vargha-Delaney A Measure for Strong Mutation (all approaches run for 10 minutes). No large effect sizes were observed.}
\label{table:resultsVDSM}
\begin{tabular}{l|rrrr}
\hline & \textbf{DSG-Sarsa} & \textbf{UCB} & \textbf{Default} & \textbf{Strong Mutation}\\ \hline
 \textbf{DSG-Sarsa}     & -     & 0.50  & 0.49   &  0.49  \\ 
 \textbf{UCB}           & 0.50  & -     & 0.49   &  0.49  \\ 
 \textbf{Default} & 0.51  & 0.51  & -      &  0.50   \\ 
 \textbf{Strong Mutation} & 0.51  & 0.51  & 0.50   & - \\ \hline
\end{tabular} 
\end{table}

We again perform statistical analysis to assess our observations, using the Friedman test and Vargha Delaney A measure. We formulate hypothesis and null hypothesis:
\begin{itemize}
\setlength{\itemsep}{1pt}
  \setlength{\parskip}{0pt}
  \setlength{\parsep}{0pt}  
\item{$H$: Generated test suites have different distributions of Strong Mutation Coverage results depending on the technique used to generate the suite.}
\item{$H0$: Observations of Strong Mutation Coverage for all techniques are drawn from the same distribution.}
\end{itemize}

The Friedman test, at p-value $<$ 0.001, suggests differences in the distributions of results for the techniques. In Table~\ref{table:resultsVDSM}, we display effect sizes. The default and Strong Mutation baselines slightly outperform UCB and DSG-Sarsa, but with a negligible effect size. While both baselines yield a slightly higher median performance, we cannot say that any technique outperforms AFFS with significance. 

\begin{center}
\begin{framed}
Optimizing Strong Mutation alone or the default baseline yields median improvement of 2.56\%/5.26\% over UCB/DSG-Sarsa. However, no technique demonstrates significant performance differences (negligible effect sizes). 
\end{framed}
\end{center}

\begin{table}[!t]
\centering
\scriptsize
\caption{Median time per generation (in seconds) for the Strong Mutation Coverage goal. The fastest technique is \textbf{bolded}.}
\label{table:SM_resultstimeGens}
\begin{tabular}{l|rrrr}
\hline
 & \textbf{DSG-Sarsa} & \textbf{UCB} & \textbf{Default} & \textbf{Strong Mutation} \\ \hline
Chart  & 6.57  &	7.90    & 7.11    &	\textbf{5.86}   \\ 
Closure & \textbf{2.44} &	2.92  &	4.63  &	3.98  \\ 
Lang & 24.69   & 17.88   &	12.65   & \textbf{12.36} \\ 
Math & 11.89   & \textbf{6.05}  & 10.69 & 9.06  \\ 
Mockito	& 0.21    &	\textbf{0.13}      &	0.21    &	0.23    \\ 
Time   & 14.49   &	19.45   &	9.86   & \textbf{8.87}  \\ \hline
Overall  & 4.98   & \textbf{3.85}  &	5.91   & 5.20  \\ \hline
\end{tabular}
\end{table}

To explain the lack of success of AFFS for this goal, we examine two factors: (1) the reward function used by reinforcement learning and its impact on overhead, and (2), the fitness functions that can be combined by AFFS for this goal.

First, we examine the impact of the overhead of reinforcement learning, particularly calculation of the reward function. In this experiment, a search budget of ten minutes is allocated to generate test suites. The amount of time that a single generation takes is not fixed, but depends on fitness calculation. If multiple fitness functions, or expensive fitness functions are used, then fewer generations of evolution will take place over that time period.

Reinforcement learning adds additional overhead on top of this calculation. An expensive reward function will further reduce the number of generations of evolution that can be completed. With exception discovery, the reward function and many of common fitness function combinations were inexpensive, resulting in AFFS techniques being faster than the default configuration. In the case of diversity, the diversity score that was used as both a fitness function and to calculate reward could have been expensive---if the test suite was large---but remained inexpensive due to feedback from other fitness functions. 

Strong Mutation Coverage is an expensive function to calculate. It requires the execution of the test suites against each mutant. The total cost of calculation depends on the number of mutants, but generally requires multiple test executions, rather than one, to calculate. We use this function not only as the reward function, but as part of many of the fitness function combinations. Although we alternate between Weak and Strong Mutation during reward calculation to control this cost, AFFS has a heavy reward calculation cost that the other approaches lack. This could have an impact on the resulting goal attainment.

In Table~\ref{table:SM_resultstimeGens}, we display the median time per generation for each approach. AFFS is again slightly faster than the baselines on average. However, the results vary by project. \textit{For three projects, the Strong Mutation baseline is significantly faster than AFFS}. As seen in Table~\ref{table:SM_coverage}, the Strong Mutation baseline also yields the highest goal attainment for those projects. The number of generations of evolution plays a role in the resulting goal attainment. For those projects, the slower performance of AFFS may have reduced effectiveness. 

\begin{center}
\begin{framed}
For three of the six projects, AFFS techniques are up to 49.94\% slower per generation than optimizing Strong Mutation alone. For these projects, optimizing Strong Mutation alone also results in improved goal attainment.
\end{framed}
\end{center}

\begin{table}[!t]
\centering
\scriptsize
\caption{Percentage of strong mutation coverage attained by test suites when the number of generations of evolution is fixed to that completed by DSG-Sarsa in 10 minutes. \textbf{Higher values are better}. The highest median is \textbf{bolded}.}
\label{table:SM_coverage_controlled}
\begin{tabular}{l|rrrr}
\hline
 & \textbf{DSG-Sarsa} & \textbf{UCB} & \textbf{Default} & \textbf{Strong Mutation} \\ \hline
Chart   & 47.00 &	47.00 & \textbf{49.50} & 44.50 \\ 
Closure & 16.00 & 16.00 & \textbf{17.00} & 16.00   \\ 
Lang    & \textbf{61.00} &	60.00 & 57.00 & 58.00  \\ 
Math    & 73.00 &	\textbf{74.00} & 70.00 & 71.00   \\ 
Mockito & \textbf{12.00} &	8.50 & 9.00  & 8.00    \\ 
Time    & \textbf{67.00} &	66.00 & 64.00 & 64.00   \\ \hline
\textbf{Overall} & 38.00 & \textbf{39.00} & 36.00 & 37.00 \\ \hline
\end{tabular}
\end{table}

To investigate this possibility, we repeated our experiment, using a fixed number of generations as the search budget instead of a fixed period of time. We used the median number of generations completed by DSG-Sarsa (generally the slower reinforcement learning technique) in ten minutes as the search budget rather than a fixed period of time. In Table~\ref{table:SM_coverage_controlled}, we indicate the median goal attainment for each technique when the number of generations of evolution is fixed. In the bottom diagram in Figure~\ref{fig:SM_coverage}, we show box plots of results for all techniques. 

For all systems, AFFS now attains equal or higher median goal attainment than the Strong Mutation baseline and, for four of the six systems, outperforms the default baseline. Overall, AFFS outperforms both baselines in median performance when the number of generations is fixed. The best technique, UCB, outperforms the default baseline in median performance by 8.33\% and the Strong Mutation baseline by 5.41\%. The box plots are still similar, but show a slight shift, with DSG-Sarsa and UCB now yielding higher third-quartile and median values than the two baselines. 

\begin{table}[!t]
\centering
\scriptsize
\caption{Results of Vargha-Delaney A Measure for Strong Mutation (number of generations fixed). No large positive effect sizes were observed.}
\label{table:resultsVDSM_limited}
\begin{tabular}{l|rrrr}
\hline & \textbf{DSG-Sarsa} & \textbf{UCB} & \textbf{Default} & \textbf{Strong Mutation}\\ \hline
 \textbf{DSG-Sarsa}     & -     & 0.50    &  0.52  & 0.52 \\ 
 \textbf{UCB}           & 0.50  & -   &  0.52  & 0.52 \\ 
 \textbf{Default}  & 0.48  & 0.48   & -      & 0.50  \\ 
 \textbf{Strong Mutation}  & 0.48  & 0.48  & 0.50   & -    \\ \hline
\end{tabular} 
\end{table}

We repeat our statistical tests as well. The effect sizes are shown in Table~\ref{table:resultsVDSM_limited}. While the effect sizes now show slight improvements from AFFS over the default and Strong Mutation baselines, the effect sizes are still negligible. There is some indication that, if more time can be allocated to the generation process, AFFS can slightly increase attainment of Strong Mutation Coverage. However, the results for all techniques are still similar. 

\begin{center}
\begin{framed}
When the budget is fixed by number of generations rather than time, AFFS techniques outperform the baselines in median performance. UCB outperforms the default by 8.33\% and Strong Mutation alone by 5.41\%. However, effect sizes still remain negligible. 
\end{framed}
\end{center}

The central hypothesis of AFFS is that certain combinations of fitness functions will provide the feedback that the existing fitness function fails to offer to the search. However, none of the functions used in our experiment offer feedback beyond that already offered by the Strong Mutation fitness function. There may be other functions that could offer this feedback, but we do not know what these are or whether they exist.

If the number of generations are fixed, AFFS may be able to offer mild improvements over the default combination or targeting Strong Mutation on its own, but these improvements are very limited. The similarity of the boxplots in Figure~\ref{fig:SM_coverage} further demonstrates the limited feedback offered by other fitness functions, as we do not see significant reductions in variance like we did with the other two goals. Improvement in attainment of Strong Mutation coverage requires further experimentation and discovery of new fitness functions.

\begin{center}
\begin{framed}
The similar performance of all techniques may indicate that the fitness functions considered by AFFS have limited impact on attainment of Strong Mutation Coverage. Other unknown functions may be more effective.
\end{framed}
\end{center}

\subsubsection{Fault Detection Effectiveness}

\begin{table}[!t]
\centering
\scriptsize
\caption{Percentage of faults detected by each approach for the Strong Mutation goal. F\#G = fixed number of generations}
\label{table:SM_faultsdetection}
\begin{tabular}{l|rrrrrr}
\hline
 & \textbf{DSG-Sarsa} & \textbf{UCB} & \textbf{Default} & \textbf{SM} & \textbf{Default (F\#G)} &\textbf{SM (F\#Gen)} \\ \hline
Chart   & 54\% &	\textbf{69\%} & \textbf{69\%} & 54\% & 65\% & 50\% \\ 
Closure & 13\% &	\textbf{25\%} & 15\% & 20\%  & 11\% & 10\%   \\ 
Lang    & 44\% & \textbf{45\%} & \textbf{45\%} & 44\% & 34\% & 30\%   \\ 
Math    & \textbf{53\%} &	46\% & 49\% & 48\% & 41\% & 41\%   \\ 
Mockito & 3\%	 &  3\%  & 3\%  & 3\%  & 3\%  & 3\%    \\ 
Time    & 50\% &	\textbf{54\%} & 50\% & 42\% & 46\% & 38\%   \\ \hline
\textbf{Overall} & 31\% & \textbf{36\%} & 32\% & 32\% & 26\% & 24\% \\ \hline
\end{tabular}
\end{table}

We examine fault detection for our two AFFS approaches and for the two benchmarks in Table~\ref{table:SM_faultsdetection}, where we list the number of faults detected per project. In the last section, we saw that the attained Strong Mutation Coverage was similar for all approaches. Here, we see fairly similar fault detection rates for the four approaches as well. The top approach was UCB, with 36\% of the faults. It outperforms both baseline by 12.50\%, and DSG-Sarsa by 16.12\%.

We again calculated the point-biserial correlation coefficient between Strong Mutation Coverage and fault detection, resulting in a coefficient of 0.31. This was the strongest correlation of our three goals to fault detection, but is still only a weak correlation. Higher Strong Mutation Coverage has a positive impact on the likelihood of fault detection, but is not---in itself---enough to ensure success. 

DSG-Sarsa was outperformed by both baselines, and was also the technique with the lowest median Strong Mutation coverage. However, (a) as no significant differences were observed between result distributions for the approaches, and (b), the weak correlation result, lower Strong Mutation Coverage does not explain its slightly weaker performance. We cannot state that AFFS will always result in improved fault detection over static fitness function choices for this goal.

When we fix the search budget in terms of the number of generations of evolution rather than the period of time, we do see that both AFFS techniques significantly outperform the two baselines. In this situation, UCB outperforms the default configuration by 38.46\% and Strong Mutation by 50.00\%. Some of the fault detection performance of the two baselines can be attributed to additional generations of evolution completed during the 10 minute search budget. When given additional time to develop suites, AFFS may yield a higher likelihood of fault detection as well. 

\begin{center}
\begin{framed}
UCB detects 12.50\% more faults than both baselines and 16.12\% more than DSG-Sarsa for the Strong Mutation goal. When the number of generations is fixed, both AFFS approaches outperform the baselines by up to 50.00\%.
\end{framed}
\end{center}

\subsubsection{Actions Selected by AFFS}

\begin{figure}[!t]
\centering
   \begin{subfigure}[t]{0.5\textwidth}
        \centering
        \includegraphics[width=2.37in]{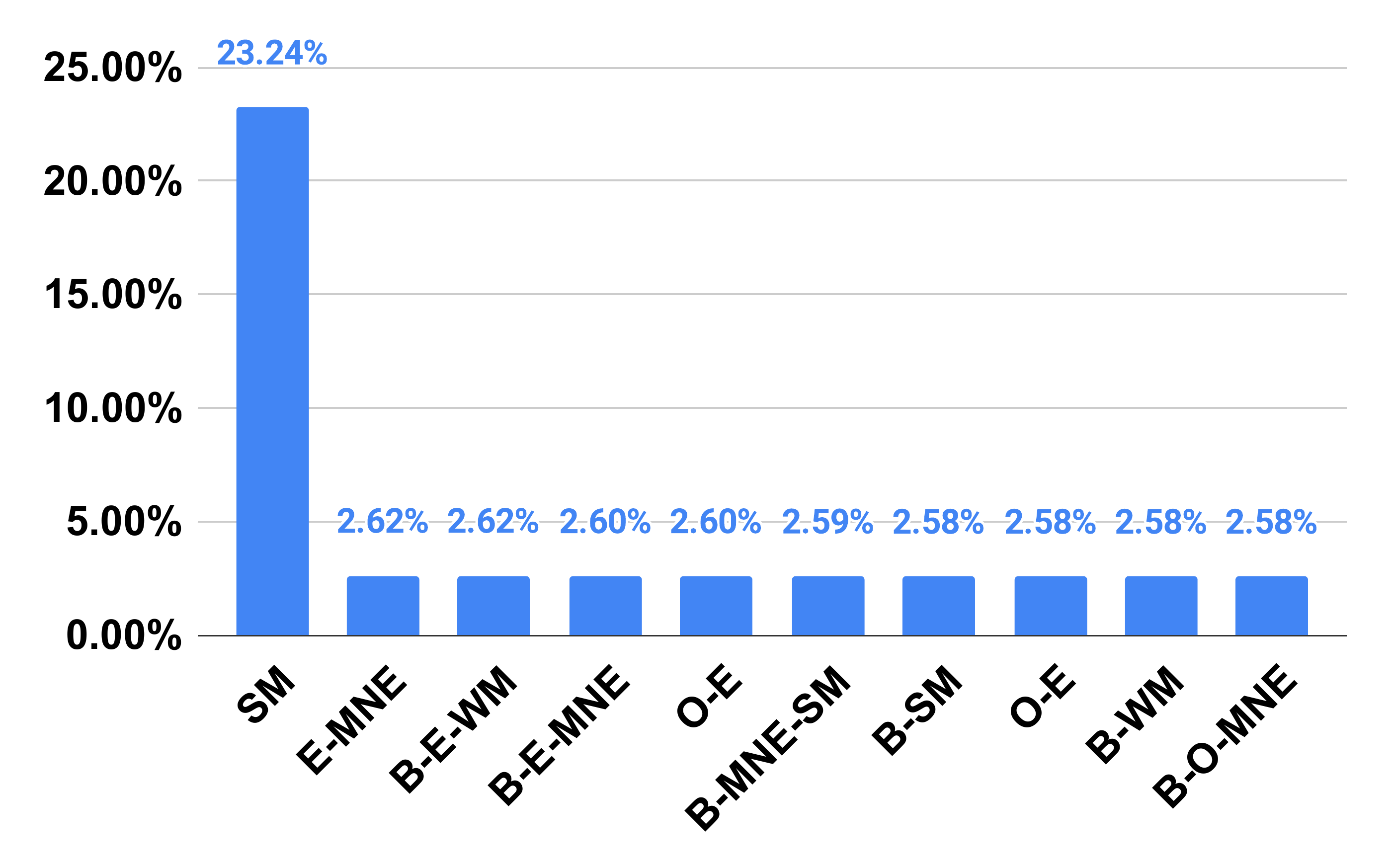} 
    \end{subfigure}%
    \begin{subfigure}[t]{0.5\textwidth}
        \centering
        \includegraphics[width=2.37in]{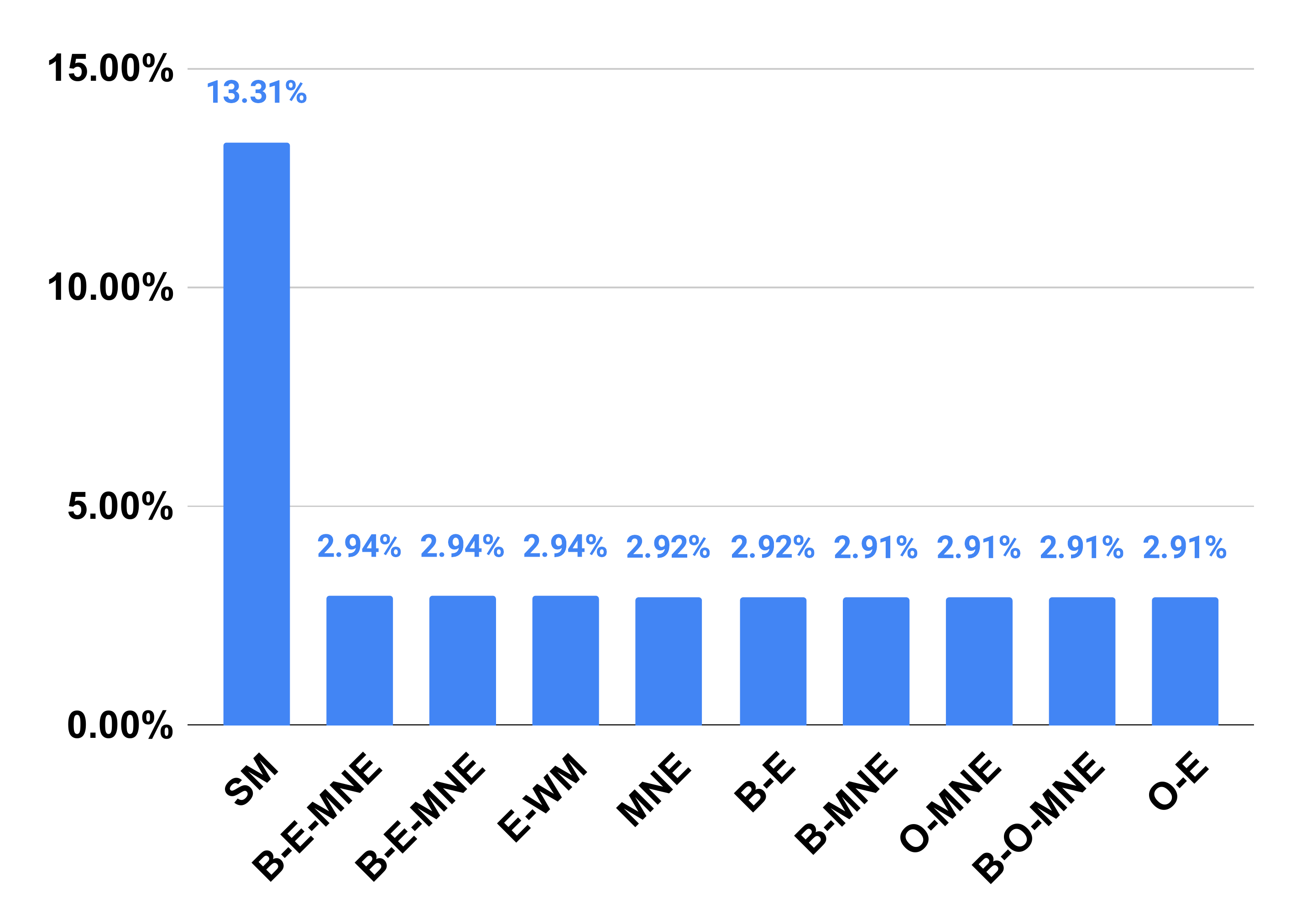} 
    \end{subfigure}%
    
    \begin{subfigure}[t]{0.5\textwidth}
        \centering
        \includegraphics[width=2.37in]{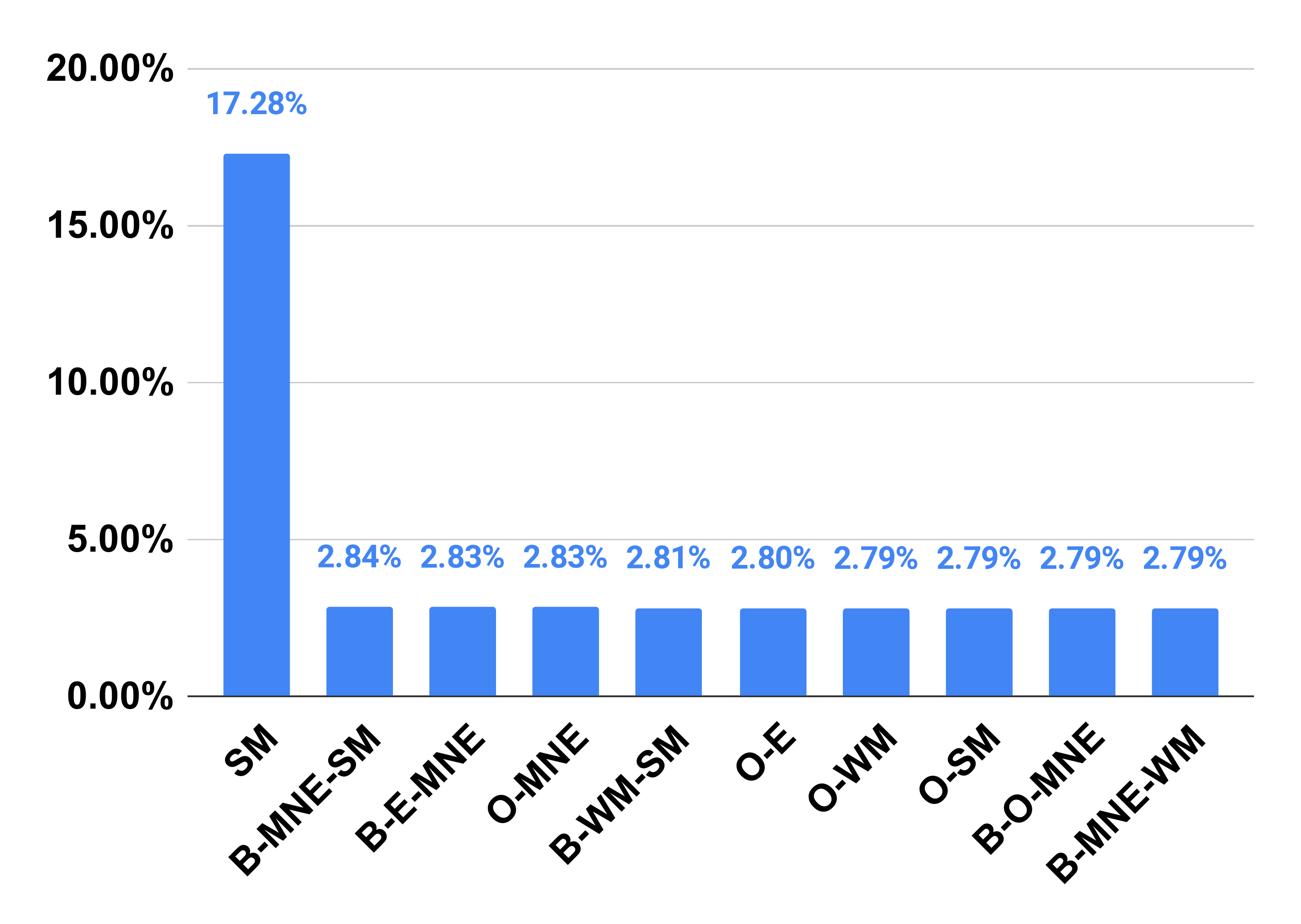} 
    \end{subfigure}%
    \begin{subfigure}[t]{0.5\textwidth}
        \centering
        \includegraphics[width=2.37in]{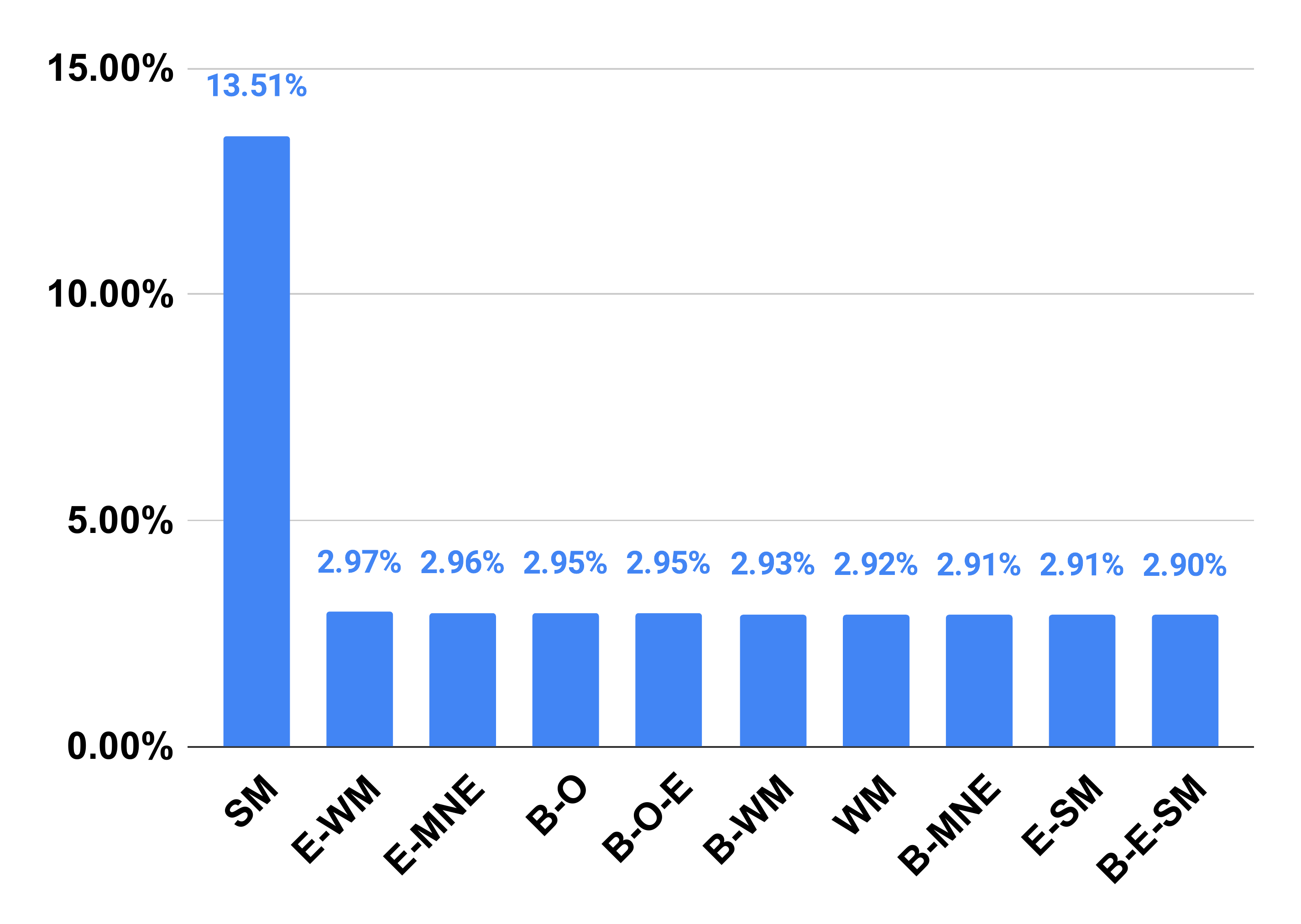} 
    \end{subfigure}%
    
    \begin{subfigure}[t]{0.5\textwidth}
        \centering
        \includegraphics[width=2.37in]{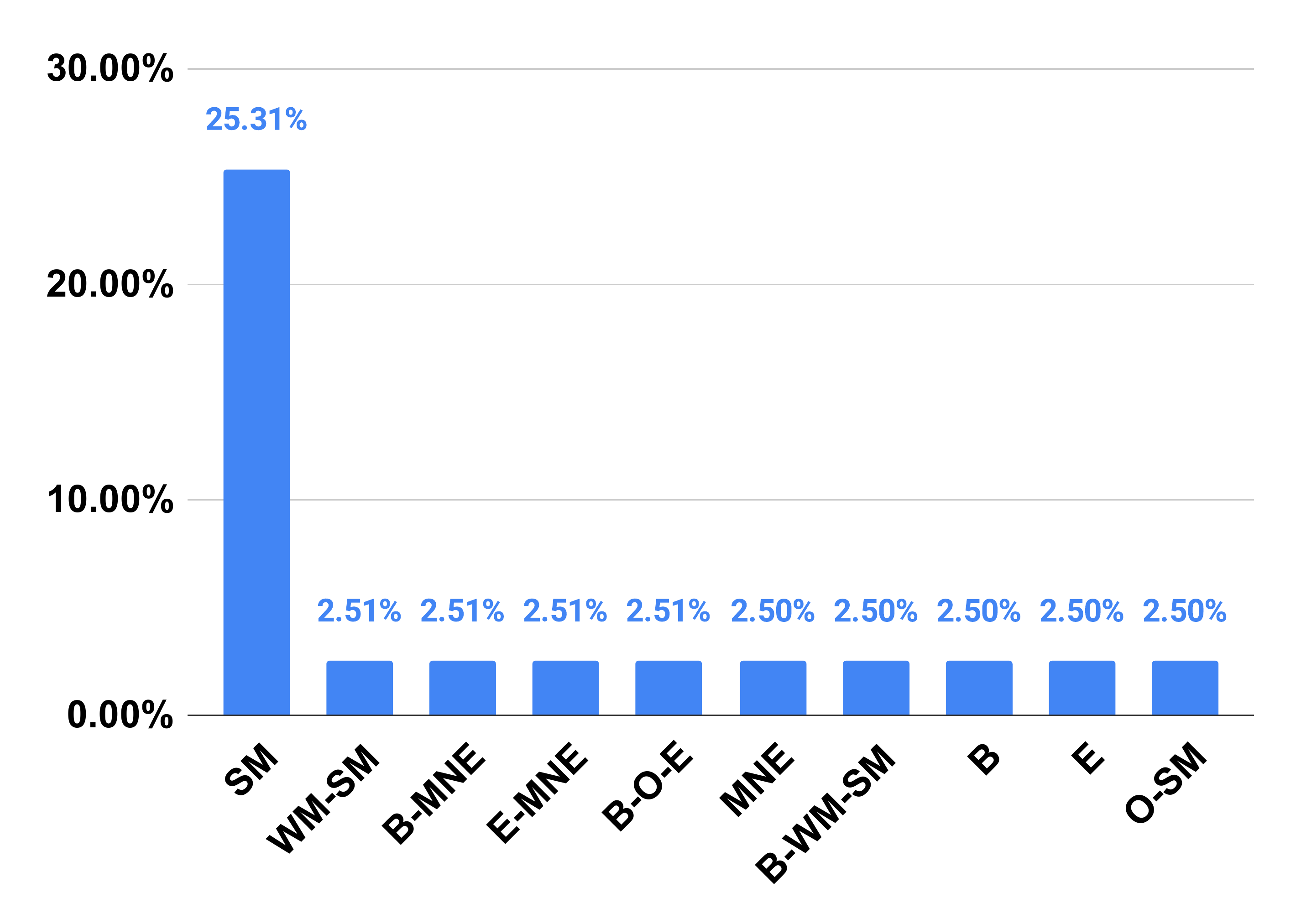} 
    \end{subfigure}%
    \begin{subfigure}[t]{0.5\textwidth}
        \centering
        \includegraphics[width=2.37in]{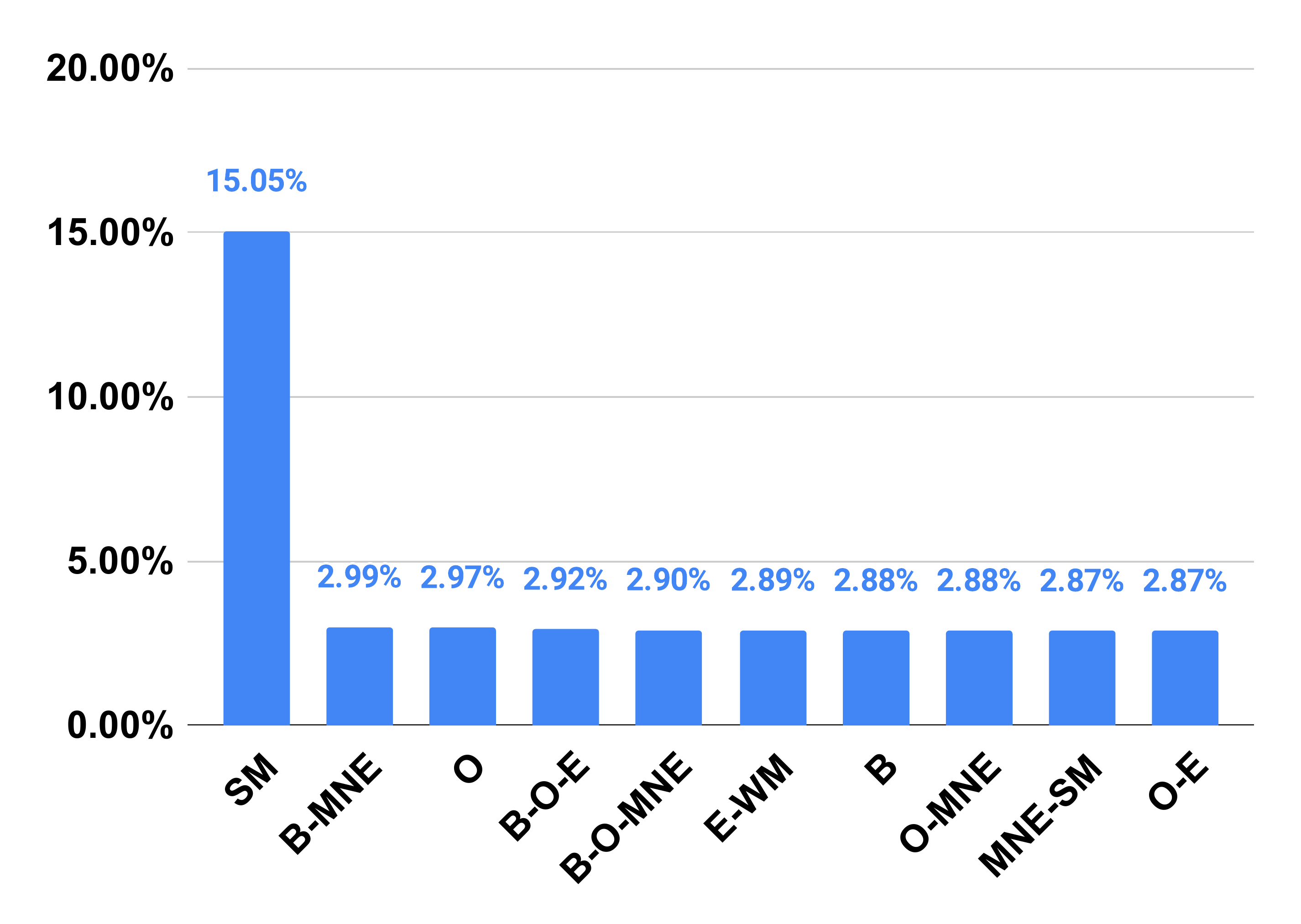} 
    \end{subfigure}%
  \caption{Top ten fitness function combinations chosen by \textbf{DSG-Sarsa} approach for the Strong Mutation goal. SM = Strong Mutation Coverage, B = Branch Coverage, O = Output Coverage, MNE = Method (No Exception), WM = Weak Mutation Coverage, E = Exception Count}
  \label{fig:SarsaSMchoices}
\end{figure}

\begin{figure}[!t]
\centering
   \begin{subfigure}[t]{0.5\textwidth}
        \centering
        \includegraphics[width=2.37in]{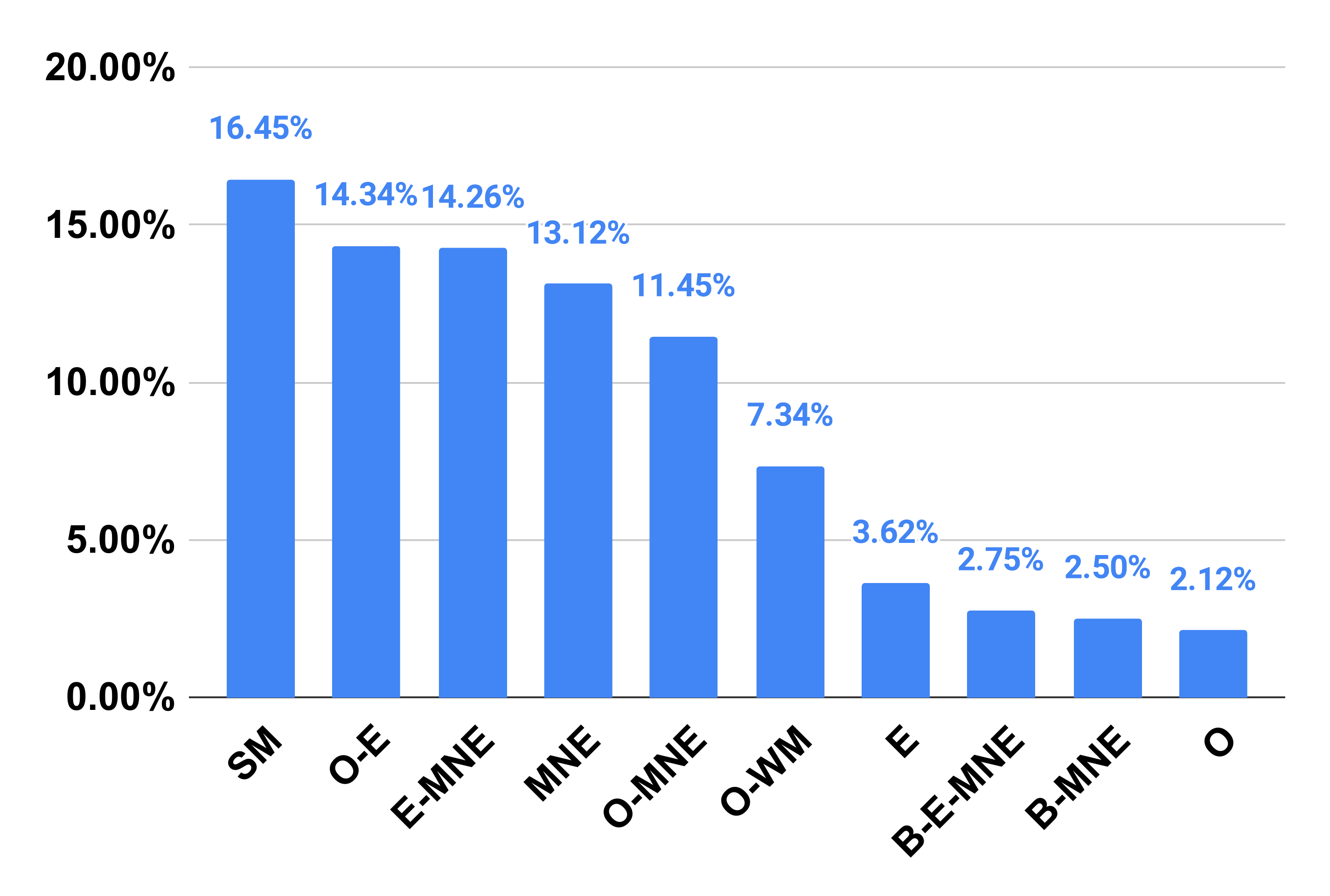} 
    \end{subfigure}%
    \begin{subfigure}[t]{0.5\textwidth}
        \centering
        \includegraphics[width=2.37in]{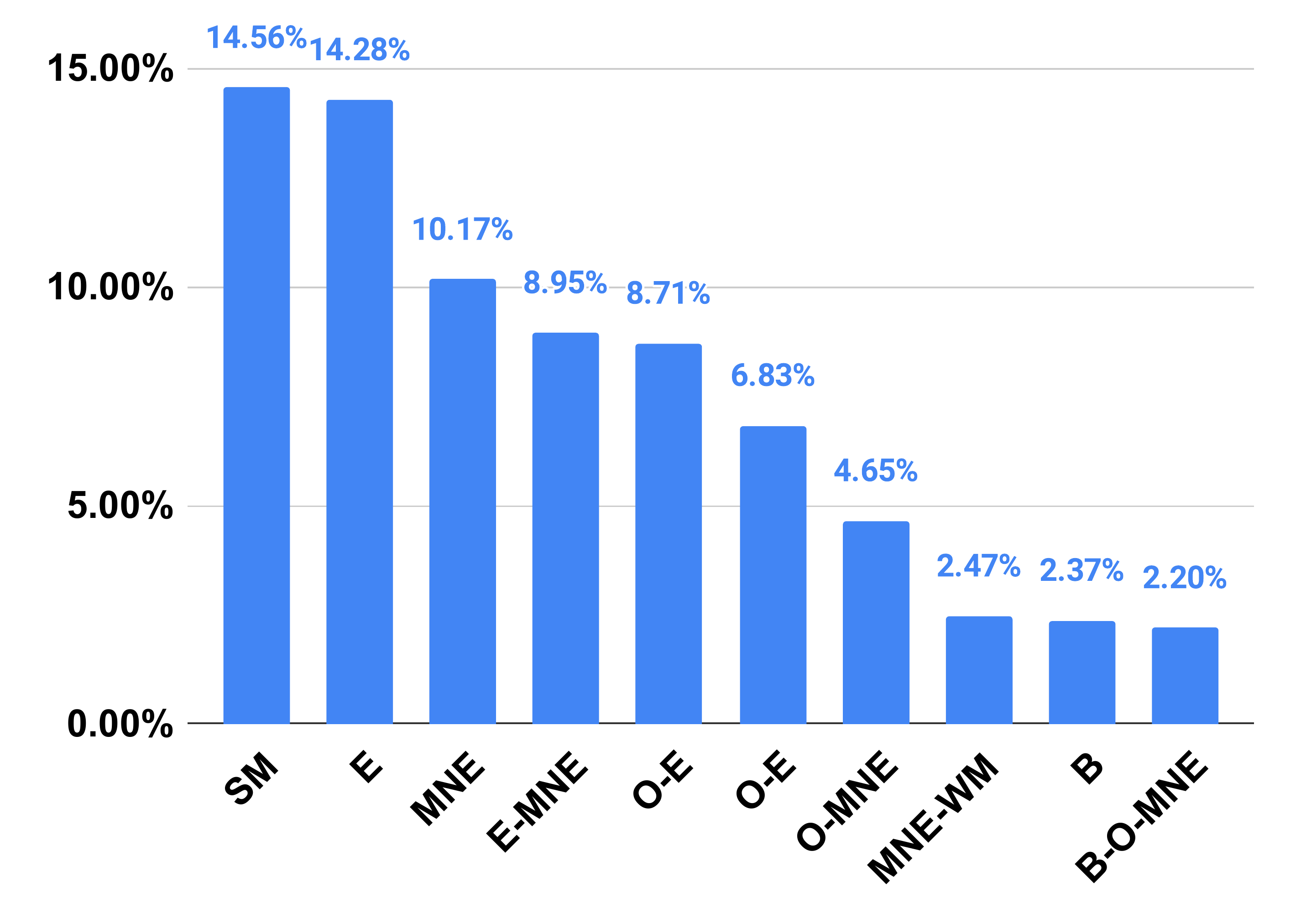} 
    \end{subfigure}%
    
    \begin{subfigure}[t]{0.5\textwidth}
        \centering
        \includegraphics[width=2.37in]{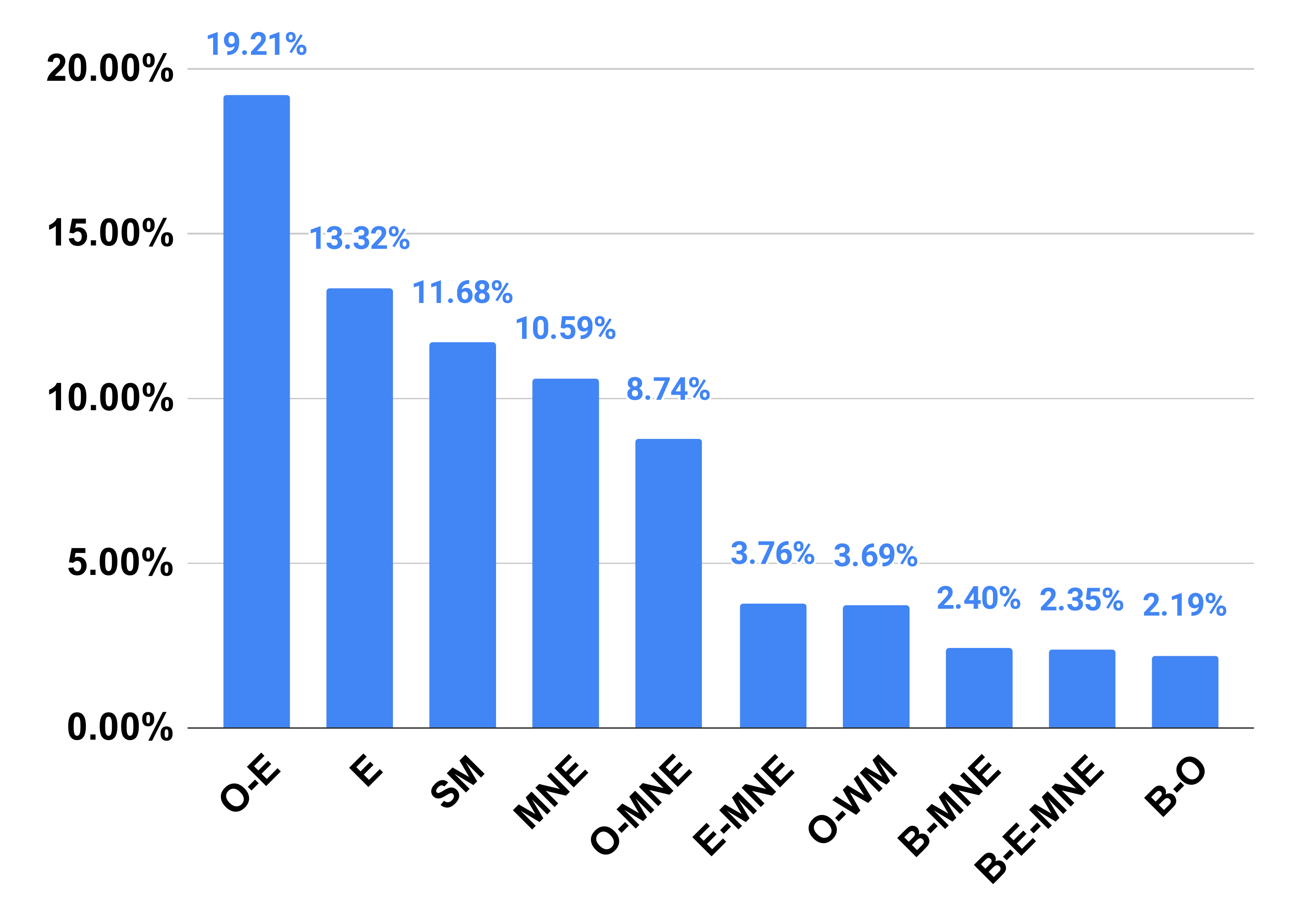} 
    \end{subfigure}%
    \begin{subfigure}[t]{0.5\textwidth}
        \centering
        \includegraphics[width=2.37in]{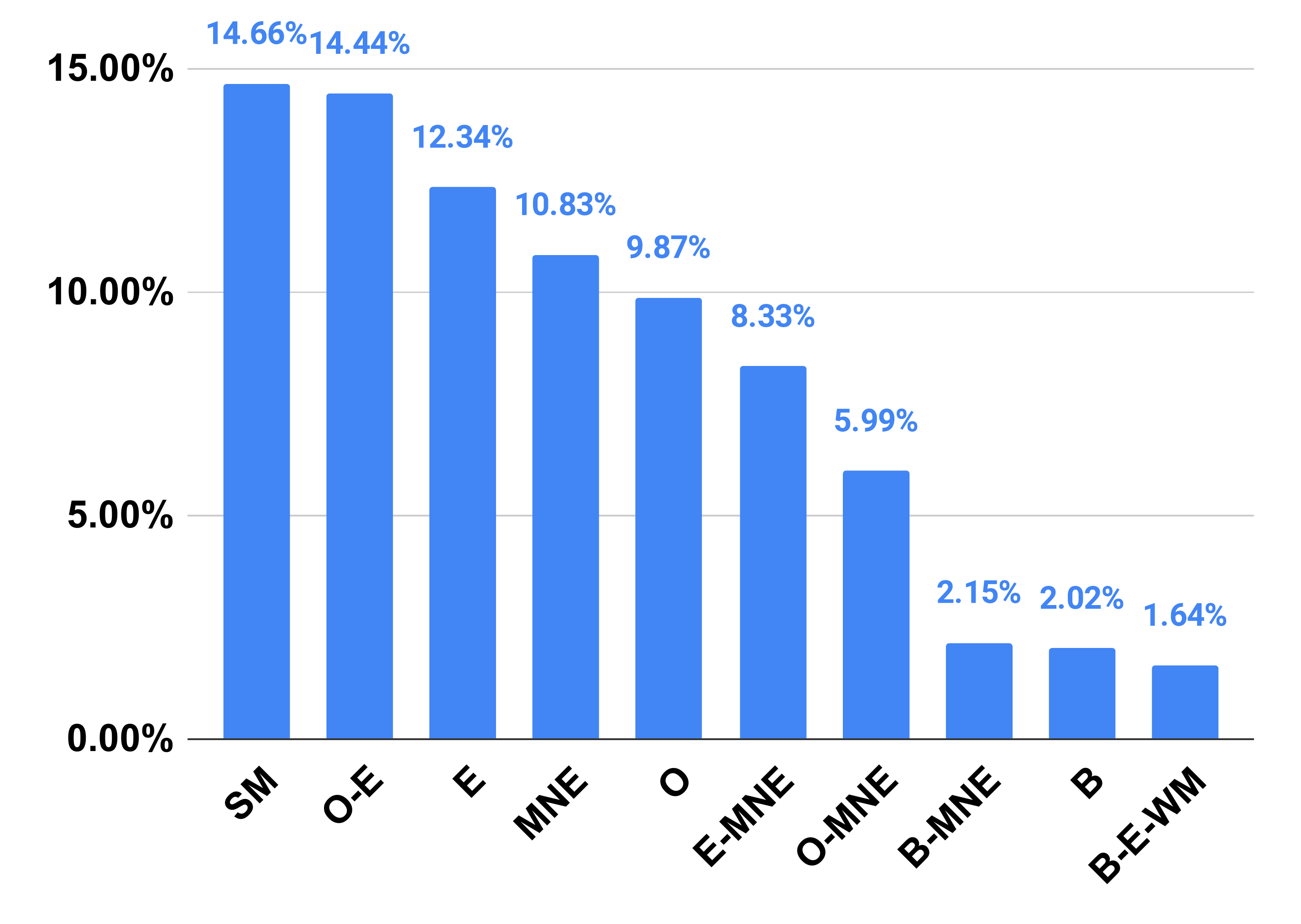} 
    \end{subfigure}%
    
    \begin{subfigure}[t]{0.5\textwidth}
        \centering
        \includegraphics[width=2.37in]{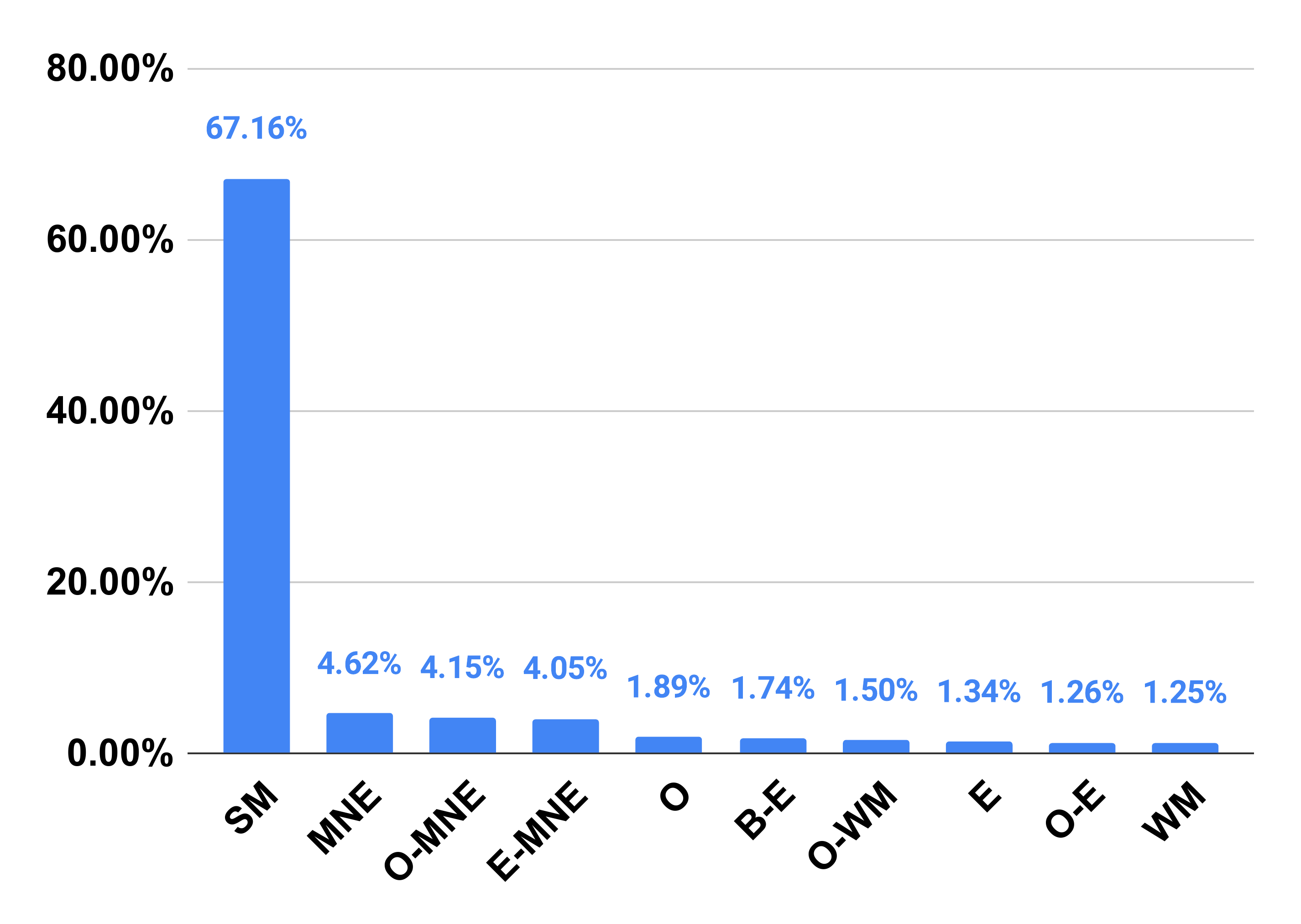} 
    \end{subfigure}%
    \begin{subfigure}[t]{0.5\textwidth}
        \centering
        \includegraphics[width=2.37in]{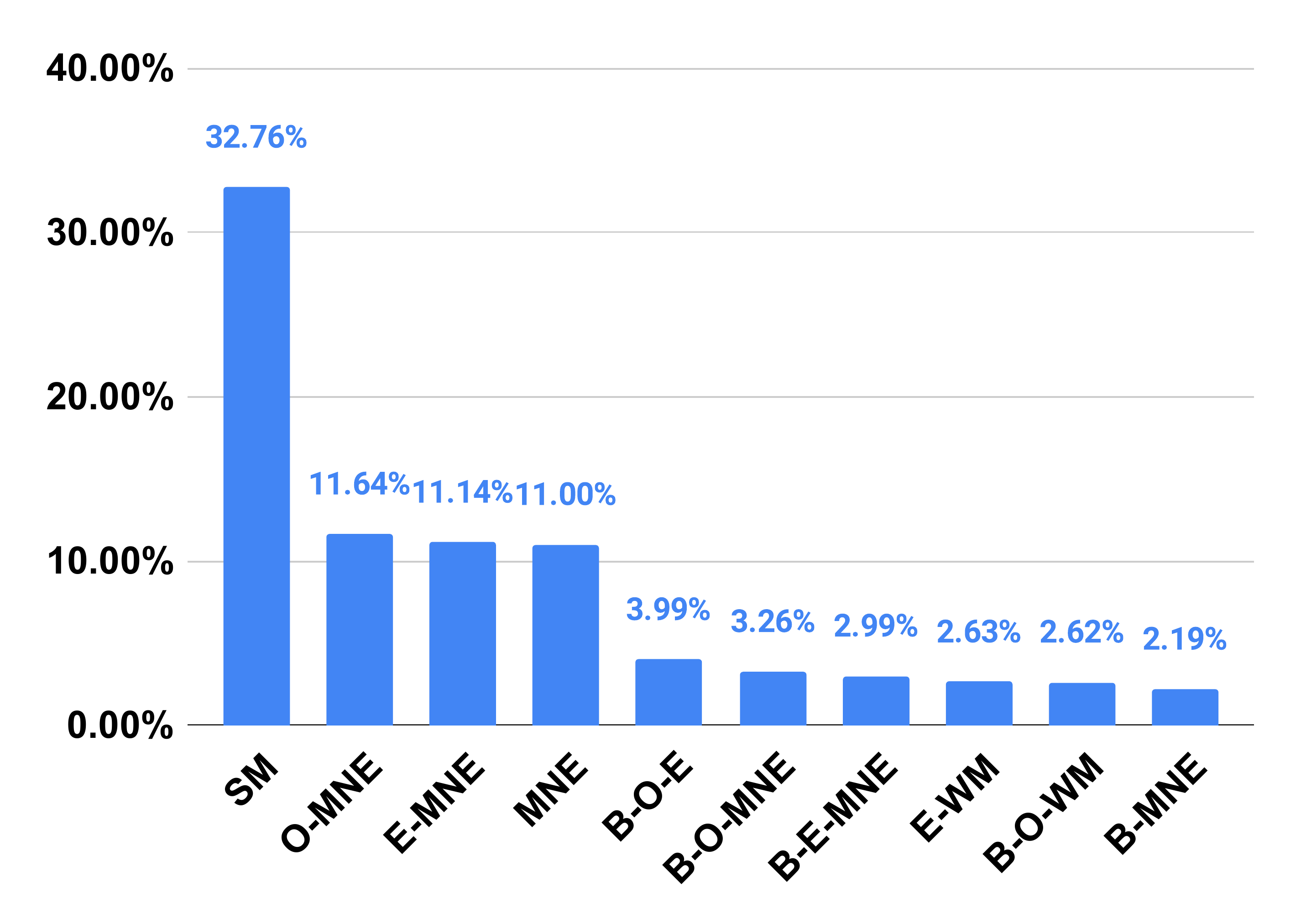} 
    \end{subfigure}%
  \caption{Top ten fitness function combinations chosen by \textbf{UCB} approach for the Strong Mutation goal. SM = Strong Mutation Coverage, B = Branch Coverage, O = Output Coverage, MNE = Method No Exception, WM = Weak Mutation Coverage, E = Exception Count}
  \label{fig:UCBSMchoices}
\end{figure}

In Figure~\ref{fig:SarsaSMchoices}, we show the top fitness function combinations chosen by DSG-Sarsa for the Strong Mutation Coverage goal. In Figure~\ref{fig:UCBSMchoices}, we do the same for UCB. For the first two goals, we saw that UCB more heavily favored exploitation of a particular action than DSG-Sarsa, which tended towards more exploration. Here, we see the reverse. DSG-Sarsa tends to choose the Strong Mutation fitness function far more often than any other option. UCB chooses Strong Mutation alone as the most common option for three of the six projects, but it spends more time exploring alternative options than DSG-Sarsa. In this case, UCB still attains slightly better median goal attainment, so DSG-Sarsa does not gain an advantage from heavier exploitation over exploration.

We see further evidence for the idea that none of the chosen fitness functions for this goal provide feedback that is sufficient to attain significant gains in Strong Mutation Coverage. The fitness function already designed for this goal, despite attaining relatively low levels of coverage, is still one of the most common optimization targets chosen by AFFS.

Weak Mutation Coverage appears often as well in the most common options, and may help improve coverage of the stronger variant. Output Coverage and the exception count also appear frequently. Both offer potential means to improve Strong Mutation Coverage, as both have a direct center around manipulation of output. Output Coverage increases the diversity of output response types, which may increase the likelihood of noticing a fault in that output. Similarly, encouraging triggering of exceptions may also increase the likelihood of a visible failure.

\begin{center}
\begin{framed}
Choices made by AFFS approaches suggest that no fitness functions significantly improved Strong Mutation Coverage. However, Output Coverage and the exception count both manipulate program output, and may lead to small improvements in Strong Mutation Coverage.
\end{framed}
\end{center}


\section{Discussion}\label{sec:disc}

In this section, we will summarize results across all three goals and discuss the impact of AFFS on multiple aspects of the test generation process.

\subsection{Impact of AFFS on Goal Attainment}

Given a high-level testing goal with no known effective fitness function or a function that is difficult to optimize, our core hypothesis was that adaptive fitness function selection would result in greater attainment of that goal than optimizing the existing fitness function. We further hypothesized that the use of AFFS may result in even greater goal attainment than the optimization of a static set of fitness functions. 

For two of the three studied testing goals, both hypotheses were confirmed, with at least medium effect size. Both EvoSuiteFIT techniques discover and retain more exception-triggering input than the baseline techniques, with DSG-Sarsa yielding better results. Additionally, both EvoSuiteFIT techniques produce more diverse test suites than static fitness function choices, with UCB outperforming DSG-Sarsa. 

For the goal of Strong Mutation Coverage, no technique demonstrates statistically significant improvements. When the search budget is a fixed number of generations rather than time, both AFFS techniques slightly outperformed the baselines in medium performance, but effect sizes remain negligible.  Given some additional time for test generation, we see some potential for improvement from using AFFS over static approaches. However, these improvements were limited.

For the first two goals in particular, these results indicate the potential of AFFS for performing test generation for difficult-to-optimize goals. In the future, we plan to explore the utility of AFFS for other goals and other types of testing. Reflecting on the experimental results, we can make the following observations:

\textbf{AFFS is an appropriate technique to apply when an effective fitness function does not already exist for the targeted goal.} In Section~\ref{sec:intro}, we gave the example of Branch Coverage as a goal with an effective fitness function, the branch distance. The branch distance offers clear guidance to the generation process and the means to attain high coverage over many classes.\footnote{As an aside, it is possible that AFFS could improve Branch Coverage in situations where the branch distance is not informative enough to guide the search, and some learned combination of fitness functions would help. We conducted a small pilot study to examine this question. In a small number of situations, AFFS did attain higher coverage than targeting Branch Coverage directly. However, in the majority of cases, targeting Branch Coverage directly yielded better results due to the overhead of reinforcement learning. Hence, our recommendation is to apply AFFS when an informative fitness function does not already exist, or if that function is known not to work well for a particular problem instance.} It is unlikely that AFFS would offer improved goal attainment, as other fitness functions are unlikely to offer additional feedback sufficient to overcome the introduced overhead of reinforcement learning. Rather, AFFS can help improve goal attainment in situations where the existing fitness function offers no or little feedback to improve fitness, like the exception count. AFFS enables the discovery of more exceptions by guiding test generation towards, for example, exploration of CUT structure. 

AFFS can also help in situations where a fitness function offers misleading feedback. Consider the Levenshtein distance used in promoting suite diversity. This fitness function rewards test suites that differ from each other, but does not assist in suggesting \textit{how} to instill this difference. We observed that targeting this function alone resulted in uncontrolled test suite growth, without a correspondingly large gain in diversity. As this fitness function grows more expensive to calculate as suites get larger, the feedback from this function actually \textit{harmed} the search process---limiting the number of generations of evolution possible during the ten-minute search budget. Rather than offering helpful feedback, the function actually led the algorithm astray. AFFS was able to produce more diverse test suites and---by keeping the test suite smaller---was substantially faster.

\textbf{AFFS requires a reward function that is fast to calculate, or requires additional time for test generation.} Reinforcement learning is an additional step in the generation algorithm. No matter how efficient it is, it will add some overhead to the process absent in the normal course of test generation. This overhead can be overcome by strategic selection of fitness functions. AFFS can be faster than the ``default'' multi-function combinations simply by virtue of calculating fewer fitness functions. However, the reward function must still be fast to calculate to gain the full benefits of using the approach. For three of the six projects studied in the Strong Mutation experiment, both AFFS techniques are significantly slower than optimizing Strong Mutation alone due to the overhead of calculating Strong Mutation Coverage as part of both fitness and reward. In cases where selecting a faster reward function is not possible, more time should be given to the test generation process. 

\textbf{The effect of AFFS is limited by the span of fitness functions available to choose from.} AFFS can only offer feedback to the search if some combination of the functions it can choose from actually offers the missing feedback. This was the case for two of our three goals. For Strong Mutation Coverage, limited improvement in median performance and variance indicate that the considered fitness functions had limited impact on goal attainment. Other functions---still unknown---may improve attainment of that goal, but there is no guarantee that such functions exist.

One may take from this the lesson that they should add as many options as possible for reinforcement learning to choose from. \textit{This is not the case.} Reinforcement learning must try and retry options, continually refining its estimations of which will best improve goal attainment. Reinforcement learning will see faster convergence and better results with fewer options to choose from to start. With too many options, AFFS will spend most of the search trying potentially suboptimal options without ever discovering the best ones. For all the goals, we actively removed some combinations that we knew or suspected would be suboptimal before even starting the experiments. We would recommend a similar process for additional testing goals---\textit{start by pruning functions and combinations that you suspect will produce weak results}.  

\subsection{Impact of AFFS on Fault Detection}

Fault detection is not a simple matter of maximizing some function, but of selecting the exact input that will trigger an observable failure~\cite{Gay15:risks}. The likelihood of fault detection is influenced by a number of factors~\cite{Gay18:fitness}. The exact relationship of those factors is not well understood, and detecting a fault is often more of a matter of blind luck than deliberate manipulation of test suites. Still, a major goal of test generation---and a major reason that we target many of these fitness functions---is to increase the likelihood that we detect faults with the generated test suites. Maximizing Branch Coverage is not the actual end goal of a tester. Rather, it is a measurable factor that may increase our likelihood of detecting a fault. Thus, it is important to examine the impact that AFFS has on fault detection. 

For the exception discovery goal, both EvoSuiteFIT techniques detect faults missed by the other techniques. UCB is able to detect more faults than all other approaches for the Strong Mutation goal, while DSG-Sarsa is outperformed by the baselines. However, when the number of generations is fixed, both AFFS approaches outperform the baselines. For the diversity goal, the random baseline outperformed all other approaches. The AFFS approaches, however, outperformed both of the other baselines. 

AFFS approaches can detect more faults than optimizing static baselines. However, this is not guaranteed. Higher goal attainment does not lead always lead to improved fault detection, and can actually mean the opposite if goal attainment does not actually have a positive correlation with the likelihood of fault detection. We do not fully understand the impact of AFFS on fault detection yet, and will examine it more closely in future work. However, we have observed several factors that may lead to a higher likelihood of fault detection. 

\textbf{AFFS results in higher attainment of goals thought to have a positive relationship with fault detection.} AFFS clearly results in improved attainment of exception discovery and test suite diversity. If hypotheses about these goals are correct, we would expect an increase in the likelihood of fault detection as well. We do see this in the exception experiment, but not in the diversity experiment. The calculated correlation coefficients for all three goals do not indicate strong connections between goal attainment and fault detection in this experiment. Still, it is a factor that may contribute to improvements in fault detection.

\textbf{Optimizing multiple fitness functions results in multifaceted test suites.} Each fitness function optimized will have an impact on the resulting test suite, shaping the test cases towards possessing the properties embodied by that fitness function. Naturally, then, optimizing multiple fitness functions can result in test suites that are multifaceted and better able to detect faults~\cite{Rojas15:Combining,Gay18:fitness}. This is not universally the case, and requires careful selection of fitness functions~\cite{Gay17:Combos}. However, this is indicated by the significant improvement in fault detection between single-function and multi-function approaches in our experiments.

\textbf{Optimizing too many fitness functions at once can introduce conflicts between functions and reduce attainment of individual functions.} Optimizing a naively-chosen combination of fitness functions can have a detrimental impact on the resulting test suite. The goals of some fitness functions will conflict with the goals of others. Optimizing one fitness function may come at a significant cost in attainment of another. EvoSuite combines the scores of fitness functions into a single score, and will favor a test suite that highly maximizes one function over a test suite that carefully balances two functions at low levels of attainment. The default combination represents a naive combination of several functions, and there may be conflicts between some of those functions. By intelligently selecting smaller combinations of functions, AFFS may better avoid such conflicts.

\textbf{Changing fitness functions as the suite evolves may result in better test suites.} AFFS is able to respond to the evolving state of the population of test suites, choosing fitness functions that are best able to improve goal attainment given the current state. This means that certain fitness functions may be applied at certain stages of test generation, but not others. This may be a better method of producing multifaceted test suites than statically applying the same fitness functions the entire time. Rather, we may see a staggered approach, where certain properties are evolved into the test suite at different stages of evolution. This may be more effective than trying to imbue many properties at once.

\subsection{Impact of Reinforcement Learning Overhead}

Reinforcement learning introduces overhead into the test generation process. As test generation is generally conducted using a fixed period of time, this overhead could result in a reduction of the number of generations of evolution that can be conducted during this period of time. If this reduction is significant, goal attainment could be reduced as well. 

\textbf{The ability to avoid calculation of unhelpful fitness functions mitigates reinforcement learning overhead.} For both exception discovery and diversity, both AFFS approaches are able to complete more generations of evolution during the search budget than the default combination. An important factor in the number of generations that can be completed is the cost of computing fitness. The more fitness functions to be calculated, the longer each generation takes. The default combination naively combines several fitness functions, some of which are likely unhelpful. The AFFS approaches learn to avoid calculating unhelpful functions, achieving speed gains that overcome the introduced overhead. 

\textbf{Feedback from effective fitness functions can help control computational costs.} The diversity fitness function grows more expensive to calculate as test case length and suite size grows. By incorporating feedback from additional fitness functions, AFFS is able to prevent uncontrolled test suite growth. As a result, it is actually faster than optimizing diversity alone, as test suites grow rapidly when diversity is the sole fitness function. 

\textbf{Expensive reward functions negatively impact AFFS.} For three of the six projects examined in the Strong Mutation experiment, both AFFS techniques are significantly slower than optimizing Strong Mutation alone due to the overhead of calculating Strong Mutation Coverage as part of both fitness and reward. When we hold the number of generations at a fixed value instead of time, AFFS is more effective. In this situation, the overhead reduces the potential positive impact of AFFS. In such cases, either a less expensive reward function should be used or more time should be allocated to AFFS.

\subsection{Actions Selected by AFFS}

The ability to adjust the fitness functions at regular intervals allows EvoSuiteFIT to make strategic choices that refine the test suite. We can see this from examining the actions chosen by UCB and DSG-Sarsa as they attempt to maximize goal attainment. We can make two key observations in this area.

\textbf{AFFS enables deeper understanding of the properties that improve goal attainment and how fitness functions can imbue those properties.} The combination of Branch Coverage, exception count, and diversity score seems particularly effective at improving test suite diversity. Ahead of time, we did not know that these three specific functions would enable diversity when used together. Individually, none of these are as effective as they are in combination. These three functions each offer feedback to each other, enabling greater diversity when used in combination. Other  function combinations similarly act in concert to improve suite diversity. AFFS enabled the discovery of these serendipitous combinations.   

Similarly, the choices made by AFFS suggest that no fitness function combination provided feedback needed to significantly improve Strong Mutation Coverage. However, Output Coverage and the exception count both encourage deviations in program output, and may lead to small improvements in Strong Mutation Coverage. Ahead of time, we did not understand their potential impact on attainment of Strong Mutation coverage, but inspecting the choices made by AFFS gave us insight into factors that could promote additional attainment of our goal.

\textbf{Fitness function combinations that are ineffective in a static context may be effective when used by AFFS to diversify a pre-evolved population of suites.} Many of the most common choices made by AFFS---particularly for the exception discovery and diversity goals---would result in poor test suites when used as the only fitness functions for the entire generation process. For example, the combination of exception count and Method Coverage (Top-Level, No Exception) was chosen very often for the exception discovery goal. Used in a static context, the produced suites are quite weak at both goal attainment (discovering exceptions) and fault detection. However, this combination is applied strategically by AFFS to suites evolved already using other functions, such as Branch Coverage. The suites are already robust at, for example, covering the code structure. Then, these combinations can be applied to reshape the suites into ones that discover new exceptions. A similar observation can be made in the other experiments. The diversity score is used quite a lot to shape existing suites, when it is a poor target in a static context. In the Strong Mutation experiment, Output Coverage and exception count offer some gain in coverage, but would yield weak coverage if used as the sole targets of generation. 

Observation of the choices made by AFFS makes it clear how the stateful evolution of test suites can be harnessed to improve goal attainment. Fitness functions shape the test suites that emerge from search-based test generation. They imbue the suites with certain emphasized properties. \textit{These properties do not need to be imbued at the same time}. Rather, fitness functions can be used to reshape a suite over time, and different functions may be best applied in different sequences or at different stages of this evolution. A future direction for this research will be to further understand this process, and how it can best be controlled to produce effective test suites. Little research in search-based test generation has looked at the controlled staggering of fitness functions, but our observations indicate the potential importance it has.

\subsection{Choice of Reinforcement Learning Approach}

We implemented two reinforcement learning approaches, UCB and DSG-Sarsa. These approaches use different mechanisms for choosing actions and associating actions with particular states. It is natural, then, to compare the two in terms of their performance. In this regard, we can make the following observation: \textbf{Overall, UCB attains a slight advantage over DSG-Sarsa. However, there are significant exceptions that rule out universal recommendation of UCB.}

In terms of goal attainment, UCB attains higher median performance for the diversity and Strong Mutation goals than DSG-Sarsa, while the reverse is true for the diversity goal. In terms of fault detection, UCB outperforms DSG-Sarsa for both the exception discovery and Strong Mutation Coverage goals. DSG-Sarsa attains better fault detection for the diversity goal, even though UCB attains better coverage. Finally, in terms of speed, UCB is faster for the diversity and Strong Mutation goals, but slower than DSG-Sarsa for the exception goal. 

We lack enough evidence to recommend one approach over the other. UCB attains a slight lead in multiple categories, but is outperformed by DSG-Sarsa in enough cases to rule out an unqualified recommendation. Overall, both approaches appear useful, and more observations will be needed to make any sort of conclusive judgement. Given the success of the two approaches, it may even make sense to execute both and pool their test cases.

\subsection{When AFFS Harms Goal Attainment}

While we observed that AFFS \textit{generally} enables greater attainment of testing goals, there are times where it not only fails to improve attainment---it actively attains worse results than all of the baselines. To gain a greater perspective on the limitations of AFFS, we manually examined situations where either DSG-Sarsa or UCB attained worse results in all, or almost all, trials than the single fitness function baseline. 

We focus on the \textit{diversity} goal in this analysis, as it offers the clearest performance differentials between AFFS and the single-function baseline. We identified the ten faults where AFFS techniques attained the worst results relative to the diversity score alone. We examined the classes-under-test and the generated suites, and attempted to identify factors that explain the worse performance of AFFS. 

For the diversity goal, the faults where AFFS attained the worst performance relative to the diversity score baseline were, in order, Gson-12, Chart-24, Lang-25, Math-35, Lang-55, Math-34, Closure 39, Math 56, Math 89, and Mockito-12. Examining the classes and tests generated for this fault, we observed two primary factors limiting performance of AFFS---the first being a factor that can affect any goal, and the second being specific to the diversity goal.

\smallskip\noindent\textbf{Fitness functions are merged into a single score:} In EvoSuite's genetic algorithm, all selected fitness functions are merged into a single score. This means that large improvements to a single function will be accepted, even if they come with a small drop in another fitness function. This creates potential conflicts between fitness functions, and allows particular functions to be dominated if they are difficult to optimize in comparison to other functions, or if they do not rise in conjunction with another function. 

In some cases, diversity and functions like code coverage may rise in conjunction with each other. For example, increasing diversity may also increase the attained code coverage. In such cases, using code coverage as one of the fitness functions may offer feedback that is not offered by optimizing diversity alone. 

In other cases, input diversity may have little impact on the code coverage. For example, only a small set of values may cause control-flow to take different paths, or only a small number of methods might accept a large selection of input values. In these cases, the ``best'' test suites may be those that cover a large span of the code, even if they lack diversity. The reward function used by AFFS will still encourage some diversity, but its impact may be lessened because the fitness functions employed prioritize large gains in coverage over diversity, and those fitness functions are the ones that ultimately evolve the test suites. 

Chart-24 offers an example of this, where the test suites generated by AFFS call a wider variety of methods than those generated targeting diversity alone. However, the latter apply a wider range of input to a smaller set of methods. The suites generated by AFFS may be ``better''. They cover more of the code, and could detect faults missed by the other suites. At the same time, they are ``worse'' with regard to the goal of diversity.

This factor can impact the results of AFFS for not just diversity, but for other testing goals as well. The success of AFFS relies on the existence of fitness functions that can improve attainment of our goal of interest. In cases where we can identify those functions, AFFS works well. However, if we cannot identify fitness functions that improve goal attainment, the end result may be worse than just trying to optimize the existing weak fitness function for that goal. In many cases, AFFS was able to identify such functions. However, for some classes, there may be no effective fitness functions for increasing diversity. 

\smallskip\noindent\textbf{Methods with limited or no input:} In multiple cases, the CUT had a large number of methods where there were no, or limited, means of interacting through input parameters. Consider, for example:
\begin{itemize}
\setlength{\itemsep}{1pt}
  \setlength{\parskip}{0pt}
  \setlength{\parsep}{0pt}  
    \item \textbf{Gson-12:} The CUT parses elements from a Json structure. The only ``input'' provided is to the constructor. EvoSuite cannot generate arbitrary Json input, so it provides an empty file to the constructor. The other methods of the class interact with this structure, and many have no input parameters.
    \item \textbf{Lang-55:} The CUT is a stopwatch. The constructor initializes the object, and it can be interacted with through methods that stop, pause, and reset the stopwatch. There are no input parameters. 
    \item \textbf{Math-34:} The CUT represents the population of a genetic algorithm in list form. The constructor initializes the population, and the methods can be used to analyze or interact with that population. Many methods have no input (e.g., getting the fittest chromosome or getting a list of chromosomes), and their result depends on the contents of the population. The primary means of introducing diversity through input are when controlling the size of the population, i.e., a method with numeric input. 
\end{itemize}

One of the primary means of improving diversity is to provide a wider range of input to method parameters. When methods lack parameters, gaining diversity becomes more difficult. Instead, other means of gaining diversity must be employed, including increasing the diversity of output---the generated assertion statements include the output of calling methods that offer concrete output, generating tests that call highly different sequences of input, and triggering exceptions---as the \texttt{try/catch} block included in the test case will give the test a very different body than many other tests. However, it can be difficult to ``discover'' test cases that exploit these routes to diversity. 

\begin{figure}[!t]
	\centering
	\begin{lstlisting}[language=Java,basicstyle={\scriptsize\ttfamily}]
  public void test0()  throws Throwable  {
      ElitisticListPopulation elitisticListPopulation0 = 
          new ElitisticListPopulation(208, 0.0);
      elitisticListPopulation0.addChromosome((Chromosome) null);
      assertEquals(208, elitisticListPopulation0.getPopulationLimit());
  }

  public void test2()  throws Throwable  {
      ElitisticListPopulation elitisticListPopulation0 = 
          new ElitisticListPopulation(208, 0.0);
      elitisticListPopulation0.getChromosomeList();
      assertEquals(208, elitisticListPopulation0.getPopulationLimit());
  }

  public void test4()  throws Throwable  {
      LinkedList<Chromosome> linkedList0 = new LinkedList<Chromosome>();
      ElitisticListPopulation elitisticListPopulation0 = 
          new ElitisticListPopulation(linkedList0, 1135, 0.0);
      // Undeclared exception!
      try { 
        elitisticListPopulation0.setPopulationLimit((-1485));
        fail("Expecting exception: IllegalArgumentException");
      
      } catch(IllegalArgumentException e) {
         //
         // population limit has to be positive
         //
         verifyException("org.apache.commons.math3.genetics.ListPopulation", 
             e);
      }
  }
	\end{lstlisting}
	\caption{Two similar test cases generated by DSG-Sarsa for Math-34, followed by one generated when targeting only diversity.}
	\label{fig:math34}
\end{figure}

In these cases, other fitness functions---i.e., code coverage---will dominate the diversity fitness function. As a result, for these classes, AFFS tends to generate a large number of highly-similar test cases, while targeting diversity-alone yields a small number of test cases that are very different from each other. Consider, for example, the first two test cases in Figure~\ref{fig:math34}. These two tests were among those generated by DSG-Sarsa for the CUT. The test suite contains many of this form, where the first and third lines (set-up, and assertion on the output) are identical. The only difference is the second line, where different methods are called. These two test cases cover different parts of the code, but are not very different in terms of the resulting diversity score. Test like these are added to the suite because they have a small positive impact on diversity, but---more importantly---because they have a large impact on other fitness functions like the code coverage. This impact comes more easily than improvements in diversity, and has a greater impact on the resulting test suite.

In contrast, targeting diversity alone prioritizes test cases like the third one in Figure~\ref{fig:math34}---longer test cases where exceptions are thrown and diversity is introduced through the available method calls with input parameters. As a result, the suites generated by AFFS are less diverse than those generated targeting diversity-only. Again, the suites generated by AFFS may be better for fault detection, but they are technically worse for the stated ``goal'' of the tester. 

\section{Threats to Validity}\label{sec:threats}

\noindent\textbf{External Validity:} Our study has focused on six systems (seven for the diversity goal)---a relatively small number. Nevertheless, we believe that such systems are representative of, at minimum, other small to medium-sized Java systems. We believe that Defects4J offers enough fault examples that our results are generalizable to other, sufficiently similar, projects. As Defects4J is used across multiple research fields, the use of this dataset also allows comparisons of our approach with other research, and allows others to replicate our experiments.

We have implemented our reinforcement learning techniques in a single test generation framework. There are many search-based methods of generating tests and these methods may yield different results. Unfortunately, no other generation framework offers the same number and variety of fitness functions. Therefore, a more thorough comparison of tool performance cannot be made at this time. By using the same framework to generate all test suites, we can compare our approach to the baselines on an equivalent basis.

Similarly, we have chosen two reinforcement learning algorithms to implement, out of the many that have been proposed. We chose these two specifically because (a) they are well-understood and widely-used, and (b) they represent different approaches to handling state (tabular versus approximate). Because these approaches have substantial differences in how they work, we believe we present a reasonable portrait of how AFFS would work. Still, different reinforcement learning techniques may lead to different outcomes.

To control experiment cost, we only generated ten test suites for each combination of fault, budget, and configuration. It is possible that larger sample sizes may yield different results. However, given the consistency of our results, we believe that this is a sufficient number of repetitions to draw stable conclusions.

\smallskip\noindent\textbf{Conclusion Validity:} When using statistical analyses, we have attempted to ensure the base assumptions behind these analyses are met. We have favored non-parametric methods, as distribution characteristics are not generally known a priori, and normality cannot be assumed. 
\section{Conclusions}\label{sec:conclusion}

Search-based test generation is guided by feedback from one or more fitness functions. Choosing informative fitness functions is crucial to meeting the goals of a tester. Unfortunately, many goals---such as forcing the class-under-test to throw exceptions, increasing test suite diversity, and attaining Strong Mutation Coverage---\textit{do not} have effective fitness function formulations. We propose that meeting such goals requires treating fitness function identification as a secondary optimization step. An \textit{adaptive} algorithm that can vary the selection of fitness functions could adjust its selection throughout the generation process to maximize goal attainment, based on the current population of test suites. To test this hypothesis, we have implemented two reinforcement learning algorithms in the EvoSuite framework, and used these algorithms to dynamically set the fitness functions used during generation for the three goals identified above. 

We have evaluated EvoSuiteFIT for each of our three goals on a set of Java case examples in terms of the ability of generated test suites to achieve the targeted goal and in terms of the ability of the generated suites to detect faults. In each case, we compare the two reinforcement learning approaches to a set of baselines. 

We have found that both EvoSuiteFIT techniques outperform all baselines with at least medium effect size for the goals of exception discovery and suite diversity---attaining improvements of up to 107.14\% in goal attainment. For the goal of Strong Mutation Coverage, no technique demonstrates significant improvements. When the search budget is a fixed number of generations rather than time, both EvoSuiteFIT techniques slightly outperform the baselines (up to 8.33\% improvement). However, the effect size is still negligible. 

Additionally, both EvoSuiteFIT techniques detect faults missed by the other techniques for the exception discovery goal (up to 259.90\% improvement). UCB is able to detect more faults for the Strong Mutation goal (12.50\% improvement), and when the number of generations is fixed, both EvoSuiteFIT approaches outperform the baselines (up to 50.00\% improvement). Both techniques are outperformed by the random baseline for the diversity goal (34.74\% difference), but outperform the other baselines. Improvements in fault detection may arise because of higher attainment of these goals, optimizing multiple fitness functions---but avoiding needlessly complex and conflicting functions---and changing fitness functions as the suite evolves. However, higher goal attainment does not ensure fault detection.

We find that AFFS is an appropriate technique to apply when an effective fitness function does not already exist for the targeted goal. However, AFFS requires a reward function that is fast to calculate, or requires additional time for test generation. Further, the effect of AFFS is limited by the span of fitness functions available to choose from. If none of the chosen functions correlate to the goal of interest, then improvements in goal attainment will be limited.

While reinforcement learning adds overhead to test generation, EvoSuiteFIT is often \textit{faster} than the default static configuration because the ability to avoid calculation of unhelpful fitness functions mitigates this overhead. Further, feedback from effective fitness functions can help control computational costs. Additionally, the ability to adjust the fitness functions at regular intervals allows EvoSuiteFIT to make strategic choices that refine the test suite and allows us to attain a deeper understanding of the properties that link to goal attainment and how fitness functions can work together to imbue those properties. Fitness function combinations that are ineffective in a static context may be effective when used by AFFS to diversify a pre-evolved population of suites.

The use of AFFS allows EvoSuiteFIT to identify combinations of fitness functions effective at achieving our testing goals, and strategically vary that set of functions throughout the ongoing generation process. We hypothesize that other goals without known effective fitness function representations could also be maximized in a similar manner. We make EvoSuiteFIT available to others for use in test generation research or practice. 

In future work, we plan apply AFFS to new goals and testing scenarios (e.g., system testing) and integrate it into metaheuristic algorithms beyond standard Genetic Algorithms. We also will perform expanded empirical studies to better understand the relationship between AFFS and fault detection and how the staggered application of fitness functions can improve goal attainment and suite effectiveness. We will also explore the generation of new fitness functions---i.e., a generative hyperheuristic rather than a selective one---and how learned policies can be transferred to new classes and systems. Finally, we will examine the application of AFFS to multiple high-level goals simultaneously. 

\bibliographystyle{spmpsci}
\bibliography{main}

\begin{thebibliography}{10}
\providecommand{\url}[1]{{#1}}
\providecommand{\urlprefix}{URL }
\expandafter\ifx\csname urlstyle\endcsname\relax
  \providecommand{\doi}[1]{DOI~\discretionary{}{}{}#1}\else
  \providecommand{\doi}{DOI~\discretionary{}{}{}\begingroup
  \urlstyle{rm}\Url}\fi

\bibitem{Pezze06:testing}
Pezze, M., Young, M.: Software Test and Analysis: Process, Principles, and
  Techniques.
\newblock John Wiley and Sons (2006)

\bibitem{Harman13:oraclesurvey}
Barr, E., Harman, M., McMinn, P., Shahbaz, M., Yoo, S.: The oracle problem in
  software testing: A survey.
\newblock IEEE Transactions on Software Engineering \textbf{41}(5), 507--525
  (2015).
\newblock \doi{10.1109/TSE.2014.2372785}

\bibitem{Anand13:Orchestrated}
Anand, S., Burke, E.K., Chen, T.Y., Clark, J., Cohen, M.B., Grieskamp, W.,
  Harman, M., Harrold, M.J., McMinn, P.: An orchestrated survey of
  methodologies for automated software test case generation.
\newblock Journal of Systems and Software \textbf{86}(8), 1978--2001 (2013)

\bibitem{McMinn04:SBTesting}
McMinn, P.: Search-based software test data generation: A survey.
\newblock Software Testing, Verification and Reliability \textbf{14}, 105--156
  (2004)

\bibitem{Harman01:SBSE}
Harman, M., Jones, B.: Search-based software engineering.
\newblock Journal of Information and Software Technology \textbf{43}, 833--839
  (2001)

\bibitem{Gay19:fitness}
Salahirad, A., Almulla, H., Gay, G.: Choosing the fitness function for the job:
  Automated generation of test suites that detect real faults.
\newblock Software Testing, Verification and Reliability \textbf{29}(4-5),
  e1701 (2019).
\newblock \doi{10.1002/stvr.1701}.
\newblock
  \urlprefix\url{https://onlinelibrary.wiley.com/doi/abs/10.1002/stvr.1701}.
\newblock E1701 stvr.1701

\bibitem{Arcuri13:Normalize}
Arcuri, A.: It really does matter how you normalize the branch distance in
  search-based software testing.
\newblock Software Testing, Verification and Reliability \textbf{23}(2),
  119--147 (2013)

\bibitem{Robillard00:DRJ}
Robillard, M.P., Murphy, G.C.: Designing robust java programs with exceptions.
\newblock In: Proceedings of the 8th ACM SIGSOFT International Symposium on
  Foundations of Software Engineering: Twenty-first Century Applications,
  SIGSOFT '00/FSE-8, pp. 2--10. ACM, New York, NY, USA (2000).
\newblock \doi{10.1145/355045.355046}.
\newblock \urlprefix\url{http://doi.acm.org/10.1145/355045.355046}

\bibitem{Neto18:Visual}
{De Oliveira Neto}, F.G., {Feldt}, R., {Erlenhov}, L., {Nunes}, J.B.D.S.:
  Visualizing test diversity to support test optimisation.
\newblock In: 2018 25th Asia-Pacific Software Engineering Conference (APSEC),
  pp. 149--158 (2018)

\bibitem{Shahbazi15:Diversity}
Shahbazi, A.: Diversity-based automated test case generation.
\newblock Ph.D. thesis, University of Alberta (2015)

\bibitem{Lind7528953}
{Lindstrom}, B., {Mrki}, A.: On strong mutation and subsuming mutants.
\newblock In: 2016 IEEE Ninth International Conference on Software Testing,
  Verification and Validation Workshops (ICSTW), pp. 112--121 (2016)

\bibitem{Rojas15:Combining}
Rojas, J.M., Campos, J., Vivanti, M., Fraser, G., Arcuri, A.: Combining
  multiple coverage criteria in search-based unit test generation.
\newblock In: M.~Barros, Y.~Labiche (eds.) Search-Based Software Engineering,
  \emph{Lecture Notes in Computer Science}, vol. 9275, pp. 93--108. Springer
  International Publishing (2015).
\newblock \doi{10.1007/978-3-319-22183-0_7}.
\newblock \urlprefix\url{http://dx.doi.org/10.1007/978-3-319-22183-0_7}

\bibitem{Gay17:ICST}
Gay, G.: The fitness function for the job: Search-based generation of test
  suites that detect real faults.
\newblock In: Proceedings of the International Conference on Software Testing,
  ICST 2017. IEEE (2017)

\bibitem{Gay17:Combos}
Gay, G.: Generating effective test suites by combining coverage criteria.
\newblock In: Proceedings of the Symposium on Search-Based Software
  Engineering, SSBSE 2017. Springer Verlag (2017)

\bibitem{Fraser14:Mutation}
Fraser, G., Arcuri, A.: Achieving scalable mutation-based generation of whole
  test suites.
\newblock Empirical Software Engineering \textbf{20}(3), 783--812 (2014)

\bibitem{Papadakis2013}
Papadakis, M., Malevris, N.: Searching and generating test inputs for mutation
  testing.
\newblock SpringerPlus \textbf{2}(1), 121 (2013).
\newblock \doi{10.1186/2193-1801-2-121}.
\newblock \urlprefix\url{https://doi.org/10.1186/2193-1801-2-121}

\bibitem{Jia15:HHICSE}
Jia, Y., Cohen, M.B., Harman, M., Petke, J.: Learning combinatorial interaction
  test generation strategies using hyperheuristic search.
\newblock In: Proceedings of the 37th International Conference on Software
  Engineering - Volume 1, ICSE '15, pp. 540--550. IEEE Press, Piscataway, NJ,
  USA (2015).
\newblock \urlprefix\url{http://dl.acm.org/citation.cfm?id=2818754.2818821}

\bibitem{Guizzo15:HMI}
Guizzo, G., Fritsche, G.M., Vergilio, S.R., Pozo, A.T.R.: A hyper-heuristic for
  the multi-objective integration and test order problem.
\newblock In: Proceedings of the 2015 Annual Conference on Genetic and
  Evolutionary Computation, GECCO '15, pp. 1343--1350. ACM, New York, NY, USA
  (2015).
\newblock \doi{10.1145/2739480.2754725}.
\newblock \urlprefix\url{http://doi.acm.org/10.1145/2739480.2754725}

\bibitem{Sutton2018}
Sutton, R.S., Barto, A.G.: Reinforcement learning: An introduction.
\newblock MIT press (2018)

\bibitem{Rojas17:wholesuite}
Rojas, J.M., Vivanti, M., Arcuri, A., Fraser, G.: A detailed investigation of
  the effectiveness of whole test suite generation.
\newblock Empirical Software Engineering \textbf{22}(2), 852--893 (2017).
\newblock \doi{10.1007/s10664-015-9424-2}.
\newblock \urlprefix\url{https://doi.org/10.1007/s10664-015-9424-2}

\bibitem{Gay20:RL}
Almulla, H., Gay, G.: Learning how to search: Generating exception-triggering
  tests through adaptive fitness function selection.
\newblock In: 13th IEEE International Conference on Software Testing,
  Validation and Verification (2020)

\bibitem{Gay20:DivRL}
Almulla, H., Gay, G.: Generating diverse test suites for {Gson} through
  adaptive fitness function selection.
\newblock In: Proceedings of the Symposium on Search-Based Software
  Engineering, SSBSE 2020. Springer Verlag (2020)

\bibitem{shamshiri15:generation}
Shamshiri, S., Just, R., Rojas, J.M., Fraser, G., McMinn, P., Arcuri, A.: Do
  automatically generated unit tests find real faults? an empirical study of
  effectiveness and challenges.
\newblock In: Proceedings of the 30th IEEE/ACM International Conference on
  Automated Software Engineering (ASE), ASE 2015. ACM, New York, NY, USA (2015)

\bibitem{Orso14:STR}
Orso, A., Rothermel, G.: Software testing: A research travelogue (2000--2014).
\newblock In: Proceedings of the on Future of Software Engineering, FOSE 2014,
  pp. 117--132. ACM, New York, NY, USA (2014).
\newblock \doi{10.1145/2593882.2593885}.
\newblock \urlprefix\url{http://doi.acm.org/10.1145/2593882.2593885}

\bibitem{Almasi17:IndustrialEval}
Almasi, M.M., Hemmati, H., Fraser, G., Arcuri, A., Benefelds, J.: An industrial
  evaluation of unit test generation: Finding real faults in a financial
  application.
\newblock In: Proceedings of the 39th IEEE/ACM International Conference on
  Software Engineering (ICSE)---Software Engineering in Practice Track (SEIP),
  ICSE 2017. ACM, New York, NY, USA (2017)

\bibitem{Ali10:SBST}
Ali, S., Briand, L.C., Hemmati, H., Panesar-Walawege, R.K.: A systematic review
  of the application and empirical investigation of search-based test case
  generation.
\newblock Software Engineering, IEEE Transactions on \textbf{36}(6), 742--762
  (2010)

\bibitem{Bianchi09:Optimization}
Bianchi, L., Dorigo, M., Gambardella, L., Gutjahr, W.: A survey on
  metaheuristics for stochastic combinatorial optimization.
\newblock Natural Computing \textbf{8}(2), 239--287 (2009).
\newblock \doi{10.1007/s11047-008-9098-4}.
\newblock \urlprefix\url{http://dx.doi.org/10.1007/s11047-008-9098-4}

\bibitem{dorigo1997ant}
Dorigo, M., Gambardella, L.M.: Ant colony system: a cooperative learning
  approach to the traveling salesman problem.
\newblock Evolutionary Computation, IEEE Transactions on \textbf{1}(1), 53--66
  (1997)

\bibitem{john1992adaptation}
Holland, J.H.: Adaptation in natural and artificial systems: an introductory
  analysis with applications to biology, control, and artificial intelligence.
\newblock MIT press (1992)

\bibitem{Feldt15:GA}
Feldt, R., Poulding, S.: Broadening the search in search-based software
  testing: It need not be evolutionary.
\newblock In: Search-Based Software Testing (SBST), 2015 IEEE/ACM 8th
  International Workshop on, pp. 1--7 (2015).
\newblock \doi{10.1109/SBST.2015.8}

\bibitem{Fraser13:AWT}
Fraser, G., Staats, M., McMinn, P., Arcuri, A., Padberg, F.: Does automated
  white-box test generation really help software testers?
\newblock In: Proceedings of the 2013 International Symposium on Software
  Testing and Analysis, ISSTA, pp. 291--301. ACM, New York, NY, USA (2013).
\newblock \doi{10.1145/2483760.2483774}.
\newblock \urlprefix\url{http://doi.acm.org/10.1145/2483760.2483774}

\bibitem{Malburg11:SBDSE}
Malburg, J., Fraser, G.: Combining search-based and constraint-based testing.
\newblock In: Proceedings of the 2011 26th IEEE/ACM International Conference on
  Automated Software Engineering, ASE '11, pp. 436--439. IEEE Computer Society,
  Washington, DC, USA (2011).
\newblock \doi{10.1109/ASE.2011.6100092}.
\newblock \urlprefix\url{http://dx.doi.org/10.1109/ASE.2011.6100092}

\bibitem{balera2019systematic}
Balera, J.M., de~Santiago~J{\'u}nior, V.A.: A systematic mapping addressing
  hyper-heuristics within search-based software testing.
\newblock Information and Software Technology \textbf{114}, 176--189 (2019)

\bibitem{drake2020recent}
Drake, J.H., Kheiri, A., {\"O}zcan, E., Burke, E.K.: Recent advances in
  selection hyper-heuristics.
\newblock European Journal of Operational Research \textbf{285}(2), 405--428
  (2020)

\bibitem{Kumari16:HHC}
Kumari, A.C., Srinivas, K.: Hyper-heuristic approach for multi-objective
  software module clustering.
\newblock Journal of Systems and Software \textbf{117}, 384 -- 401 (2016).
\newblock \doi{https://doi.org/10.1016/j.jss.2016.04.007}.
\newblock
  \urlprefix\url{http://www.sciencedirect.com/science/article/pii/S0164121216300231}

\bibitem{Burke2019Revisited}
Burke, E.K., Hyde, M.R., Kendall, G., Ochoa, G., {\"O}zcan, E., Woodward, J.R.:
  A Classification of Hyper-Heuristic Approaches: Revisited, pp. 453--477.
\newblock Springer International Publishing, Cham (2019).
\newblock \doi{10.1007/978-3-319-91086-4_14}.
\newblock \urlprefix\url{https://doi.org/10.1007/978-3-319-91086-4_14}

\bibitem{Katehakis87:Bandit}
Katehakis, M.N., Veinott~Jr, A.F.: The multi-armed bandit problem:
  decomposition and computation.
\newblock Mathematics of Operations Research \textbf{12}(2), 262--268 (1987)

\bibitem{Jia15:HHSBST}
Jia, Y.: Hyperheuristic search for sbst.
\newblock In: Proceedings of the Eighth International Workshop on Search-Based
  Software Testing, SBST '15, pp. 15--16. IEEE Press, Piscataway, NJ, USA
  (2015).
\newblock \urlprefix\url{http://dl.acm.org/citation.cfm?id=2821339.2821343}

\bibitem{Gay14:coverage}
Gay, G., Whalen, M.W., Heimdahl, M.P., Staats, M.: The risks of coverage
  directed test case generation.
\newblock Currently under submission, draft available from
  http://greggay.com/pdf/14risks.pdf (2014)

\bibitem{Sutton98:Reinforcement}
Sutton, R.S., Barto, A.G.: Reinforcement learning: An introduction, vol.~1.
\newblock MIT press Cambridge (1998)

\bibitem{Busoniu11:Approx}
{Buşoniu}, L., {Ernst}, D., {De Schutter}, B., {Babuška}, R.: Approximate
  reinforcement learning: An overview.
\newblock In: 2011 IEEE Symposium on Adaptive Dynamic Programming and
  Reinforcement Learning (ADPRL), pp. 1--8 (2011).
\newblock \doi{10.1109/ADPRL.2011.5967353}

\bibitem{SBST17_competition}
Fraser, G., Arcuri, A.: Evosuite at the sbst 2017 tool competition.
\newblock In: 10th International Workshop on Search-Based Software Testing
  (SBST'17) at ICSE'17, pp. 39--42 (2017)

\bibitem{ssbse18_tutorial}
Fraser, G.: A tutorial on using and extending the evosuite search-based test
  generator.
\newblock In: Search-Based Software Engineering, pp. 106--130. Springer (2018)

\bibitem{Gay18:DBC}
Gay, G.: To call, or not to call: Contrasting direct and indirect branch
  coverage in test generation.
\newblock In: Proceedings of the 11th International Workshop on Search-Based
  Software Testing, SBST 2018. ACM, New York, NY, USA (2018)

\bibitem{Alshahwan14:OutputCoverage}
Alshahwan, N., Harman, M.: Coverage and fault detection of the
  output-uniqueness test selection criteria.
\newblock In: Proceedings of the 2014 International Symposium on Software
  Testing and Analysis, ISSTA 2014, pp. 181--192. ACM, New York, NY, USA
  (2014).
\newblock \doi{10.1145/2610384.2610413}.
\newblock \urlprefix\url{http://doi.acm.org/10.1145/2610384.2610413}

\bibitem{Iqbal19:transfer}
Iqbal, M.S., Kotthoff, L., Jamshidi, P.: Transfer learning for performance
  modeling of deep neural network systems.
\newblock In: 2019 $\{$USENIX$\}$ Conference on Operational Machine Learning
  (OpML 19), pp. 43--46 (2019)

\bibitem{Navarro01:strings}
Navarro, G.: A guided tour to approximate string matching.
\newblock ACM Comput. Surv. \textbf{33}(1), 31–88 (2001).
\newblock \doi{10.1145/375360.375365}.
\newblock \urlprefix\url{https://doi.org/10.1145/375360.375365}

\bibitem{Crawford2013:Tuning}
Crawford, B., Soto, R., Monfroy, E., Palma, W., Castro, C., Paredes, F.:
  Parameter tuning of a choice-function based hyperheuristic using particle
  swarm optimization.
\newblock Expert Systems with Applications \textbf{40}(5), 1690 -- 1695 (2013).
\newblock \doi{https://doi.org/10.1016/j.eswa.2012.09.013}.
\newblock
  \urlprefix\url{http://www.sciencedirect.com/science/article/pii/S0957417412010676}

\bibitem{Ochoa09:HH}
Ochoa, G., Vazquez-Rodriguez, J.A., Petrovic, S., Burke, E.: Dispatching rules
  for production scheduling: A hyper-heuristic landscape analysis.
\newblock In: 2009 IEEE Congress on Evolutionary Computation, pp. 1873--1880
  (2009).
\newblock \doi{10.1109/CEC.2009.4983169}

\bibitem{Zamli_Alkazemi_2016}
Zamli, K.Z., Alkazemi, B.Y., Kendall, G.: A tabu search hyper-heuristic
  strategy for t-way test suite generation.
\newblock Appl. Soft Comput. \textbf{44}(C), 57–74 (2016).
\newblock \doi{10.1016/j.asoc.2016.03.021}.
\newblock \urlprefix\url{https://doi.org/10.1016/j.asoc.2016.03.021}

\bibitem{Zamli17:HHCIT}
Zamli, K.Z., Din, F., Kendall, G., Ahmed, B.S.: An experimental study of
  hyper-heuristic selection and acceptance mechanism for combinatorial t-way
  test suite generation.
\newblock Information Sciences \textbf{399}, 121 -- 153 (2017).
\newblock \doi{https://doi.org/10.1016/j.ins.2017.03.007}.
\newblock
  \urlprefix\url{http://www.sciencedirect.com/science/article/pii/S0020025517305820}

\bibitem{din8004298}
{Din}, F., {Alsewari}, A.R.A., {Zamli}, K.Z.: A parameter free choice function
  based hyper-heuristic strategy for pairwise test generation.
\newblock In: 2017 IEEE International Conference on Software Quality,
  Reliability and Security Companion (QRS-C), pp. 85--91 (2017)

\bibitem{Din_zamli_2018}
Din, F., Zamli, K.Z.: Hyper-heuristic based strategy for pairwise test case
  generation.
\newblock Advanced Science Letters \textbf{24}(10), 7333--7338 (2018).
\newblock \doi{doi:10.1166/asl.2018.12938}.
\newblock
  \urlprefix\url{https://www.ingentaconnect.com/content/asp/asl/2018/00000024/00000010/art00068}

\bibitem{Ahmed_2020}
Ahmed, B.S., Enoiu, E., Afzal, W., Zamli, K.Z.: An evaluation of monte
  carlo-based hyper-heuristic for interaction testing of industrial embedded
  software applications.
\newblock Soft Computing  (2020).
\newblock \doi{10.1007/s00500-020-04769-z}.
\newblock \urlprefix\url{http://dx.doi.org/10.1007/s00500-020-04769-z}

\bibitem{Guizzo15:HHITO}
Guizzo, G., Vergilio, S.R., Pozo, A.T.R.: Evaluating a multi-objective
  hyper-heuristic for the integration and test order problem.
\newblock In: 2015 Brazilian Conference on Intelligent Systems (BRACIS), pp.
  1--6 (2015).
\newblock \doi{10.1109/BRACIS.2015.11}

\bibitem{Guizzo_Vergilio_2017}
Guizzo, G., Vergilio, S.R., Pozo, A.T., Fritsche, G.M.: A multi-objective and
  evolutionary hyper-heuristic applied to the integration and test order
  problem.
\newblock Appl. Soft Comput. \textbf{56}(C), 331–344 (2017).
\newblock \doi{10.1016/j.asoc.2017.03.012}.
\newblock \urlprefix\url{https://doi.org/10.1016/j.asoc.2017.03.012}

\bibitem{Guizzo2017AHF}
Guizzo, G., Bazargani, M., Paix{\~a}o, M., Drake, J.H.: A hyper-heuristic for
  multi-objective integration and test ordering in google guava.
\newblock In: SSBSE (2017)

\bibitem{Mariani_Guizzo_2016}
Mariani, T., Guizzo, G., Vergilio, S.R., Pozo, A.T.: Grammatical evolution for
  the multi-objective integration and test order problem.
\newblock In: Proceedings of the Genetic and Evolutionary Computation
  Conference 2016, GECCO ’16, p. 1069–1076. Association for Computing
  Machinery, New York, NY, USA (2016).
\newblock \doi{10.1145/2908812.2908816}.
\newblock \urlprefix\url{https://doi.org/10.1145/2908812.2908816}

\bibitem{Guizzo2018APS}
Guizzo, G., Vergilio, S.R.: A pattern-driven solution for designing
  multi-objective evolutionary algorithms.
\newblock Natural Computing pp. 1--14 (2018)

\bibitem{Ferreira7895294}
{Ferreira}, T.N., {Lima}, J.A.P., {Strickler}, A., {Kuk}, J.N., {Vergilio},
  S.R., {Pozo}, A.: Hyper-heuristic based product selection for software
  product line testing.
\newblock IEEE Computational Intelligence Magazine \textbf{12}(2), 34--45
  (2017)

\bibitem{Ferreira7744315}
{do Nascimento Ferreira}, T., {Kuk}, J.N., {Pozo}, A., {Vergilio}, S.R.:
  Product selection based on upper confidence bound moea/d-dra for testing
  software product lines.
\newblock In: 2016 IEEE Congress on Evolutionary Computation (CEC), pp.
  4135--4142 (2016)

\bibitem{STRICKLER20161232}
Strickler, A., {Prado Lima}, J.A., Vergilio, S.R., Pozo, A.T.: Deriving
  products for variability test of feature models with a hyper-heuristic
  approach.
\newblock Applied Soft Computing \textbf{49}, 1232 -- 1242 (2016).
\newblock \doi{https://doi.org/10.1016/j.asoc.2016.07.059}.
\newblock
  \urlprefix\url{http://www.sciencedirect.com/science/article/pii/S1568494616303994}

\bibitem{Filho3131152}
Filho, H.L.J., Lima, J.A.P., Vergilio, S.R.: Automatic generation of
  search-based algorithms applied to the feature testing of software product
  lines.
\newblock In: Proceedings of the 31st Brazilian Symposium on Software
  Engineering, SBES’17, p. 114–123. Association for Computing Machinery,
  New York, NY, USA (2017).
\newblock \doi{10.1145/3131151.3131152}.
\newblock \urlprefix\url{https://doi.org/10.1145/3131151.3131152}

\bibitem{Filho3266275}
Filho, H.L.J., Ferreira, T.N., Vergilio, S.R.: Multiple objective test set
  selection for software product line testing: Evaluating different
  preference-based algorithms.
\newblock In: Proceedings of the XXXII Brazilian Symposium on Software
  Engineering, SBES ’18, p. 162–171. Association for Computing Machinery,
  New York, NY, USA (2018).
\newblock \doi{10.1145/3266237.3266275}.
\newblock \urlprefix\url{https://doi.org/10.1145/3266237.3266275}

\bibitem{Filho8477803}
{Luiz Jakubovski Filho}, H., {Nascimento Ferreira}, T., {Regina Vergilio}, S.:
  Incorporating user preferences in a software product line testing
  hyper-heuristic approach.
\newblock In: 2018 IEEE Congress on Evolutionary Computation (CEC), pp. 1--8
  (2018)

\bibitem{Moghadam19:Perf}
{Helali Moghadam}, M., {Saadatmand}, M., {Borg}, M., {Bohlin}, M., {Lisper},
  B.: Machine learning to guide performance testing: An autonomous test
  framework.
\newblock In: 2019 IEEE International Conference on Software Testing,
  Verification and Validation Workshops (ICSTW), pp. 164--167 (2019)

\bibitem{bauersfeld2012reinforcement}
Bauersfeld, S., Vos, T.: A reinforcement learning approach to automated gui
  robustness testing.
\newblock In: Fast abstracts of the 4th symposium on search-based software
  engineering (SSBSE 2012), pp. 7--12 (2012)

\bibitem{Bauersfeld6494948}
{Bauersfeld}, S., {Vos}, T.E.J.: Guitest: a java library for fully automated
  gui robustness testing.
\newblock In: 2012 Proceedings of the 27th IEEE/ACM International Conference on
  Automated Software Engineering, pp. 330--333 (2012).
\newblock \doi{10.1145/2351676.2351739}

\bibitem{Grechanik2012}
{Grechanik}, M., {Fu}, C., {Xie}, Q.: Automatically finding performance
  problems with feedback-directed learning software testing.
\newblock In: 2012 34th International Conference on Software Engineering
  (ICSE), pp. 156--166 (2012)

\bibitem{joffe2019directing}
Joffe, L., Clark, D.: Directing a search towards execution properties with a
  learned fitness function.
\newblock IEEE (2019)

\bibitem{Romano2011AnAF}
Romano, D., Penta, M.D., Antoniol, G.: An approach for search based testing of
  null pointer exceptions.
\newblock 2011 Fourth IEEE International Conference on Software Testing,
  Verification and Validation pp. 160--169 (2011)

\bibitem{Goffi2931061}
Goffi, A., Gorla, A., Ernst, M.D., Pezz\`{e}, M.: Automatic generation of
  oracles for exceptional behaviors.
\newblock In: Proceedings of the 25th International Symposium on Software
  Testing and Analysis, ISSTA 2016, p. 213–224. Association for Computing
  Machinery, New York, NY, USA (2016).
\newblock \doi{10.1145/2931037.2931061}.
\newblock \urlprefix\url{https://doi.org/10.1145/2931037.2931061}

\bibitem{Blasi18:Spec}
Blasi, A., Goffi, A., Kuznetsov, K., Gorla, A., Ernst, M.D., Pezz\`{e}, M.,
  Castellanos, S.D.: Translating code comments to procedure specifications.
\newblock In: Proceedings of the 27th ACM SIGSOFT International Symposium on
  Software Testing and Analysis, ISSTA 2018, p. 242–253. Association for
  Computing Machinery, New York, NY, USA (2018).
\newblock \doi{10.1145/3213846.3213872}.
\newblock \urlprefix\url{https://doi.org/10.1145/3213846.3213872}

\bibitem{Nasser66299}
Albunian, N.M.: Diversity in search-based unit test suite generation.
\newblock In: T.~Menzies, J.~Petke (eds.) Search Based Software Engineering,
  pp. 183--189. Springer International Publishing, Cham (2017)

\bibitem{Feldt7515474}
{Feldt}, R., {Poulding}, S., {Clark}, D., {Yoo}, S.: Test set diameter:
  Quantifying the diversity of sets of test cases.
\newblock In: 2016 IEEE International Conference on Software Testing,
  Verification and Validation (ICST), pp. 223--233 (2016)

\bibitem{LinhaiMA2018162}
Ma, L., Wu, P., Chen, T.Y.: Diversity driven adaptive test generation for
  concurrent data structures.
\newblock Information and Software Technology \textbf{103}, 162 -- 173 (2018).
\newblock \doi{https://doi.org/10.1016/j.infsof.2018.07.001}.
\newblock
  \urlprefix\url{http://www.sciencedirect.com/science/article/pii/S0950584918301356}

\bibitem{vogel2019does}
Vogel, T., Tran, C., Grunske, L.: Does diversity improve the test suite
  generation for mobile applications?
\newblock In: International Symposium on Search Based Software Engineering, pp.
  58--74. Springer (2019)

\bibitem{Souza2016}
Souza, F.C.M., Papadakis, M., Le~Traon, Y., Delamaro, M.E.: Strong
  mutation-based test data generation using hill climbing.
\newblock In: Proceedings of the 9th International Workshop on Search-Based
  Software Testing, pp. 45--54. ACM (2016)

\bibitem{Dianxiang3395599}
Xu, D., Shrestha, R., Shen, N.: Automated strong mutation testing of xacml
  policies.
\newblock In: Proceedings of the 25th ACM Symposium on Access Control Models
  and Technologies, SACMAT ’20, p. 105–116. Association for Computing
  Machinery, New York, NY, USA (2020).
\newblock \doi{10.1145/3381991.3395599}.
\newblock \urlprefix\url{https://doi.org/10.1145/3381991.3395599}

\bibitem{Mark2025144}
Harman, M., Jia, Y., Langdon, W.B.: Strong higher order mutation-based test
  data generation.
\newblock In: Proceedings of the 19th ACM SIGSOFT Symposium and the 13th
  European Conference on Foundations of Software Engineering, ESEC/FSE ’11,
  p. 212–222. Association for Computing Machinery, New York, NY, USA (2011).
\newblock \doi{10.1145/2025113.2025144}.
\newblock \urlprefix\url{https://doi.org/10.1145/2025113.2025144}

\bibitem{Just14:Defects4J}
Just, R., Jalali, D., Ernst, M.D.: Defects4{J}: A database of existing faults
  to enable controlled testing studies for {J}ava programs.
\newblock In: Proceedings of the 2014 International Symposium on Software
  Testing and Analysis, ISSTA 2014, pp. 437--440. ACM, New York, NY, USA
  (2014).
\newblock \doi{10.1145/2610384.2628055}.
\newblock \urlprefix\url{http://doi.acm.org/10.1145/2610384.2628055}

\bibitem{Gay18:fitness}
Gay, G.: Choosing the fitness function for the job: Automated generation of
  test suites that detect real faults.
\newblock Under revision, Journal of Software Testing, Verification, and
  Reliability \textbf{X}(Y), 1--20 (2018).
\newblock Draft available from \url{http://greggay.com/pdf/18fitness.pdf}

\bibitem{Vargha00:Measure}
Vargha, A., Delaney, H.D.: A critique and improvement of the cl common language
  effect size statistics of mcgraw and wong.
\newblock Journal of Educational and Behavioral Statistics \textbf{25}(2),
  101--132 (2000).
\newblock \doi{10.3102/10769986025002101}.
\newblock \urlprefix\url{https://doi.org/10.3102/10769986025002101}

\bibitem{Gay15:risks}
Gay, G., Staats, M., Whalen, M., Heimdahl, M.: The risks of coverage-directed
  test case generation.
\newblock Software Engineering, IEEE Transactions on \textbf{PP}(99) (2015).
\newblock \doi{10.1109/TSE.2015.2421011}

\end{thebibliography}

\end{document}